\documentclass[twocolumn]{bmcart}

\def\includegraphics{}

\usepackage[english]{babel}

\usepackage{amsmath}
\usepackage{amssymb}
\usepackage{comment}
\usepackage{mathrsfs}
\usepackage{stmaryrd}
\usepackage[normalem]{ulem}
\usepackage{soul}
\usepackage{graphicx}
\graphicspath{{./figs/}}
\usepackage[inline]{enumitem}
\usepackage{lipsum}

\usepackage{tensind}
\tensordelimiter{?}

\usepackage{hyperref}
\hypersetup{
  colorlinks=true,        
  linkcolor=blue,         
  citecolor=cyan,         
}

\startlocaldefs

\newcommand{\msun}{M_\odot}

\newcommand{\ie}{i.e.,~}
\newcommand{\eg}{e.g.,~}
\newcommand{\cf}{cf.,~}
\newcommand{\plainie}{i.e.~}

\endlocaldefs

\begin{document}

\begin{frontmatter}

\begin{fmbox}

\dochead{Research}

\title{Entropy-limited hydrodynamics: a novel approach to relativistic
  hydrodynamics}

\author[
    addressref={aff1},
    corref={aff1},
    noteref={},
    email={guercilena@th.physik.uni-frankfurt.de}
]{\inits{FG}\fnm{Federico} \snm{Guercilena}}
\author[
    addressref={aff2,aff3},
    noteref={}
]{\inits{DR}\fnm{David} \snm{Radice}}
\author[
    addressref={aff1,aff4},
    noteref={}
]{\inits{LR}\fnm{Luciano} \snm{Rezzolla}}

\address[id=aff1]{
    \orgname{Institut f\"ur Theoretische Physik, Goethe Universit\"at},
    \street{Max-von-Laue-Str. 1},
    \postcode{60438}
    \city{Frankfurt am Main},
    \cny{Germany}
}
\address[id=aff2]{
    \orgname{Institute for Advanced Study},
    \street{1 Einstein Dr.},
    \postcode{08540}
    \city{Princeton, NJ},
    \cny{USA}
}
\address[id=aff3]{
    \orgname{Department of Astrophysical Sciences, Princeton University},
    \street{4 Ivy Lane},
    \postcode{08544}
    \city{Princeton, NJ},
    \cny{USA}
}
\address[id=aff4]{
    \orgname{Frankfurt Institute for Advanced Studies},
    \street{Ruth-Moufang-Str. 1},
    \postcode{60438}
    \city{Frankfurt am Main},
    \cny{Germany}
}

\begin{abstractbox}

\begin{abstract}
We present entropy-limited hydrodynamics (ELH): a new approach for the
computation of numerical fluxes arising in the discretization of
hyperbolic equations in conservation form. ELH is based on the
hybridisation of an unfiltered high-order scheme with the first-order
Lax-Friedrichs method. The activation of the low-order part of the scheme
is driven by a measure of the locally generated entropy inspired by the
artificial-viscosity method proposed by Guermond et al.
\cite{Guermond11}. Here, we present ELH in the context of high-order
finite-differencing methods and of the equations of general-relativistic
hydrodynamics. We study the performance of ELH in a series of classical
astrophysical tests in general relativity involving isolated, rotating
and nonrotating neutron stars, and including a case of gravitational
collapse to black hole. We present a detailed comparison of ELH with the
fifth-order monotonicity preserving method MP5 \cite{suresh_1997_amp},
one of the most common high-order schemes currently employed in
numerical-relativity simulations. We find that ELH achieves comparable
and, in many of the cases studied here, better accuracy than more
traditional methods at a fraction of the computational cost (up to
${\sim}\,50\,\%$ speedup). Given its accuracy and its simplicity of
implementation, ELH is a promising framework for the development of new
special- and general-relativistic hydrodynamics codes well adapted for
massively parallel supercomputers.
\end{abstract}


\end{abstractbox}

\end{fmbox}

\end{frontmatter}

\section{Introduction}
\label{sec:intro}

Large-scale general-relativistic hydrodynamical numerical simulations have been
shown to be a very powerful tool for the study of astrophysical systems
involving compact objects such as black holes and neutron stars \cite{Font08,
ShibataTaniguchilrr-2011-6, Rezzolla_book:2013, Marti2015, Shibata_book:2016,
Baiotti2016, Paschalidis2016}. The realisation of such simulations requires
dealing with very different physical, mathematical and computational issues.
One of the most challenging of such issues, and one that could lead to
significant differences on the outcome of resolution-limited simulations, is
the choice of the numerical method for the solution of the hydrodynamics
equations (one such difference would be \eg the dephasing of the gravitational
waveforms in binary neutron star merger simulations, where different numerical
schemes can lead to different dynamics of the matter bulk, see \eg
\cite{Baiotti2011,Read2013,Radice2013c,Thierfelder2011,Bernuzzi2012}).

The most commonly used methods in this context are collectively known as
high-resolution shock-capturing (HRSC) techniques (see
\cite{Rezzolla_book:2013} for an introduction). Belonging to this class,
which contains both finite-differences and finite-volumes schemes, are
\eg slope limiting methods (\eg \cite{Roe1985}), the piece-wise parabolic
method (PPM) \cite{Colella84,Marti96}, the fifth-order monotonicity
preserving (MP5) method \cite{suresh_1997_amp}, essentially/weighted
non-oscillatory (ENO/WENO) methods \cite{eno,Liu1994, Jiang1996,
  DePietri2016, Bernuzzi2016} and many others.

HRSC methods are very effective in dealing with shocks and suppressing
spurious oscillations, and have been employed with varying degree of
success in astrophysical simulations. In recent times much work has gone
into improving these schemes (\eg by innovative mesh refinement
techniques such as in \cite{DeBuhr2015}) or moving beyond them; one
promising alternative paradigm is that of discontinuous Galerkin methods
(see, \eg \cite{Radice2011, Bugner2015, Zanotti2015b, Kidder2016,
  Miller2016}). Many of such schemes, however, potentially suffer from a
few shortcomings: (i) they are complex to derive and implement, or to
extend and modify (\eg to increase the formal order of accuracy); (ii)
they often depend on many coefficients that require some degree of
optimisation (\eg the WENO methods); (iii) they can lead to load
imbalance in parallel implementations as a result of their complexity.

In this work we propose a different approach that is able to address some
of these shortcomings [especially the points (i) and (iii) previously
  mentioned], which we refer to as ``entropy-limited hydrodynamics''
(ELH) and formulate it in a finite-differences framework. The underlying
concept is relatively straightforward: the hydrodynamical fluxes are
computed using an unfiltered, high-order stencil, to which a contribution
from a low-order, stable numerical flux (the Lax-Friedrichs flux) is
added in order to ensure stability. To determine which gridpoints are in
need of the low-order contribution, we employ a ``shock detector'', which
not only marks region of the computational domain requiring the limiter,
but also determines the relative weights of the high and low-order
fluxes.

The use of a hybrid numerical flux to achieve both accuracy and stability
places our method in the class of flux-limiting schemes (see, \eg the
classification in \cite{Leveque92}), which have long been a feature in
the panorama of numerical schemes for hydrodynamics. In this context the
Lax-Friedrichs flux is a common choice for low-order methods, being
monotone and dissipative (a different example of combining high- and
low-order methods, in the context of the reconstruction method, would be
\eg \cite{tchekhovskoy_2007_wham}).

To drive the activation of the Lax-Friedrichs method, a criterion to flag
generically problematic points of the computational domain is
needed. Such a criterion is offered by the entropy viscosity function
described by Guermond et al. (we refer primarily to \cite{Guermond11},
but see also Refs. \cite{Guermond08,Zingan2013}), in which the local
production of entropy is used to identify shocks. Since entropy is
produced only in the presence of shocks, this results in a stable method
able to recover high-order in regions of smooth flow. We extend the
definition of the entropy viscosity from the classical to the
relativistic case, and employ it to drive the flux limiting scheme rather
than as a weight to additional viscous terms in the hydrodynamical
equations. As a result, and in contrast to the approach by Guermond et
al. \cite{Guermond11}, we do not modify the underlying equations
of relativistic hydrodynamics by introducing additional entropy-related
terms.

In the following we describe the method and the details of our
implementation, then report the results of tests we conducted in order to
gauge its behaviour against a standard HRSC method, namely, MP5
\cite{suresh_1997_amp}. The paper is structured as follows: in section
\ref{sec:overview} we briefly summarise the equations of relativistic
hydrodynamics and the finite-differences framework we employ. In section
\ref{sec:ELH} the ELH method and its implementation are described, while
the results of the numerical tests are collected in section
\ref{sec:tests}. We present our conclusions and outlook in section
\ref{sec:conclusions}.

In the following we use the spacetime signature $(-,+,+,+)$, with Greek
indices running from 0 to 3 and Latin indices from 1 to 3. We also employ
the Einstein convention for the summation over repeated indices. Unless
otherwise stated, all quantities are expressed in a geometrized system of
units in which $c=G=1$.

\section{Relativistic hydrodynamics: theory and numerics overview}
\label{sec:overview}
\medskip
\subsection{Relativistic hydrodynamics}

We summarize here the mathematical framework of relativistic
hydrodynamics. In the interest of simplicity, the discussion is limited
to special relativity, while the general-relativistic case, which is
relevant for the neutron-star tests of section \ref{sec:3Dtests}, can be
found in appendix \ref{sec:gr_hydrodynamics}.

Since most of our interest is in modelling neutron-star matter, we assume
it to be described by a perfect fluid, therefore with a corresponding
energy-momentum tensor given by
\begin{equation}
    ?T_\mu\nu? = \rho h u_\mu u_\nu + p?\eta_\mu\nu?\,,
    \label{eq:Tmunu}
\end{equation}
where $?\eta_\mu\nu?$ is the Minkowski metric, $\rho$ is the rest-mass
density, $u^\mu$ is the fluid four-velocity, $p$ is the pressure,
$h=1+\epsilon+p/\rho$ is the specific enthalpy and $\epsilon$ the
specific internal energy \cite{Rezzolla_book:2013}. The equations of
motion for the fluid are the conservation of the stress-energy tensor
\begin{equation}
    \partial_\mu ?T^\mu\nu? = 0\,,
    \label{eq:consTmunu}
\end{equation}
and conservation of rest mass
\begin{equation}
    \partial_\mu (\rho u^\mu) = 0\,,
    \label{eq:continuity}
\end{equation}
These two sets of equations are closed by an equation of state (EOS) $p =
p(\rho, \epsilon)$, and we here assume a simple ideal-fluid (or Gamma-law) EOS:
\begin{equation}
    p = \rho\epsilon(\Gamma-1)\,,
    \label{eq:Gammalaw}
\end{equation}
with $\Gamma$ the adiabatic index.

Equations \eqref{eq:consTmunu} and \eqref{eq:continuity} can be cast in
conservation form and therefore written symbolically as \cite{Marti2015}
\begin{equation}
    \partial_t \boldsymbol{U} + \partial_i \boldsymbol{F}^i = \boldsymbol{S}\,,
    \label{eq:valencia}
\end{equation}
by defining the conserved variables $\boldsymbol{U}$ as
\begin{align}
    \boldsymbol{U} &:=
    \left(
    \begin{array}{c}
    D \\ S_j \\ \tau
    \end{array}
    \right) \nonumber\\
    &:=\left(
    \begin{array}{c}
    \rho W \\ \rho h W^2 v_j \\ \rho h W^2-p-\rho W
    \end{array}
    \right)\,.
    \label{eq:conservatives}
\end{align}
These are functions of the ``primitive'' variables
$(\rho,v_i,p,\epsilon)$. In these expressions and in the
following the fluid three-velocity measured by the normal (or
Eulerian) observers is defined as $v^i := u^i/W$ and the Lorentz
factor is $W := (1 - v^i v_i)^{-\frac{1}{2}} = u^t$. The fluxes
$\boldsymbol{F}^i$ are
\begin{align}
    \boldsymbol{F}^i &=
    \left(
    \begin{array}{c}
    v^iD \\ S_jv^i+p\delta^i_j \\ S^i-Dv^i
    \end{array}
    \right)\,,
\end{align}
where $\delta^i_j$ is the Kronecker delta.

Note that the source functions $\boldsymbol{S}$ are identically zero in special
relativity, but this is no longer the case in a generic spacetime, where
metric-dependent terms appear both in the fluxes and sources (see appendix
\ref{sec:gr_hydrodynamics}).

\subsection{Numerical methods}
\label{sec:num_methods}

The ELH method proposed here has been implemented in the code
\verb+WhiskyTHCEL+ as a variant of the \verb+WhiskyTHC+ code
\cite{Radice2012a,Radice2013b,Radice2013c} based on the
\verb+Einstein Toolkit+
\cite{loeffler_2011_et,ET2013,EinsteinToolkit:web}. \verb+WhiskyTHC+
implements both finite-difference and finite-volume methods applied to a
characteristic-variables decomposition with a Lax-Friedrichs
flux-splitting for upwinding. It also crucially provides a positivity
preserving limiter to cope with large rest-mass density jumps, \eg as
those appearing across the surface of compact stars. In the following we
summarise the main components of the underlying algorithm and refer the
interested reader to \cite{Radice2012a} and \cite{Radice2013c} for a more
detailed description.

Given a discrete mesh with $\Delta$ being the spatial grid
spacing, the finite-difference algorithm we employ provides an estimate
for the right-hand-side of an evolution equation in flux-conservative
form as
\begin{equation}
    \partial_t U|_i = -\frac{f_{i+1/2} - f_{i-1/2}}{\Delta} + S_i\,,
    \label{eq:dflux}
\end{equation}
\ie as a difference between numerical fluxes at the cell interfaces, plus
the sources contribution (to simplify the notation, here $U$ is any
one of the components of \eqref{eq:conservatives}).

The numerical fluxes $f_{i\pm 1/2}$ are obtained via a reconstruction
operator, \ie an operator yielding a high-order approximation of a
generic function at a given point using its volume averages (see, \eg
\cite{Leveque92, Rezzolla_book:2013, Marti2015}). Out of the variety of
reconstruction operators available in the literature and implemented in
\verb+WhiskyTHC+, we focus here on the fifth-order and seventh-order
unfiltered stencils
\begin{align}
    ^5\mathcal{S}^- :=
    \frac{1}{60}(&2f_{i-2} - 13f_{i-1}+47f_{i}
    +\nonumber\\
    &27f_{i+1} - 3f_{i+2})
    \label{eq:U5}
\end{align}
and
\begin{align}
    ^7\mathcal{S}^- :=
    -&\frac{1}{140}f_{i-3} + \frac{5}{84}f_{i-2}
    -\frac{101}{420}f_{i-1}+\frac{319}{420}f_{i}\nonumber\\
    +&\frac{107}{210}f_{i+1}
    -\frac{19}{210}f_{i+2} + \frac{1}{105}f_{i+3}\,,
    \label{eq:U7}
\end{align}
returning the value of the flux at $x_{i+1/2}$, and which we refer to as
U5 and U7, respectively (we have only written here the left-biased
operators $\mathcal{S}^-$, since the right-biased ones $\mathcal{S}^+$
are symmetric)\footnote{Note that the operators \eqref{eq:U5} and
  \eqref{eq:U7} return approximations of the function $h$ defined by
  $f_i=:\int_{x_{i-1/2}}^{x_{i+1/2}} h(x')\, d x'$ at $x_{i+1/2}$. In
  this sense, they act on volume averages, the point-wise flux being the
  volume average of $h$. The values $h_{i\pm1/2}$ should appear in
  \eqref{eq:dflux} instead of $f_{i\pm1/2}$. We have here simplified the
  notation, but a full discussion can be found in \cite{Radice2012a}.}.

Furthermore, we select the MP5 scheme as our benchmark against which to
test the properties of the EL method. MP5 is built on top of the U5
stencil, but the resulting fluxes are limited so as to preserve
monotonicity near discontinuities \cite{suresh_1997_amp,mignone_2010_hoc,
  Radice2012a}. MP5 offers a good compromise between robustness and
accuracy and it has been successfully employed in several realistic
scenarios in which it also achieved high convergence-order
\cite{Radice2013b,Radice2013c,Radice2015}. It has therefore become a
\emph{de facto} standard in our work, hence we use it as a reference.

To ensure the stability of the scheme, the reconstruction must be
appropriately upwinded, \ie a right- (left-) biased operator has to be
applied to the left-(right-) going part of the flux. We therefore split
the flux $f$ in a right-going flux $f^+$ and a left-going one $f^-$, so
that $f=f^++f^-$. The splitting is performed using the Lax-Friedrichs or
Rusanov flux splitter \cite{Shu97}, \ie
\begin{equation}
    f^\pm := f(U)\pm\kappa U\,,
    \qquad \kappa := \max\left|f'(U)\right|\,,
    \label{eq:flux_split}
\end{equation}
where the maximum is taken over the stencil of the reconstruction
operator.

The reconstruction procedure outlined above can be applied on each
equation in the system \eqref{eq:valencia} (this is also called a
components split) or to its local characteristic variables (in which case
it is referred to as a characteristics split).

A further ingredient in our algorithm is a so-called
``positivity-preserving'' limiter \cite{Hu2013, Radice2013c}. The basic
idea is to split the numerical flux in Eq. \eqref{eq:dflux} in two
contributions, as:
\begin{equation}
    f_{i+1/2} = \theta f^{\rm HO}_{i+1/2}+(1-\theta)f^{\rm LF}_{i+1/2}\,,
    \label{eq:hybrid}
\end{equation}
where $f^{\rm HO}$ is the original high-order flux, while the Lax-Friedrichs
flux $f^{\rm LF}$ has the standard form
\begin{equation}
    f^{\rm LF}_{i+1/2} := \frac{1}{2}\left(f_i + f_{i+1}\right)
    -\frac{\kappa}{2}\left(
    U_i - U_{i+1}\right)
\end{equation}
and $\kappa$ is defined as in Eq. \eqref{eq:flux_split}. For a single Euler
step, the result of the evolution of $U$ can be explicitly written as
\begin{equation}
    U^{n+1}_i = \frac{1}{2}\left(U^n_i + 2\lambda f_{i-1/2}\right)
    +\frac{1}{2}\left(U^n_i - 2\lambda f_{i+1/2}\right)\,,
    \label{eq:splitevol}
\end{equation}
where the fluxes are defined as in Eq. \eqref{eq:hybrid} and $\lambda$
depends on the maximum propagation speed of the system as well as the CFL
factor.  The value of $\theta$ is defined as the one that makes both
terms of Eq.  \eqref{eq:splitevol} positive. Applied to the continuity
equation this guarantees that the density never becomes negative (see
\cite{Wu2017} for a way to generally ensure the physicality of the fluid
state-vector in a generic spacetime)\footnote{Further details can be
  found in \cite{Radice2013c}. Here, $\theta$ is computed following
  the algorithm in section 2.2 of \cite{Hu2013}, but replacing the
  rest-mass density $\rho$ with its conserved relativistic counterpart
  $D$.}.

We note that this algorithm does not free us from having to employ an
artificial floor (or atmosphere) to treat (ideally) vacuum regions: these
are filled with a uniform and tenuous fluid with rest-mass density
$\rho_{\rm atmo}$. Whenever in the subsequent evolution the rest-mass
density of a gridpoint falls below the floor value $\rho_{\rm atmo}$, it
is reset to the floor value, its three-velocity is set to zero and the
specific internal energy is set to the corresponding value coming from
the EOS. In neutron star simulations we fix the floor at $\rho_{\rm
  atmo}=10^{-16}$ $\msun^{-2}$, \plainie the typical value of $\rho_{\rm
  atmo}/\rho_{\rm max}$ is $\sim\,10^{-13}$.

Details of the algorithms we employ to evolve the spacetime and couple it to
the fluid evolution are given in appendix \ref{sec:mclachlan}.

The last step in the algorithm is the actual time evolution. Since after the
spatial discretization, the original PDEs to be solved are in the form of a
coupled systems of ODEs, this is taken care of in a method-of-lines (MOL)
fashion by means of a fixed step Runge-Kutta time integrator. We employ either
the standard fourth-order Runge-Kutta method (RK4) or a third-order (RK3, see
\cite{gottlieb2009}) method with strong stability preserving (SSP) properties.

\section{Entropy-limited hydrodynamics}
\label{sec:ELH}
\medskip
\subsection{Description of the scheme}
\label{sec:viscosityandpplim}

The scheme we propose consists of two building blocks: 1) a function detecting
shocks; 2) a limiter scheme of the high-order fluxes. We will start discussing
the latter.

As customary in flux-limiting schemes, we modify the high-order
approximation of the flux by combining (or ``hybridising'') it with a
(local) Lax-Friedrichs flux contribution as in \eqref{eq:hybrid}.

Of course, the hybridisation of the high-order flux with the
Lax-Friedrichs should be activated only in regions of the flow that are
problematic. In order to flag such regions we introduce a regularisation
function that we refer to as the ``viscosity'' $\nu$. Hence, we \emph{redefine}
the parameter $\theta\in[0,1]$ in Eq. \eqref{eq:hybrid} in terms of the
quantity
\begin{equation}
    \theta := \min\left[\tilde{\theta},
    1 - \frac{1}{2}(\nu_i + \nu_{i+1})\right]\,,
    \label{eq:thetatilde}
\end{equation}
so that the contribution of the Lax-Friedrichs flux grows linearly with the
viscosity. The value of the coefficient $\tilde{\theta}$ is the one mentioned
in the previous section to guarantee the positivity of the rest-mass density.
With the choice \eqref{eq:thetatilde} for the limiting coefficient $\theta$,
additional dissipation is inserted when $\nu$ attains large values as well as
in near-vacuum regions. On the other hand, in regions where the flow is smooth
and away from near vacuum, $\theta$ is close to unity, ensuring the use of the
high-order flux and preserving the high accuracy of the method.

We still need to associate the viscosity $\nu$ to some property of the
flow. To this scope we take inspiration from the work of Guermond and
collaborators \cite{Guermond11} and associate $\nu$ to the specific
entropy $s$. In general, the precise functional form of $s$ will depend
on the EOS, but for the simpler case of a perfect fluid with an
ideal-fluid EOS [\cf Eq. \eqref{eq:Gammalaw}], the specific entropy can
be shown to be equal to (apart from constant multiplicative factors) to
\cite{Rezzolla_book:2013}
\begin{equation}
\label{eq:entropydef}
    s = \log\left(\frac{\epsilon}{\rho^{\Gamma - 1}}\right)\,.
\end{equation}
Of course, the specific entropy must satisfy the second law of
thermodynamics, so that we can introduce the \emph{entropy residual}, or
entropy-production rate, $\mathcal{R}$ as
\begin{equation}
    \mathcal{R} := \partial_\mu(\rho s u^\mu) \geq 0\,.
    \label{eq:Rinequality}
\end{equation}
As a result, the computation of the entropy residual $\mathcal{R}$,
effectively represents the first step in defining the viscosity and hence
the root to limiter parameter $\theta$.

The expected behaviour of the entropy residual is that it cannot decrease
in time and that is spatially restricted to very small regions in the
neighbourhood of shocks, ideally expressed a delta function peaked at the
location $\boldsymbol{x}_s$ of shocks, \ie $\mathcal{R} \propto
\delta(\boldsymbol{x}-\boldsymbol{x}_s)$. A physical justification for
this latter expectation is rather simple to motivate. Euler equations
generally apply to perfect fluids, and while they can capture non-ideal
features (\ie shocks), the description of the latter is only
approximate. As long as the flow is smooth and the perfect-fluid
approximation holds, all phenomena are reversible and there can be no
production of entropy. However, in those regimes where the perfect-fluid
approximation breaks down and non-ideal effects appear, namely, at the
location of shocks, the entropy production is nonzero and the entropy
jumps locally to a higher value. Since shocks are regions of dimension
$N-1$ in spatial manifolds with $N$ spatial dimensions, the entropy
residual $\mathcal{R}$ must be a Dirac delta peaked at shock locations
for it to provide a finite contribution.

To seal the strict connection between the entropy viscosity $\nu_e$ and
the entropy residual and to embody the property that this quantity should
be a function of the spatial discretization, we define it as
\begin{equation}
    \nu_e := {c_e \Delta}|\mathcal{R}| \,.
    \label{eq:nuentropy}
\end{equation}
The absolute value is used for the entropy residual since the inequality
\eqref{eq:nuentropy} is not guaranteed to be satisfied during the
numerical integration. In fact, $\mathcal{R}$, having to approximate a
delta function, is expected to show an oscillatory behaviour with
potential negative values in practical numerical applications. In
expression \eqref{eq:nuentropy}, $\Delta$ is the spacing of the mesh,
$c_e$ is a positive tunable constant with dimensions of $[\rm
  time]^3\times[\rm temperature]\times[\rm mass]^{-1}$, so that $\nu_e$
is effectively dimensionless. We note that despite $\nu_e$ not having
the dimensions of a viscosity, we still refer to as the ``entropy
viscosity'', mostly for convenience and in analogy with the very similar
quantity defined in \cite{Guermond11}.

An additional benefit of this definition is the ability of the resulting
scheme in differentiating automatically between shocks and contact
discontinuities. This follows from the fact that at contact
discontinuities there is no entropy production and therefore the
viscosity there would be zero as well \cite{Rezzolla_book:2013}.

A potential problem of the definition \eqref{eq:nuentropy} is that it can
lead to an unbounded value since the entropy residual $\mathcal{R}$ is
not physically upper limited. In our case the value of $\theta$ cannot
however exceed unity, and so the viscosity must not exceed this value as
well. To enforce this requirement and cut-off excessively large values of
the entropy viscosity we set the entropy viscosity to be used in the
limiter \eqref{eq:thetatilde} as
\begin{equation}
    \nu := \min[\nu_e, c_{\rm max}]\,.
    \label{eq:nu}
\end{equation}
where $\nu_e$ is given by Eq. \eqref{eq:nuentropy}. Here, $c_{\rm max}$
is a tunable dimensionless coefficient, which, together with the other
tunable coefficient $c_e$, we have assumed to be equal to one in all of
the tests presented here.  As we will comment in section \ref{sec:tests},
the results are not very sensitive to the choices made for these
coefficients.

\subsection{Numerical implementation}
\label{sec:implementation}

In our numerical implementation we compute the entropy residual
\eqref{eq:Rinequality} by first rewriting its definition in a way that
involves only derivatives of the specific entropy $s$
\begin{align}
    \mathcal{R} &= \partial_\mu(s \rho u^\mu)
      = s \partial_\mu (\rho u^\mu) + \rho u^\mu \partial_\mu s
      = \rho u^\mu \partial_\mu s\,,
    \label{eq:Rs}
\end{align}
where the continuity equation \eqref{eq:continuity} was used to obtain
the final expression in \eqref{eq:Rs}. By expressing the 4-velocity
$u^\mu$ in terms of the fluid three-velocity $v^i$, we
finally write the residual as
\begin{equation}
    \mathcal{R} =
    \rho W\left(\partial_t s + v^i \partial_i s\right)\,.
    \label{eq:R3+1}
\end{equation}

The spatial derivatives in Eq. \eqref{eq:R3+1} are approximated with a
standard centered finite-difference stencil of order $p+1$, where $p$ is
the order of the stencil used to approximate the physical fluxes. This
restriction arises from the need to ensure that the viscosity converges
to zero fast enough not to spoil the overall convergence of the scheme at
the nominal order. The time derivative in \eqref{eq:R3+1} is also
approximated by finite differencing. In particular, at every iteration we
use the current value of the specific entropy and the values at the two
previous timesteps to compute a second-order approximation of $\partial_t
s$ via a one-sided stencil, \ie as
\begin{equation}
  (\partial_t s)^n = \frac{1}{2\Delta t}
  \left(3s^n - 4s^{n-1} + s^{n-2}\right) +
  \mathcal{O}\left((\Delta t)^2\right)\,.
  \label{eq:dt_s}
\end{equation}

A few remarks are useful at this point. First, the time derivative of the
specific entropy in Eq. \eqref{eq:R3+1} is computed with a low-order
method and this could in principle be a limiting factor for the
convergence properties of the overall scheme. In practice, however, we
find that the space discretization error dominates over the error on the
time derivative in the tests we have performed, so that the scheme
achieves high-order convergence as expected despite the use of a
low-order approximation for $\partial_t s$.  Second, the high-order flux
$f^{\rm HO}$ is computed component by component. In fact since the
reconstruction operators U5 and U7 [\eqref{eq:U5} and \eqref{eq:U7}] are
linear, they commute with the matrices used to perform the characteristic
decomposition, and there is therefore no difference in this case between
component-by-component and characteristic decomposition. This contributes
(along with other intrinsic differences in the formulation of the
schemes) to a significant speed-up of the code (up to $\sim\,50\%$,
depending on the setup of the grid on the computing nodes) with respect
to MP5, since there is no need to compute the system eigenvectors and
apply the resulting matrix. By contrast, the MP5 reconstruction is
nonlinear and does not commute with the characteristic decomposition. As
a result, when using MP5 we always switch to characteristic variables,
since this is known to reduce spurious numerical oscillations in
high-order methods \cite{suresh_1997_amp}.

Two further operations are applied on the viscosity before it is used in
Eq. \eqref{eq:thetatilde}. Firstly, since the viscosity is found to be
very close to zero in regions of very low rest-mass density, we improve
the behaviour close to atmosphere values by simply setting the viscosity
to some small and constant value $\nu_{v}$ at a given point $x_{ijk}$
whenever the rest-mass density at the given point, and at all nearest
neighbours, is below a certain threshold $\rho_{v}$, \ie if
\begin{align*}
\rho_{i+l,j+m,k+n}<\rho_{v} \qquad \forall\; l,m,n=-1,0,1
\end{align*}
then
\begin{align*}
\nu_{ijk}=\nu_{v}\,.
\end{align*}
In practice, therefore, \eqref{eq:nu} needs to be slightly revised and
the expression for the entropy viscosity we actually implement in the
numerical code is
\begin{equation}
  \label{eq:nu_fin}
    \nu :=
    \begin{cases}
        \nu_{v}&\quad \text{if}\ \rho<\rho_{v}\\
        \min[\nu_e, \nu_{\rm max}]&\quad \text{elsewhere}\,.
    \end{cases}
\end{equation}
In all of the numerical tests presented we have used $\nu_{v}=10^{-12}$
and $\rho_{v}=10^{-11}\,\msun^{-2}$ (\ie the threshold is 5 orders of
magnitude larger then the atmosphere floor,
$\rho_{\rm atmo}=10^{-16}\,\msun^{-2}$). Secondly, and following the
original implementation in Ref. \cite{Guermond11}, we introduce a
smoothing step which removes unwanted oscillations in the viscosity. This
is accomplished by applying a five-point stencil of the form
\begin{equation}
    \bar{\nu}_{ijk} := \sum_{l=-2}^2 \sum_{m=-2}^2 \sum_{n=-2}^2
    a_l\, a_m\, a_n\, \nu_{i+l,j+m,k+n}\,,
    \label{eq:smoothing}
\end{equation}
where the coefficients $a_l$ have values $a_0=0.58$, $a_{\pm 1}=0.06$ and
$a_{\pm 2}=0.15$. This stencil in Eq. \eqref{eq:smoothing} is constructed
so to approximate the convolution of the viscosity with a Gaussian kernel
of characteristic cutoff length scale 4 times the grid spacing, in such a
way that the residual of the transfer function of the target filter and
of its approximation is minimised over a broad range of wavelengths (see
\cite{Sagaut99} for details).

In addition to dampening oscillations in the viscosity mentioned in the
previous section, the necessity of the smoothing step stems from the fact
that the viscosity is computed once at the beginning of every new
timestep before its value is used in \eqref{eq:thetatilde}. Since the
viscosity is kept constant during the series of Runge-Kutta internal
steps, it ``lags behind'' in time with respect to the solution. This
issue is addressed by the smoothing procedure, but in practice we have
found that this does not represent a problem in our tests. The smoothing
\eqref{eq:smoothing} also prevents the viscosity to plunge to very small
values where it should instead be non negligible. This can happen, \eg
close to stellar surfaces as a result of oscillations in the
solution. The application of the smoothing operator removes this problem
by joining seamlessly the values of the viscosity in the neighbouring
points.

\section{Numerical tests}
\label{sec:tests}

We report in this section the results of some of the tests obtained with
the ELH method described in the previous sections. In all tests we compare
the ELH results with those obtained using the monotonicity
preserving, fifth-order scheme (MP5). In particular, unless otherwise
stated, we couple the ELH method to the fifth-order U5 stencil
\eqref{eq:U5}, to make a fair and sensible comparison between methods of
the same order. In few cases, however, we will also report results
obtained with the seventh-order stencil U7 \eqref{eq:U7}. We will refer
to the corresponding schemes as to EL5 and EL7, respectively. Finally,
it is useful to remark that in all of the following tests no attempt was
made to tune the coefficients $c_e$ and $c_{\rm max}$ introduced in
section \ref{sec:viscosityandpplim}, and that have been set to unity
here. Despite this very simple choice, the ELH method is stable and
accurate in all cases considered, as the following sections make
clear. At the same time, we consider it possible (if not likely) that the
results could be further improved after a careful exploration of the
changes in the solution upon a change of $c_e$ and $c_{\rm max}$; we will
leave this exploration to a future work.

\subsection{Special-relativistic tests}
\label{sec:srtests}

We begin with a series of mostly one-dimensional tests, performed in
special-relativistic hydrodynamics, so that the metric $?g_\mu\nu?$ is
fixed to the flat Minkowski metric $?\eta_\mu\nu?$ and no spacetime
evolution is performed. Also, since we are mostly interested in the
behaviour of the scheme in realistic astrophysical applications, we focus
on just a handful of one-dimensional test cases: a smooth nonlinear wave
and three shock-tube tests.

\subsubsection{Smooth nonlinear wave}
\label{sec:sw}

We first show the accuracy of the scheme in the case of a smooth solution
and measure rigorously its convergence order so as to show that the
entropy-driven limiter does not spoil the convergence properties of the
high-order method upon which it is built. This test has been discussed in
\cite{Radice2012a} (which adapted it from \cite{Zhang2006}). In short, we
consider a one-dimensional, large-amplitude, smooth, nonlinear wave with
initial rest-mass density profile given by
\begin{align}
    \rho_0(x) =
    \begin{cases}
        1+\exp\left[-1/(1-x^2/L^2)\right]&\text{  if }|x|<1\\
        1&\text{  elsewhere}\,,
    \end{cases}
\end{align}
where $L=0.3$. The initial data employs a polytropic EOS,
$p=K\rho^{\tilde{\gamma}}$, with $K=100$ and $\tilde{\gamma}=5/3$, and we then
evolve it with the ideal-fluid EOS \eqref{eq:Gammalaw} with $\Gamma=5/3$. Since
in this test discontinuities are absent (so that $\tilde{\gamma}=\Gamma$) and
there are no stability issues, we use as time integrator the standard
fourth-order RK4 method with a timestep of $\sim\,0.13$ times the grid spacing.

\begin{figure}[!t]
    \centerline{\includegraphics[width=0.9\columnwidth]{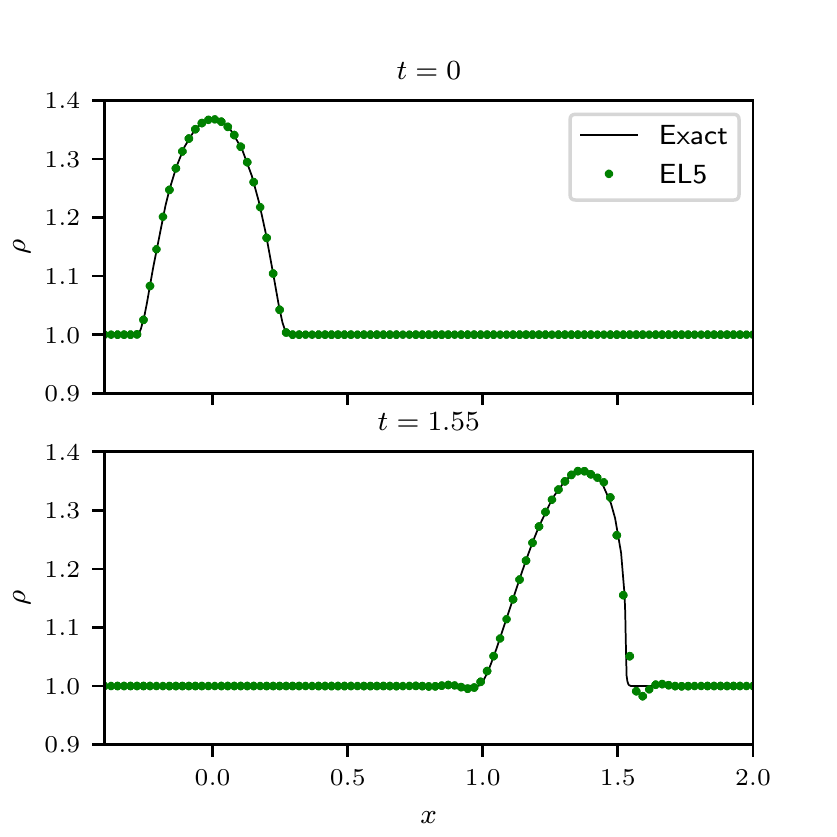}}
    \caption[Smooth wave test: rest-mass density profiles]{Rest-mass density
    profiles for the smooth nonlinear wave test. The EL5 data shown corresponds
    to the coarsest resolution of $100$ gridpoints over the domain.}
    \label{fig:sw_dp}
\end{figure}

The analytic solution of the test (shown in Fig. \ref{fig:sw_dp} with a
black solid line) is represented by a wave profile propagating towards
the right and ``steepening'' in the direction of its motion. At time
$t_c\simeq1.6$, the wave develops a shock, leading to a sharp
discontinuity. Up to the formation of the caustic, the analytic solution
can be computed using the method of characteristics \cite{Anile1990} on a
Lagrangian grid. We obtain an accurate enough approximation by computing
it on a very fine grid of $10^5$ gridpoints and interpolating the
solution using cubic splines on the Eulerian grid. This solution, which
we refer to as the ``exact'' solution, is then used as the reference
against which the numerical solutions are compared.

We perform this test with both the EL5 and EL7 schemes of the ELH method
to validate that high-order schemes can be employed with great ease in our
approach by simply swapping a lower-order stencil for a higher-order one;
this operation is far more demanding in standard finite-volume HRSC
schemes.

\begin{figure}[!t]
    \centerline{\includegraphics[width=0.9\columnwidth]{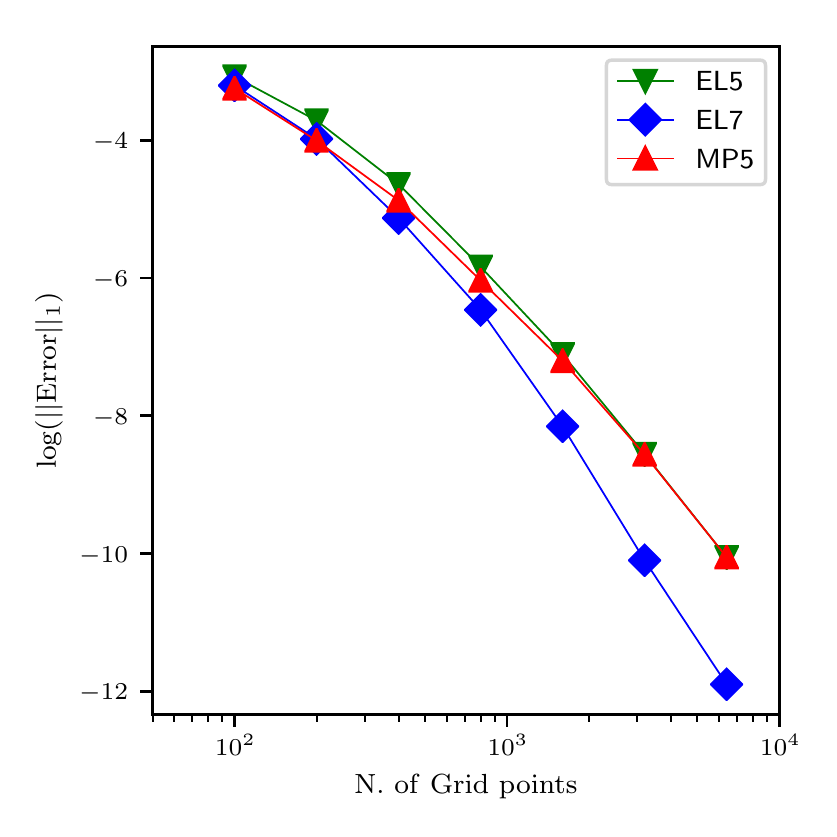}}
    \caption[Smooth wave test: $L_1$-norm of error]{$L_1$-norm of the error on
    the rest-mass density for the smooth nonlinear-wave test at time $t=0.8$.}
    \label{fig:sw_error}
\end{figure}

Figure \ref{fig:sw_error} shows the $L_1$-norm of the error with respect
to the analytic solution at time $t=0.8$ for the various schemes and at
various resolutions. The latter are parametrized by the number of
gridpoints used on the $x$-axis, and we have considered seven different
resolutions, each twice as fine as the preceding one, going from $100$
gridpoints up to $6400$. The different lines in Fig. \ref{fig:sw_error}
show that at the lowest resolutions all schemes show very similar errors,
MP5 being the most accurate by a small margin. As the resolution is
increased, however,the gap in accuracy between EL5 and MP5 decreases and
disappears at very high resolutions. The error curve of EL7, being a
higher-order scheme, decreases much more rapidly with resolution, so that
its error at the highest resolution of $6400$ gridpoints is two orders of
magnitude lower than for the fifth-order schemes.

\begin{figure}[!t]
    \centerline{\includegraphics[width=0.9\columnwidth]{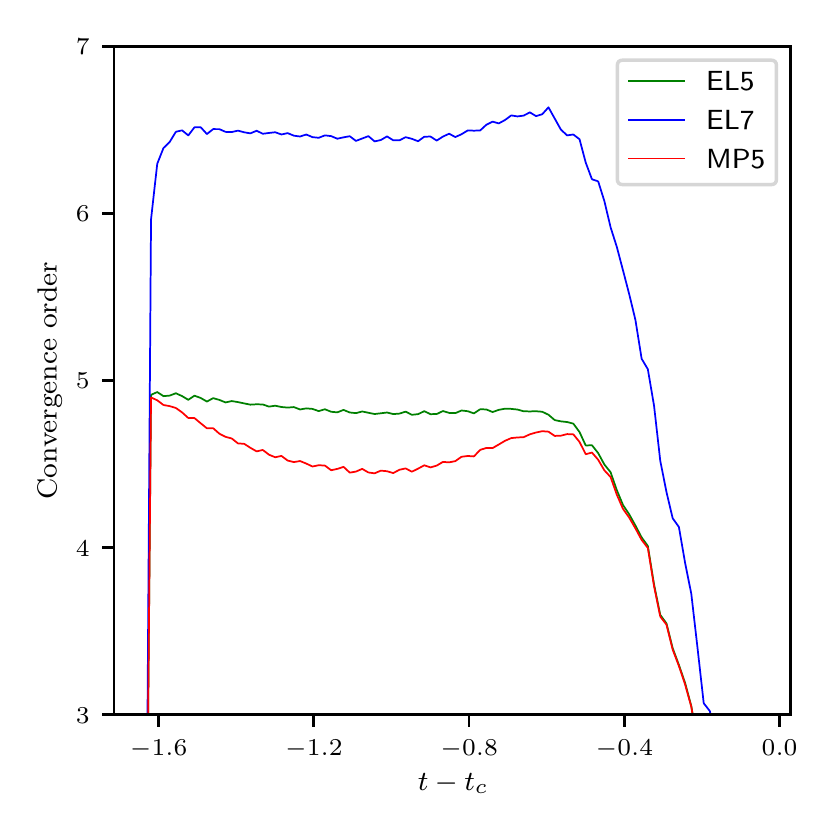}}
    \caption[Smooth wave test: convergence order]{Convergence order computed on
    the smooth nonlinear-wave test as a function of time to caustic formation.}
    \label{fig:sw_convergence}
\end{figure}

We also compute the convergence order of the various schemes using the
data at resolutions of $1600$ and $3200$ gridpoints and after comparing
it with the ``exact'' solution. The result is shown in
Fig. \ref{fig:sw_convergence} as a function of time to the development of
the shock. The computed order should be equal to the
nominal order of each scheme as long as the solution is smooth, gradually
degrading to first order as the caustic is approached. Every scheme
matches this description, in particular EL5, whose convergence order is
almost exactly five. EL7 similarly appears to saturate just below its
nominal convergence order of seven. Deviations from the nominal
convergence order of each scheme are due to contaminations from other
error sources, which become increasingly significant at high resolution;
these are: the truncation error due to the time-integrator, the accuracy
of the inversion from conservative to primitive variables, or the
low-order approximation for the evolution of the entropy [\cf
  Eq. \eqref{eq:dt_s}]. What is relevant here is that the EL method does
not interfere with the convergence properties of the underlying stencil
and can exploit their accuracy.

\subsubsection{Shock-tube tests}
\label{sec:shock}

We choose as a first shock-tube test the special-relativistic version of the
classical Sod test \cite{Sod1978}. In this case, the adiabatic index for
both the polytropic initial data EOS and the ideal-fluid evolution EOS is
$\Gamma=1.4$ and the right ($R$) and left ($L$) initial states are
\begin{align}
&(\rho_R,v_R,p_R)=(0.125,0,0.1)\,,\nonumber\\
&(\rho_L,v_L,p_L)=(1,0,1)\,.
\end{align}
The analytic solution consists in a left-going rarefaction wave and a
right-going shock wave separated by a right-going contact
discontinuity. We perform the test with a variety of spatial resolutions
ranging from $\Delta x=0.01$ to $\Delta x=3.125 \times 10^{-4}$, and a
timestep $\Delta t=0.1 \, \Delta x$.

\begin{figure*}[!t]
    \centerline{\includegraphics[width=1.8\columnwidth]{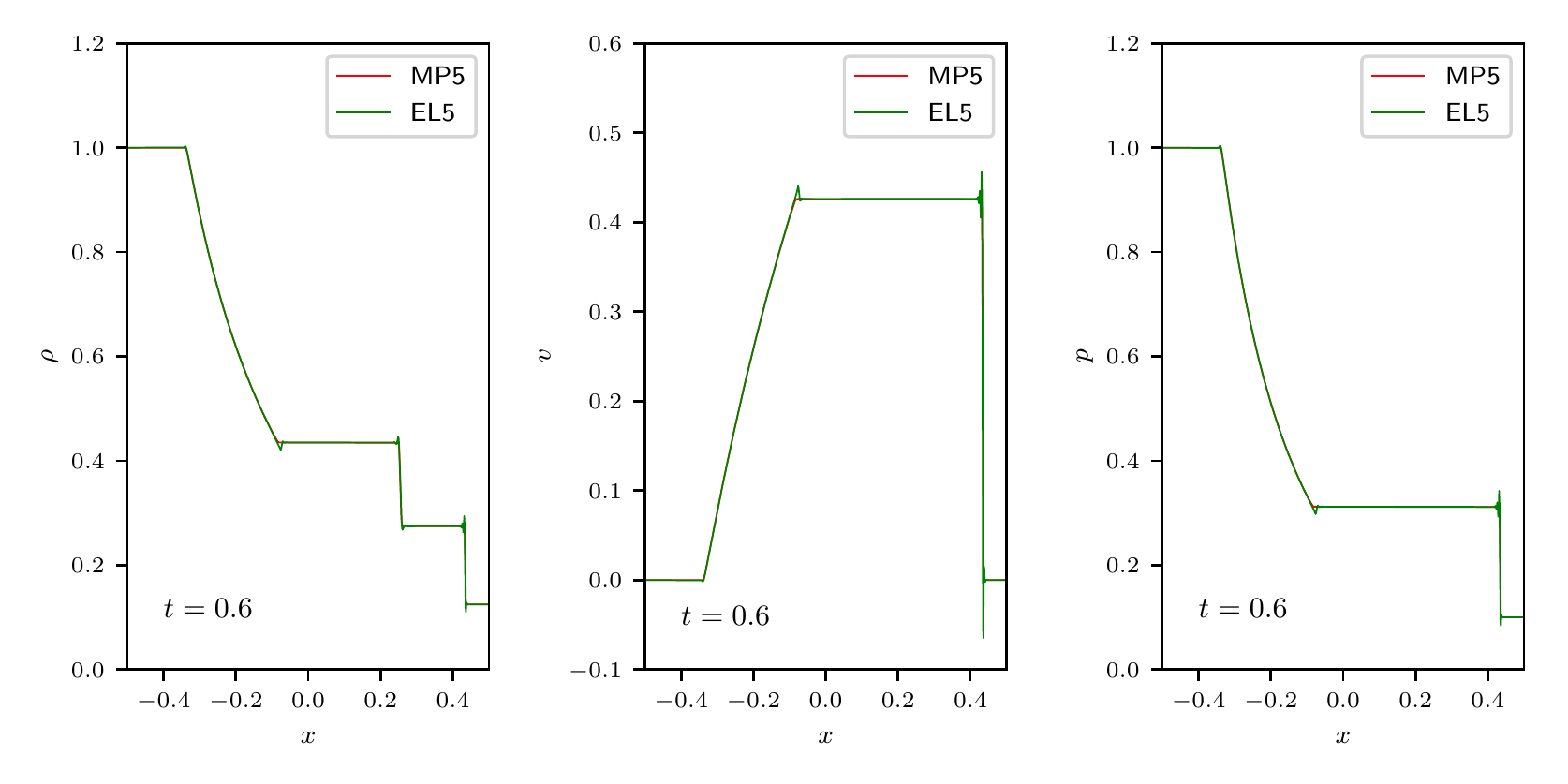}}
    \caption[Sod test: solution profiles]{Profiles of the rest-mass density
    (left), velocity (center) and pressure (right) for the special-relativistic
    Sod test at $t=0.6$. The solution is computed on a grid of $800$ points.
    The EL5 scheme correctly captures the features of the solution despite
    oscillations at the discontinuities.}
    \label{fig:sod_p}
\end{figure*}

\begin{figure}[!t]
    \centerline{\includegraphics[width=0.9\columnwidth]{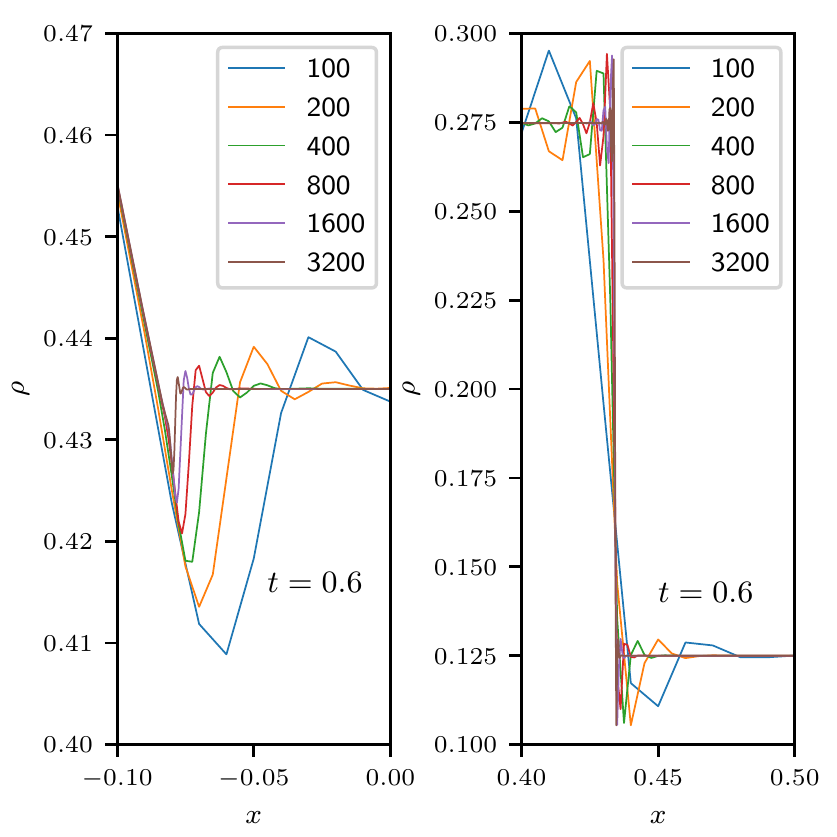}}
    \caption[Sod test: solution at different resolutions]{Rest-mass density
    profiles, zoomed on the right edge of the rarefaction wave (left) and on
    the shock (right) for the special-relativistic Sod test at $t=0.6$,
    computed with EL5 at different resolutions (parametrized by the number of
    points on the $x$ axis). The oscillations in the solution can be seen
    converging away with resolution.}
    \label{fig:sod_osc}
\end{figure}

Figure \ref{fig:sod_p} shows the test results at time $t=0.6$ for both
the EL5 and MP5 schemes at resolution $\Delta x=1.25 \times
10^{-3}$. Both schemes capture the main features of the solution, with
the shocks being captured within $\sim\,3$ gridpoints, as are the
constant states in the pressure and velocity. However, the EL5 scheme
displays some oscillations downstream of the shock as well as over- and
undershoots around the location of the discontinuities and in the
transition between the rarefaction wave and the surrounding flat regions,
while the MP5 scheme is able to resolve the solution avoiding such
artefacts. This is not surprising since MP5 is a monotonicity preserving
scheme (the number of local maxima and minima cannot increase by effect
of this method, therefore over- and undershoots cannot occur by
construction) while EL5 is not. We should remark, however, that this
property of MP5 is valid only for scalar equations in one spatial
dimension, and it basically accounts for the differences in the behaviour
of the two schemes. It should also be stressed that the EL5 scheme is
indeed stable and that the oscillations that are present in the solution
converge away with resolution (see Fig. \ref{fig:sod_osc}).

\begin{figure}[!t]
    \centerline{\includegraphics[width=0.9\columnwidth]{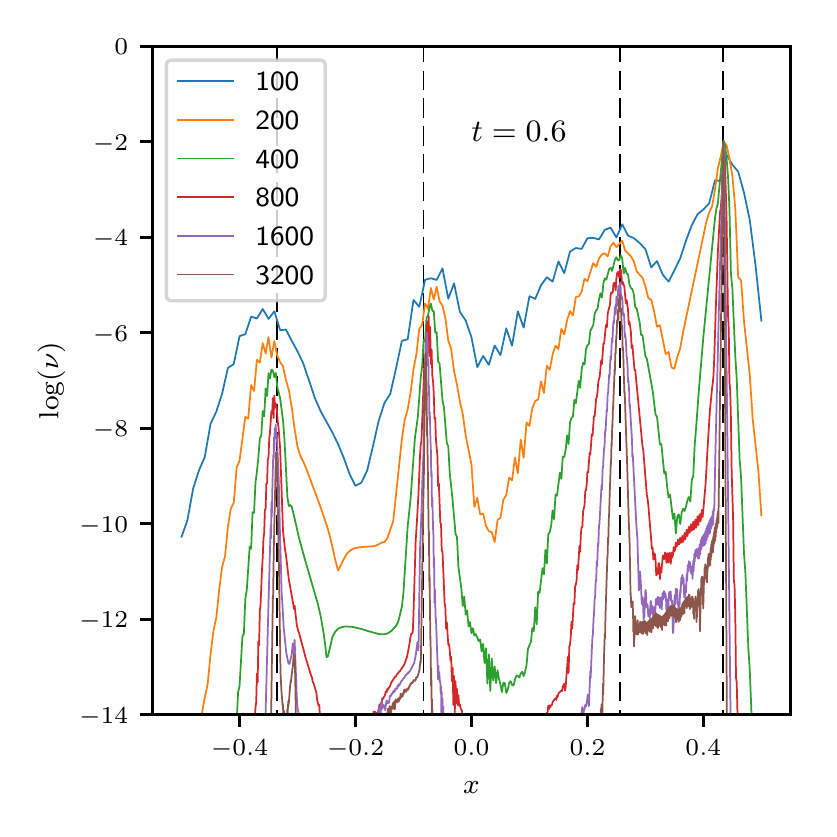}}
    \caption[Sod test: viscosity at different resolutions]{Profiles of the
    viscosity in logarithmic scale for the special-relativistic Sod test at
    $t=0.6$, computed with EL5 at different resolutions (parametrized by the
    number of points on the $x$ axis). The four peaks correspond to the four
    different features of the solution, \ie from left to right, the edges of
    the rarefaction wave, the contact discontinuity and the shock (vertical
    dashed lines highlight their location). As the resolution increases, they
    tend to delta functions.}
    \label{fig:sod_nu}
\end{figure}

The behaviour of the viscosity is displayed in Fig. \ref{fig:sod_nu}. It
presents four well distinct peaks, each corresponding to the four nonlinear
waves generated by the Riemann problem and corresponding to the edges of the
rarefaction wave, where the solution is continuous but non-smooth, of the
contact discontinuity and of the shock. The viscosity is higher in
correspondence with the latter, and decreases by several orders of magnitude
for the other three.  It can also be seen clearly how the peaks in the
viscosity sharpen as the resolution is increased, mirroring the decreasing size
of the aforementioned features (see Fig. \ref{fig:sod_osc}), and seemingly
tending towards a delta function at infinite resolution, as expected.

The second shock-tube test we select is a more extreme ``blast-wave''
test \cite{Marti03}. In this case, the adiabatic index used is
$\Gamma=5/3$ and the right and left initial states are
\begin{align}
&(\rho_R,v_R,p_R)=(10^{-3},0,1)\,,\nonumber\\
&(\rho_L,v_L,p_L)=(10^{-3},0,10^{-5})\,.
\end{align}
The exact solution consists in a right-going shock wave, followed by a
contact discontinuity and a left-going rarefaction wave. We employ the
same resolutions and timestep choices as for the Sod test.

\begin{figure*}[!t]
    \centerline{\includegraphics[width=1.8\columnwidth]{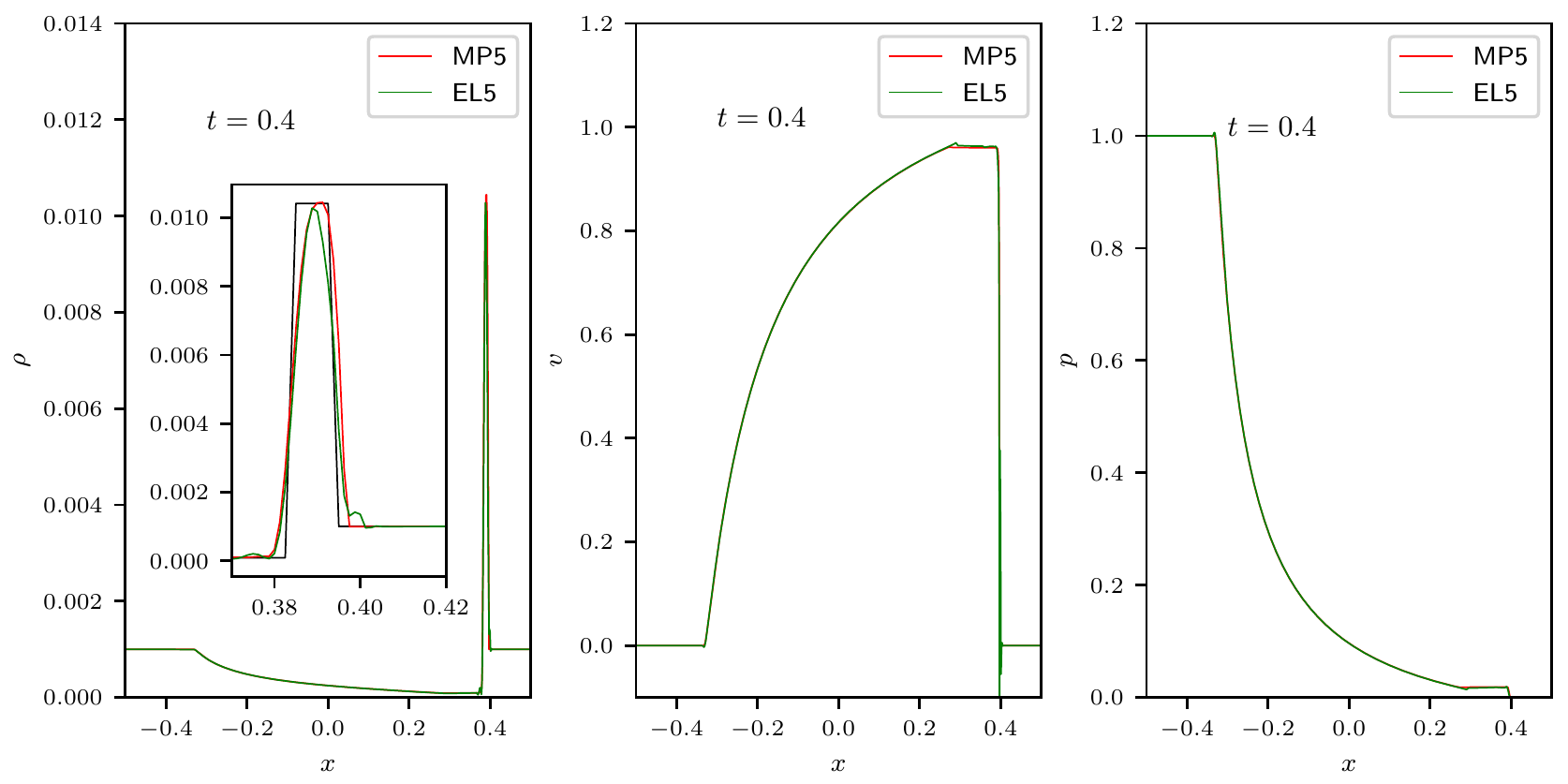}}
    \caption[Blast-wave test: solution profiles]{Profiles of the rest-mass
    density (left), velocity (center) and pressure (right) for the
    special-relativistic blast-wave test at $t=0.4$. The solution is computed
    on a grid of $800$ points. The inset in the density panel magnifies the
    blast wave, showing also the exact solution in black. Inversion failures
    due to oscillations when using the EL5 scheme spoil the quality of the
    solution.}
    \label{fig:blast_p}
\end{figure*}

Figure \ref{fig:blast_p} is similar to Fig. \ref{fig:sod_p}, but relative
to the blast-wave test at time $t=0.4$. We should remark that this is a
very extreme test (the pressure has an initial jump of five orders of
magnitude) in which the contact discontinuity and shock wave move at
essentially the same speed, yielding a very narrow constant rest-mass
density state between the two. The oscillations in the EL5 scheme data
are in this case more severe than in the Sod shock-tube test, especially
around the shock location. As a result, the solution with the EL5 scheme
tends to a general decrease of the pressure between the rarefaction wave
and the shock wave, whose relative value is however of $\lesssim 7\,\%$
at most; the MP5 scheme performs better and has a relative error in
pressure that is $\sim\,1\%$. In both cases, the agreement with the exact
solution improves with resolution.

Finally, we perform a three-dimensional shock-tube problem, involving
non-grid-aligned shocks \ie the relativistic-explosion test. The initial
data in this case is given by
\begin{align}
    \begin{cases}
    (\rho,v_i,p)=(1,0,1) &\text{ if }r\le0.4\,,\nonumber\\
    (\rho,v_i,p)=(0.125,0,0.1) &\text{ otherwise}\,,
    \end{cases}
\end{align}
where $r$ is the distance from the origin. The computational domain is a
cube of side 1 centered on the origin, and we use a grid spacing of
$\Delta x=0.01$ and a timestep $\Delta t=0.1 \, \Delta x$. The adiabatic
index for this test is again $\Gamma=1.4$. The feature of the solution
are similar to those of the Sod test \ie an ingoing rarefaction wave and
an outgoing shock, separated by an outgoing contact discontinuity. Note
however that because of the spherical symmetry of the test (compared to
the planar symmetry in the Sod case), the regions at the two sides of the
contact discontinuity are no longer constant states in rest-mass density,
velocity and pressure, but display a smooth radial dependence.

\begin{figure}[!t]
    \centerline{\includegraphics[width=0.9\columnwidth]{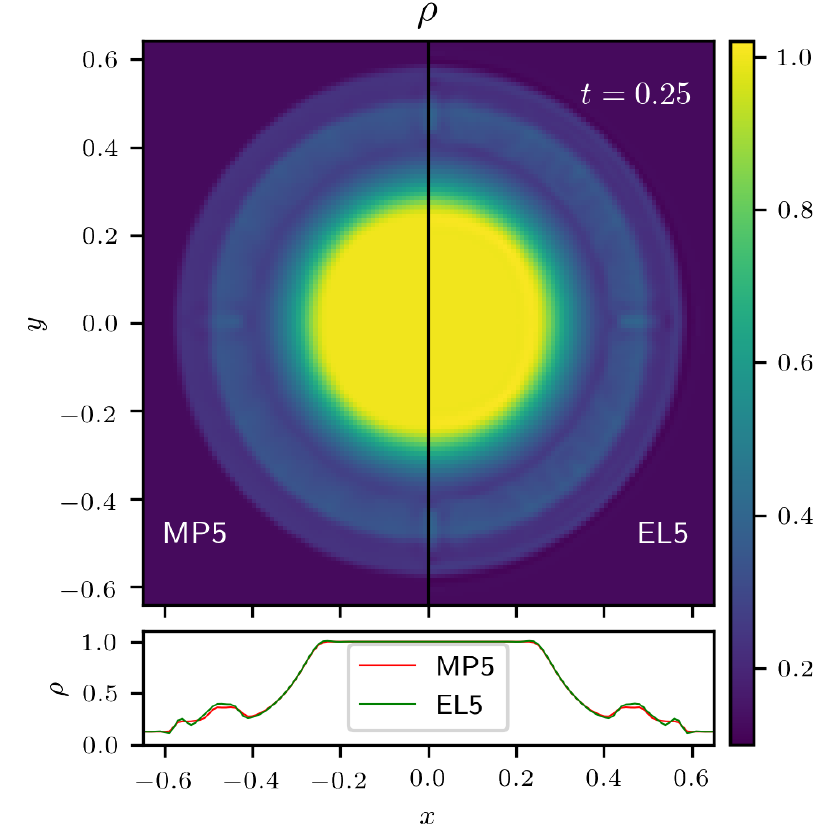}}
    \caption[Relativistic-explosion test: density on $(x,y)$ plane]{Rest-mass
    density for the relativistic-explosion test at time $t=0.25$. In the top
    panel the distribution on the $(x,y)$ plane is plotted, MP5 on the left
    side and EL5 on the right. In the bottom panel, the rest mass density is
    plotted on the $x$ axis. Both schemes capture very well the solution.}
    \label{fig:relexp_p}
\end{figure}

Figure \ref{fig:relexp_p} shows the rest-mass density for this test at
time $t=0.25$, on the $(x,y)$ plane as well as on the $x$ axis. Both EL5
and MP5 perform very similarly, with differences being barely noticeable
in the two-dimensional plot. The curves on the $x$ axis reveal that while
both schemes capture the features of the solution, as in the Sod test,
the EL5 scheme is slightly more oscillatory.

Overall, these shock-tube tests demonstrate how the entropy-driven
hybridisation of the high-order stencil is sufficient to stabilise the
scheme even for discontinuous initial data and it is remarkable that such
a simple scheme can achieve good accuracy.

\subsection{Three-dimensional general-relativistic tests: neutron stars}
\label{sec:3Dtests}

We next test the EL5 scheme against a series of three-dimensional tests
mostly based on the evolution of single, isolated neutron stars in
general relativity (with the exception of grazing-collision test of
section \ref{sec:BTOV}). In each test we employ for the evolution the
ideal-fluid EOS \eqref{eq:Gammalaw} with $\Gamma=2$. The neutron star
initial data is constructed using a polytropic EOS
$p=K\rho^{\tilde{\gamma}}$ also with $\tilde{\gamma}=2$ and $K=100$
$\msun^{-2}$.

\subsubsection{Isolated star in the Cowling approximation}
\label{sec:TOVC}

The first test we perform is the evolution of a stable nonrotating (or
TOV, from Tolmann-Oppenheimer-Volkoff) neutron star in a fixed spacetime
(\ie adopting the Cowling approximation) with the goal of assessing the
properties of the EL5 scheme over long timescales. Despite its conceptual
simplicity (a TOV is just a static solution of the Einstein-Euler
equations) the test can be rather challenging. This is because in this
test the location of the stellar surface, which is the hardest feature to
simulate due to the steep gradient in the hydrodynamics variables, is
essentially stationary; as a result, errors can accumulate and grow,
affecting the accuracy of the simulation. This behaviour is to be
contrasted with the typical situation encountered when evolving
inspiralling binary neutron stars, where the stellar surfaces move very
supersonically with respect to the floor and most of the errors at the
surface are absorbed into the shocks.

For this test we build and evolve a TOV model with central rest-mass
density $1.28 \times 10^{-3}\,\msun^{-2}$, yielding a (baryon) rest
mass of $1.5\,\msun$ and a radius of $\sim\,10\,\msun$. We perform the
test on a single refinement level with outer boundaries placed at
$16\,\msun$ and a resolution of $\Delta^i=0.2\,\msun\simeq0.3$ ${\rm km}$.
The timestep is set to $0.15$ times the grid spacing, and the time integrator
is RK3.

\begin{figure}[!t]
    \centerline{\includegraphics[width=0.9\columnwidth]{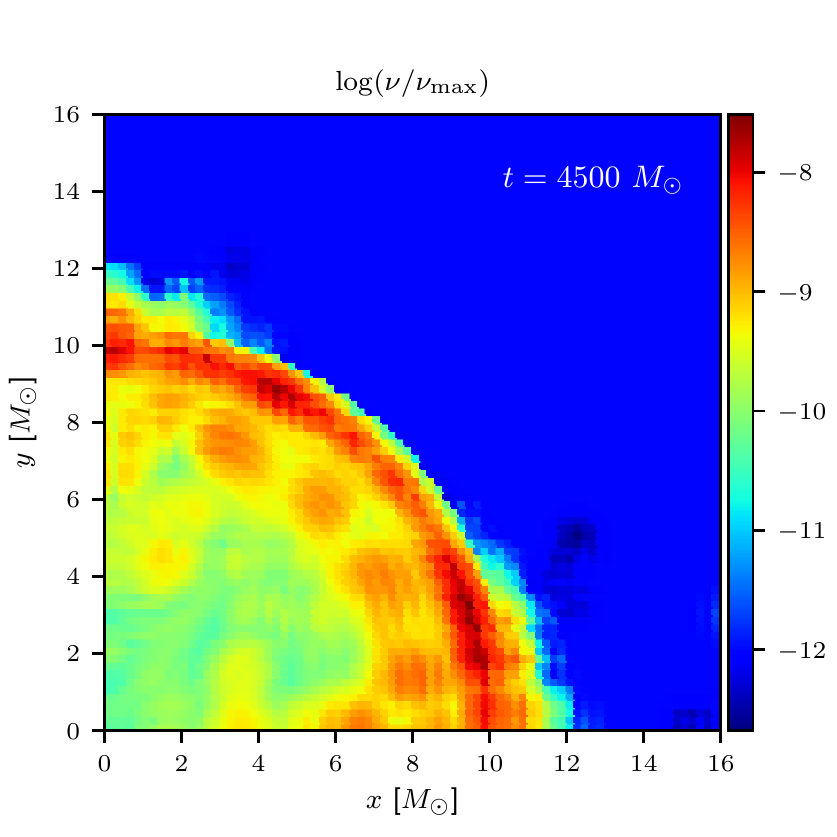}}
    \caption[Cowling TOV test: viscosity on $(x,y)$ plane]{Two-dimensional
    viscosity distribution relative to its upper limit on the equatorial
    $(x,y)$ plane at time $t=4500\,\msun$ for the Cowling TOV test. The
    viscosity peaks at the stellar surface, identified as a shock by the
    scheme, and drops in the interior.}
    \label{fig:tovc2D_nu}
\end{figure}

Figure \ref{fig:tovc2D_nu} reports the distribution of the viscosity on the
equatorial plane, which clearly shows a local annular peak around the location
of the stellar surface, where the hydrodynamical variables experience the most
violent variations, leading to large values of the viscosity. In the external
low-density fluid, the viscosity is set to a small constant value almost
everywhere as detailed in section \ref{sec:implementation}. The inner part of
the neutron star is expected to be isentropic, as it consists of a shock-free
perfect fluid. Indeed, in the stellar interior the viscosity is nonzero but
also $10^2$ to $10^3$ times smaller than at the surface and does not
significantly affect the evolution. Quite generally, these features of the
viscosity profile are typical in all the tests we considered, whenever a sharp
matter/vacuum interface is present.

\begin{figure}[!t]
    \centerline{\includegraphics[width=0.9\columnwidth]{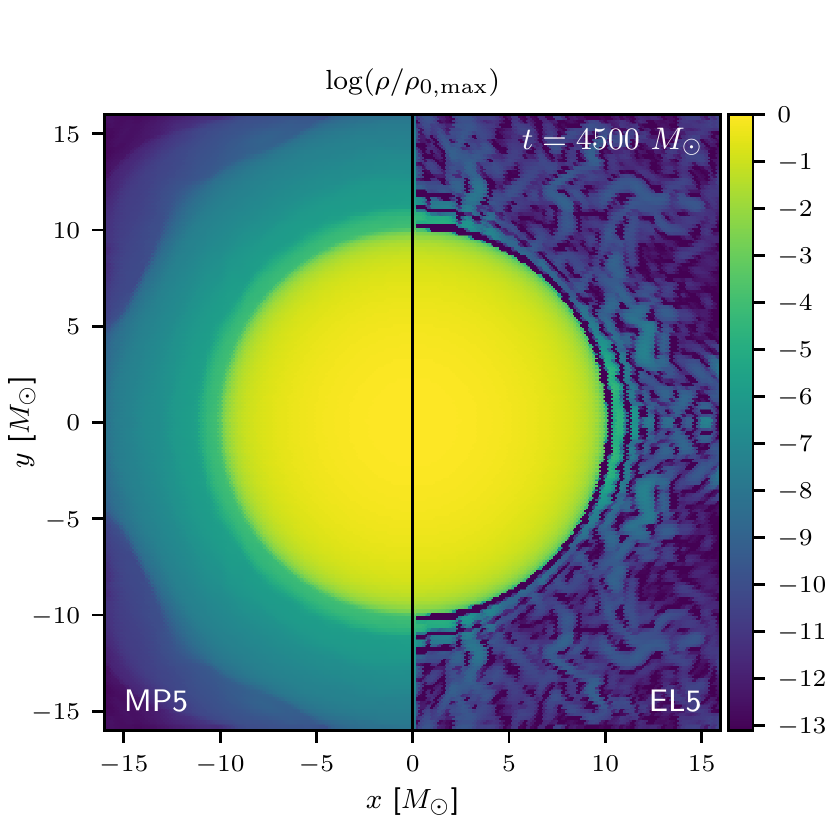}}
    \caption[Cowling TOV test: density on $(x,y)$ plane]{Two-dimensional
    rest-mass density distribution relative to the initial data maximum value
    on the equatorial $(x,y)$ plane at time $t=4500\,\msun$ for the Cowling TOV
    test; the left part of the panel (\ie $x<0$) refers to the MP5 scheme,
    while the right (\ie $x>0$) part to the EL5 scheme. Oscillations are
    visible with the EL5 scheme at the stellar surface, but the exterior fluid
    is visibly less dense than in the MP5 case.}
    \label{fig:tovc_2Drho}
\end{figure}

The general behaviour of the EL5 scheme, when compared to the MP5 scheme, is
well illustrated by Fig. \ref{fig:tovc_2Drho}, showing the rest-mass density
distribution on the equatorial plane for the two schemes (the left part of the
panel, \ie for $x<0$, refers to the MP5 scheme, while the right part, \ie for
$x>0$, to the EL5 scheme\footnote{The use of a higher-order stencil in the EL
approach, \eg EL7, does not yield to improvements in the solution; the
treatment of the low-density regions is far more delicate and the mass
conservation is degraded.}). As it can be seen from the figure, both schemes
accurately capture the solution in the stellar interior, but significant
differences arise at the surface and in the exterior. The MP5 scheme shows a
rather diffusive behaviour, with a smooth transition to the external ``vacuum''
(\ie to a region close to the rest-mass density floor) and extended low-density
tails. The EL5 scheme, on the other hand, produces a sharper edge. Oscillations
in the solution can be seen just outside of the star, resulting in shell-like
structures around the surface, which are particularly noticeable in the
coordinate axes directions. The stellar exterior is much closer to the vacuum
with the EL5 scheme and, in contrast to MP5, it also displays small-scale
dynamics at very low densities.

\begin{figure}[!t]
    \centerline{\includegraphics[width=0.9\columnwidth]{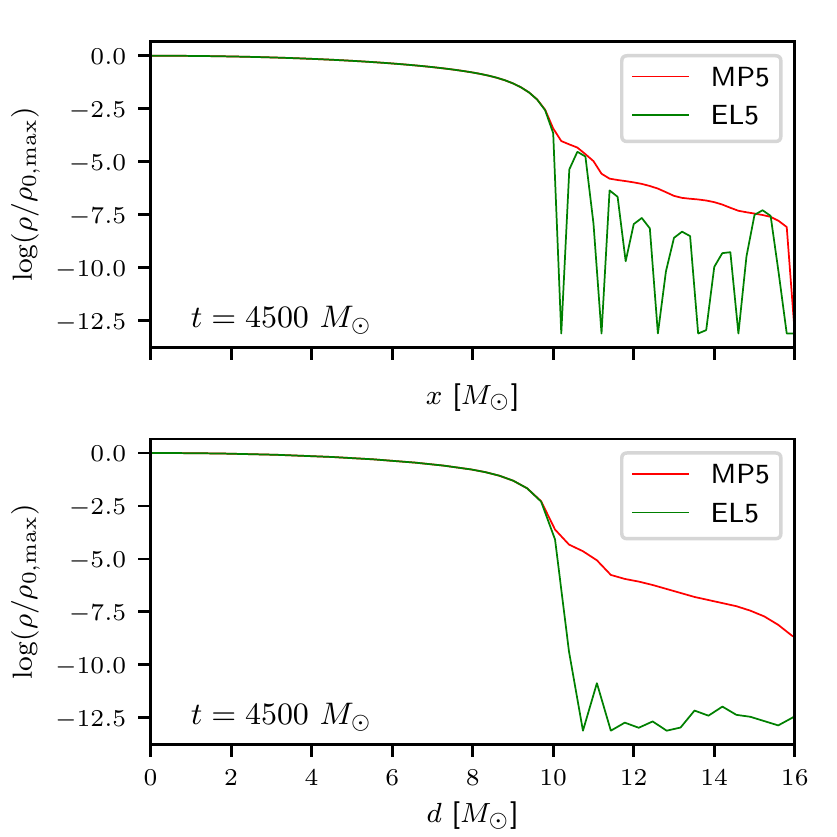}}
    \caption[Cowling TOV test: 1D density profiles]{One-dimensional rest-mass
    density profiles in the $x$ (top) and $d$ direction (bottom) at time
    $t=4500\,\msun$ relative to the initial data maximum value for the Cowling
    TOV test. The oscillations seen in the EL5 data are a direction dependent
    artefact, absent in the diagonal direction.}
    \label{fig:tovc_dp}
\end{figure}

The properties of the oscillations present in the solution computed with the
EL5 scheme are made clearer in Fig. \ref{fig:tovc_dp}, which shows the
rest-mass density profiles along different radial cuts. Along the $x$
direction, the oscillations in the EL5 data have large amplitude and a similar
behaviour is observed along the $y$ and $z$ axes. On the other hand, on the
three-dimensional diagonal (\ie along the $x=y=z$ line), the EL5 scheme manages
to capture the sharp transition between the stellar interior and the outside
vacuum almost perfectly, without significant oscillations or other artefacts.
By contrast, the use of the MP5 scheme leads to smooth, rest-density profiles
that are only slowly decaying in all directions\footnote{Of course, for both
schemes the amount of rest-mass outside the star is minute, being only
$10^{-7}$ of the initial rest-mass for the EL5 scheme and $\sim\,10^{-5}$ for
the MP5 scheme.}.

The direction-dependent behaviour shown in Fig. \ref{fig:tovc_dp} for the
EL5 scheme is due to the well-known anisotropy of the phase error common
to finite-differencing schemes \cite{Vichnevetsky82, Lele92}. The MP5
scheme is able to mask this behaviour, but at the price of sacrificing
the ability to sharply define stellar surface. We expect that the
performance of the EL5 scheme could be improved through the use of
multidimensional stencils (\ie employing a multidimensional interpolation
in the reconstruction step), as opposed to the current approach in which
the stencil is simply oriented in the direction of the flux to be
reconstructed.

\begin{figure*}[!t]
    \centerline{\includegraphics[width=1.8\columnwidth]{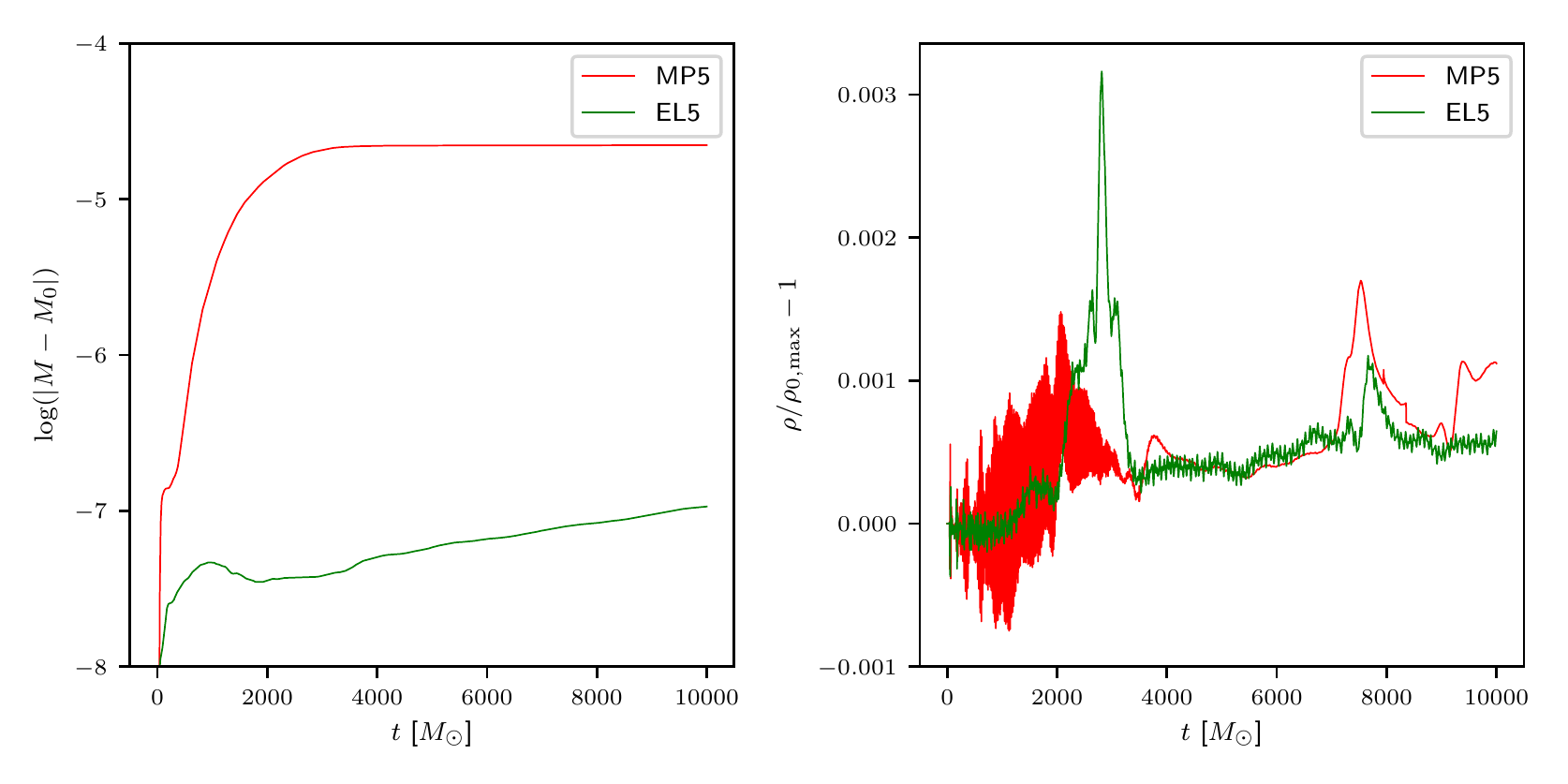}}
    \caption[Cowling TOV test: total mass and central density
    evolution]{Deviation of the total rest mass (left panel) and central
    rest-mass density (right panel) from the initial values for the Cowling TOV
    test.}
    \label{fig:tovc_mr}
\end{figure*}

The quantitative differences between the two schemes are better captured
in Fig. \ref{fig:tovc_mr}, where the evolution of the total rest mass and
of the central rest-mass density are shown. We recall that the total rest
mass (or baryon mass), is defined as
\begin{equation}
    M := \int \rho W \sqrt{\gamma}\, d^3x\,,
\end{equation}
where the integral is performed over the whole computational domain. From the
continuity equation \eqref{eq:GRcontinuity} follows that it should be conserved
in absence of a net total flow of matter in or out of the domain. The numerical
schemes we employ are conservative (see \eg \cite{Leveque92}), and therefore
preserve the value of the rest mass to the one determined by the initial data.
Nonetheless, violations of this conservation can take place in at least three
different ways. First, winds originating at the stellar surface (physically, as
\eg in binary neutron star merger, or spuriously as in a stationary case such
as the present one) can yield a net loss of mass when they reach the outer
boundary and leave the computational domain. Second, matter can be spuriously
created or destroyed, in a way that is hard to control, because of
floating-point or interpolation errors at the boundaries of refinement levels
(this is not the case for this particular test clearly, since we employ a
single grid, but it is relevant for the following ones). Finally, when a value
of the density is floored mass is spuriously created or destroyed. It is
therefore important to characterize the interplay between the numerical scheme
and these grid related effects.

The left panel of Fig. \ref{fig:tovc_mr} shows deviations, in absolute
value, of the rest mass from the initial value for the two schemes. The
EL5 scheme is evidently much better at conserving mass in this test than
MP5, leading to a cumulative deviation of $\sim\,10^{-7}\,\msun$ which is
almost three orders of magnitude smaller than the MP5 value.

The central rest-mass density also undergoes an evolution (right panel of
Fig. \ref{fig:tovc_mr}), with oscillations triggered by the treatment of
the stellar surface. Both schemes perform at a similar level of accuracy,
with relative variations from the initial value no greater than about
$0.3\%$ (even though spurious peaks are present in both data series at
various times). The short term behaviour of the two schemes is noticeably
different, and the frequency content in the two data series appears
different, with the MP5 scheme seeming to show more pronounced
high-frequency modes. However, at later times both schemes appear to
relax and oscillations decrease significantly in amplitude.

\begin{figure}[!t]
    \centerline{\includegraphics[width=0.9\columnwidth]{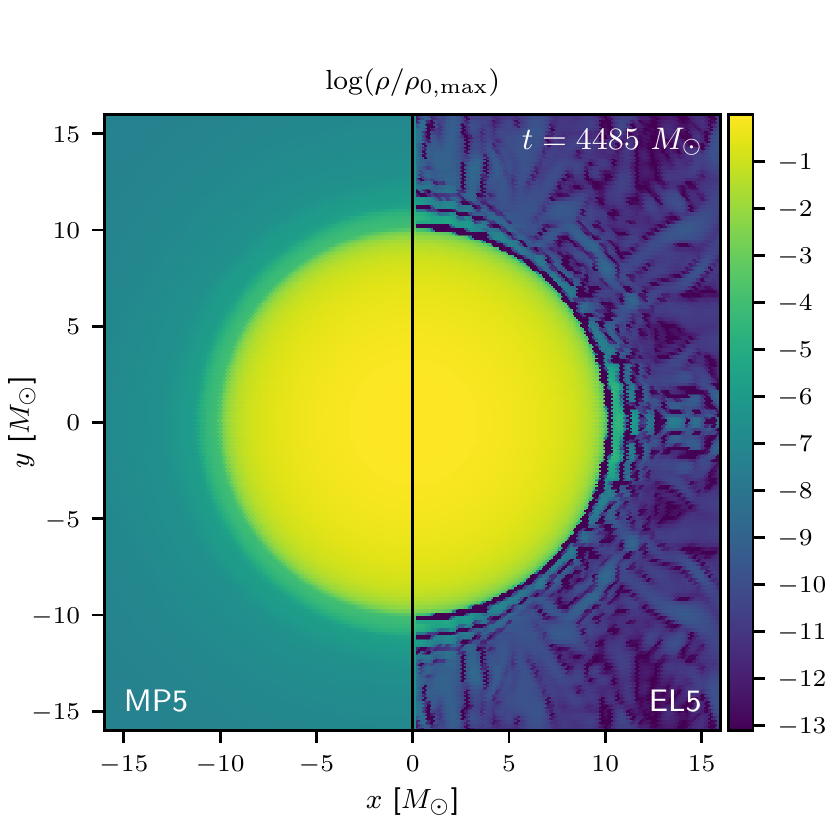}}
    \caption[Dynamical TOV test: density on $(x,y)$ plane]{Two-dimensional
    rest-mass density distribution relative to the initial data maximum value
    on the equatorial $(x,y)$ plane at time $t=4485\,\msun$ for the dynamical
    TOV test. The matter tails are even more extended in MP5 case compared with
    the Cowling test, EL5 instead preserves its behaviour at the stellar
    surface and exterior.}
    \label{fig:tovd_2Drho}
\end{figure}

\begin{figure*}[!t]
    \centerline{\includegraphics[width=1.8\columnwidth]{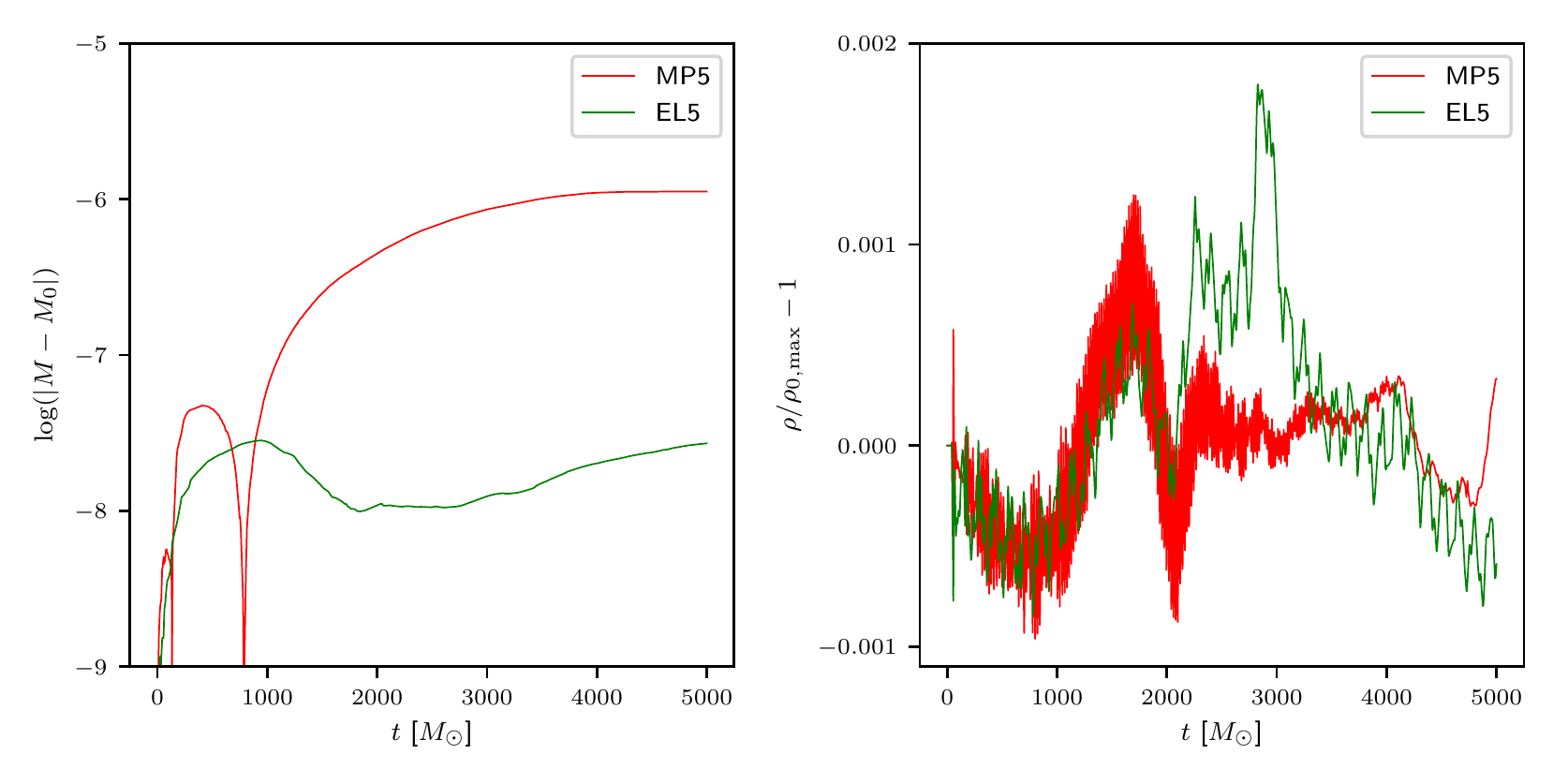}}
    \caption[Dynamical TOV test: total mass and central density
    evolution]{Deviation of the total rest mass (left panel) and central
    rest-mass density (right panel) from the initial values for the dynamical
    TOV test.}
    \label{fig:tovd_mr}
\end{figure*}

\subsubsection{Isolated star in a dynamical spacetime}
\label{sec:TOVD}

We then proceed to relax the Cowling approximation and test the entropy-limited
method coupled with a dynamically evolved spacetime. As first step, we evolve
the same initial data for a isolated stable star as in the previous section
(\ie with central density $1.28 \times 10^{-3}\,\msun^{-2}$, baryon mass of
$1.5\,\msun$ and radius $\sim\,10\,\msun$). We perform the test on a grid
consisting of three refinement levels centered on the star with sides lengths
$16$, $32$ and $60$ $\msun$ from finest to coarsest, and with a constant
refinement factor of $2$. The spatial resolution of the innermost and finest
level is set to $\Delta^i=0.2\,\msun\simeq0.3$ ${\rm km}$, and the timestep to
$0.15$ times the grid spacing. This factor is largest possible to guarantee the
positivity of the rest-mass density (see discussion in section
\ref{sec:viscosityandpplim} and Ref. \cite{Radice2013c} for details). The
atmosphere value of the density is set to $10^{-16}$ $\msun^{-2}$, that is,
almost 13 orders of magnitude smaller than the maximum value. As a time
integrator we select the third-order SSP Runge-Kutta RK3. Unless stated
differently, we employ the same grid setup for each one of the following
single star tests.

Figure \ref{fig:tovd_2Drho} shows the distribution of rest-mass density
on the equatorial plane for this test, again with the MP5 and EL5 schemes
shown on the left and right parts of the panel, respectively. It can be
appreciated how the MP5 scheme produces rest-mass tails which are even
more dense and extended than in the Cowling case, making the near vacuum
solution obtained by the EL5 scheme all the more striking.

Another difference from the Cowling test can be seen in the conservation of the
rest mass (left panel of Fig. \ref{fig:tovd_mr}). In this case too, the EL5
scheme is able to conserve the initial value to an accuracy roughly two orders
of magnitude better than the MP5 scheme. Furthermore, it is interesting to
notice how the behaviour of the EL5 scheme is much more smooth and predictable;
MP5 by contrast displays both spurious losses and gains of mass, which lead to
the zero crossings clearly visible in the figure. This is due to interpolation
errors arising during the restriction and prolongation operations between
different refinement levels. These errors are more severe with MP5 due to the
presence of long tails of low density matter in the stellar exterior, as we
checked by varying the extent of the refinement levels. In contrast, the EL5
scheme is less affected since the exterior of the star (especially away from
the coordinate axes) is nearly vacuum.

The evolution of the central rest-mass density, as shown in the right
panel of Fig. \ref{fig:tovd_mr}, is similar to the one shown in the
previous section for the Cowling approximation, with both schemes varying
no more than $0.2\%$ from the initial value, but with MP5 displaying
oscillations at much higher frequency.

To further investigate this point, we compute the power spectral density
(PSD) of the density evolution, in order to quantitatively gauge the
differences between the two schemes. The PSD is computed over the first
$5000\,\msun$ of data and with the use of a Hann window function. Before
computing the PSD, any linearly growing component of the signal is
removed via a least-squares fit.

\begin{figure}[!t]
    \centerline{\includegraphics[width=0.9\columnwidth]{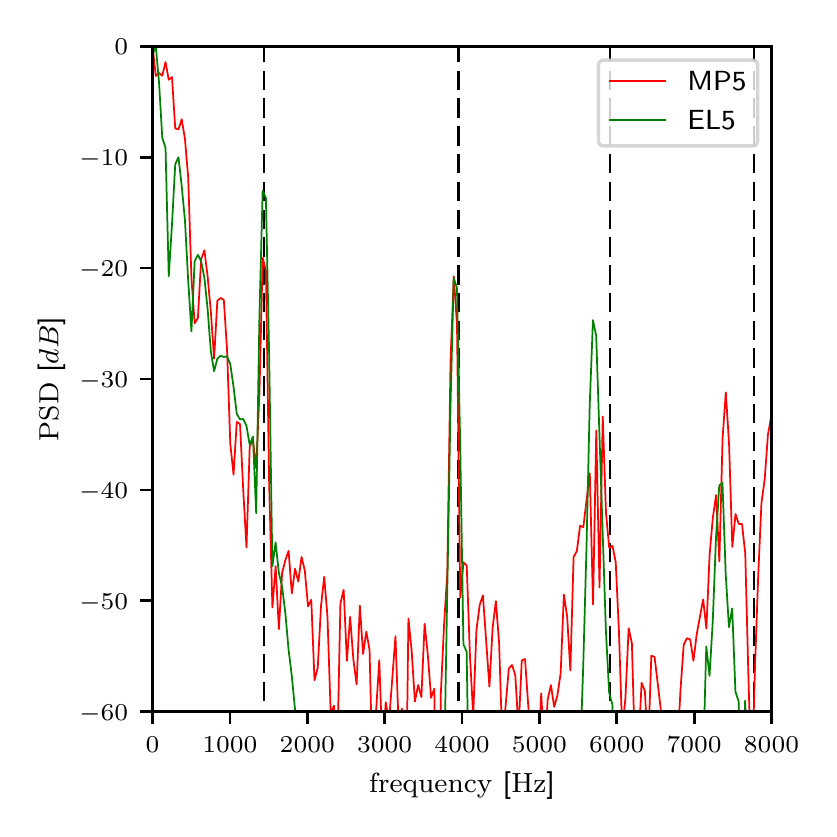}}
    \caption[Dynamical TOV test: PSD of central density]{PSD of the central
    rest-mass density evolution and physical eigenfrequencies of the stellar
    model for the dynamical TOV test. Note the good agreement between the first
    eigenfrequencies and the peaks in the data.}
    \label{fig:tovd_PSD}
\end{figure}

The PSDs for both schemes are shown in Fig. \ref{fig:tovd_PSD} along with
the oscillation frequencies of this stellar model as computed
perturbatively following the methods discussed in Refs. \cite{Yoshida01,
  Takami:2011}, and shown as vertical dashed lines. For both schemes, the
PSD is dominated by a low-frequency component due the well-known secular
changes in the central rest-mass density \cite{Font02c} and that
disappear with resolution. However, peaks are clearly visible above the
noise. The lowest-frequency peaks correspond to the fundamental
oscillation mode of the star and its first overtone, while the following
ones are higher overtones and are progressively more offset from the
corresponding perturbative eigenfrequencies. The peaks in the EL5 data
appear to be more clearly identifiable and less broad than in the MP5
case. Above $\sim\,8000~{\rm Hz}$, and not shown in Fig. \ref{fig:tovd_PSD},
the MP5 scheme shows significant high-frequency components, clearly
visible in the first part of the corresponding curve in
Fig. \ref{fig:tovd_mr}, but with a rather disordered spectrum. These same
frequencies are instead greatly suppressed in the EL5 scheme. Overall,
also this test highlights that the EL5 scheme captures quite well the
physical behaviour of the system as expected from perturbative methods
and is free from some of the artefacts which appear instead in the
evolution with the MP5 scheme.

\subsubsection{Perturbed isolated star}
\label{sec:KTOV}

The next test we perform is a slight modification of previous setup, \ie we
evolve the same isolated neutron star model, but applying a small velocity
perturbation to the initial solution. The perturbation consists of a radially
outgoing velocity growing linearly in radius to a maximum value of $0.005$.

We employ this scenario, more realistic than the simple smooth-wave test of
section \ref{sec:sw}, to measure the convergence order of the EL5 and MP5
methods. We performed three sets of simulations at resolutions $0.24$, $0.12$
and $0.06$ $\msun$ on the finest level, extracting the evolution of the
rest-mass density over time from each one. The initial perturbation is added so
that the density evolution is not dominated by the truncation error, but
displays a cleaner behaviour. Otherwise, as the resolution is increased, the
density evolution would show additional high-frequency modes, which would make
the dependence on resolution discontinuous, making it difficult to compute the
instantaneous convergence order.

\begin{figure}[!t]
    \centerline{\includegraphics[width=0.9\columnwidth]{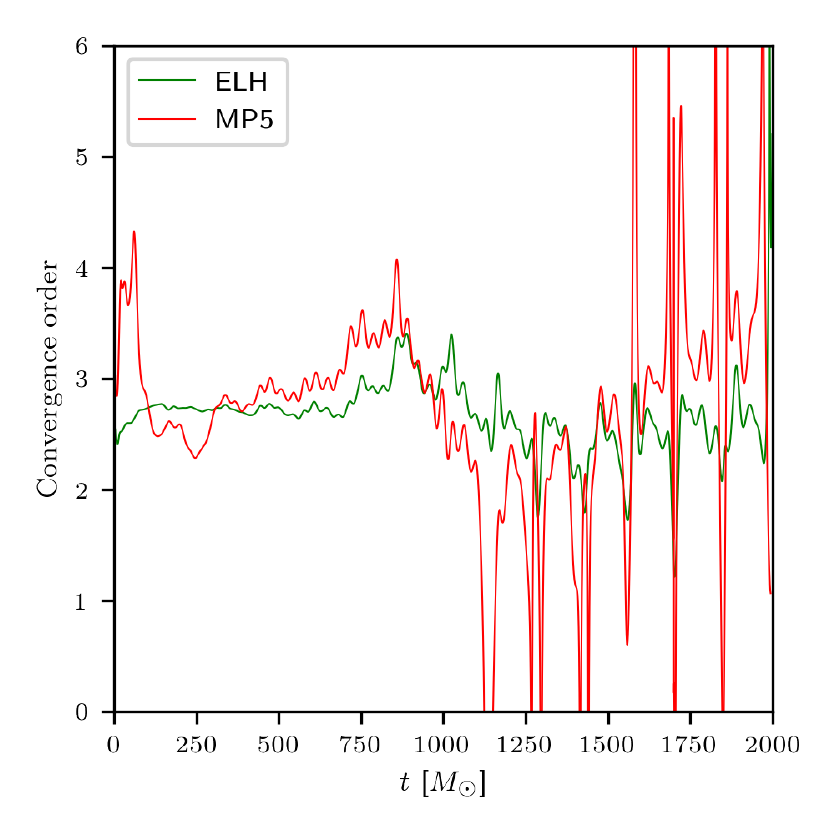}}
    \caption[Perturbed TOV test: convergence order]{Instantaneous convergence
    order measured in a perturbed TOV simulation, computed from the $L_1$-norm
    of the rest-mass density.}
    \label{fig:ktov_convergence}
\end{figure}

Using the values of the $L_1$-norm of the rest-mass density over the
domain at the three resolutions we also computed the instantaneous
convergence order $\msun$ as shown in
Fig. \ref{fig:ktov_convergence}. Because this is the instantaneous
convergence order and because the underlying system is oscillating, the
curves are somewhat noisy (especially for MP5); however, when taking the
running average, both schemes generally show a convergence order just
below three, consistent with the results in
Refs. \cite{Radice2013b,Radice2013c}. It is also however apparent how EL5
maintains a fairly constant order of convergence through time, while the
behaviour of MP5 is more irregular, especially at later times.

While the hydrodynamics schemes are both formally fifth-order accurate,
other parts of the algorithm operate at different degrees of accuracy. In
particular, the time integrator is third-order accurate, which most
likely accounts for the convergence order being closer to three than to
five. The result is also consistent with the ones found for the MP5 scheme
in \cite{Radice2013b,Radice2013c,Radice2015}. Overall, this test
highlights how both the ELH and MP5 schemes perform fairly consistently
over time, with no major loss of accuracy.

\subsubsection{Migration test}
\label{sec:MIG}

Another important test in our series is the migration of a TOV star
moving from a solution on the unstable branch of equilibrium solutions to
a stable one. We recall that for any given EOS, increasingly massive but
stable TOV models can be constructed by considering increasingly large
values of the central rest-mass density. This can continue until a
maximum mass is reached, at which point, an increase of the central
rest-mass density corresponds to a decrease of the mass of the
star. Models on this second branch of the mass/ central-rest-mass density
curve are unstable, and if a perturbation is present will evolve to
either a stable configuration or collapse to a black hole. This is
precisely the physical scenario that the migration test simulates: we
construct a model on the unstable branch of the mass/density curve and
force its ``migration'' to a stable configuration by applying a suitable
velocity perturbation.

This is a common test for numerical relativity codes (see, \eg \cite{Font02c,
Baiotti04, Baiotti03a, Cordero2009, Thierfelder2011}), and has been studied in
detail in \cite{Radice:10}. In particular, we build a nonrotating stellar model
on the unstable branch of the equilibrium solutions and with central rest-mass
density of $7 \times 10^{-3}\,\msun^{-2}$ (yielding a total rest mass of
$1.6\,\msun$ and a radius of $6\,\msun$). The migration is then triggered by
injecting a radially outgoing velocity perturbation where the velocity grows
linearly in radius, reaching a maximum value of $0.01$. The star then undergoes
a series of violent expansions and contractions as it migrates to the stable
branch and then settles on the new equilibrium. During each contraction and
expansion strong shocks are formed, and the shocked matter is ejected at large
velocities.

\begin{figure}[!t]
    \centerline{\includegraphics[width=0.9\columnwidth]{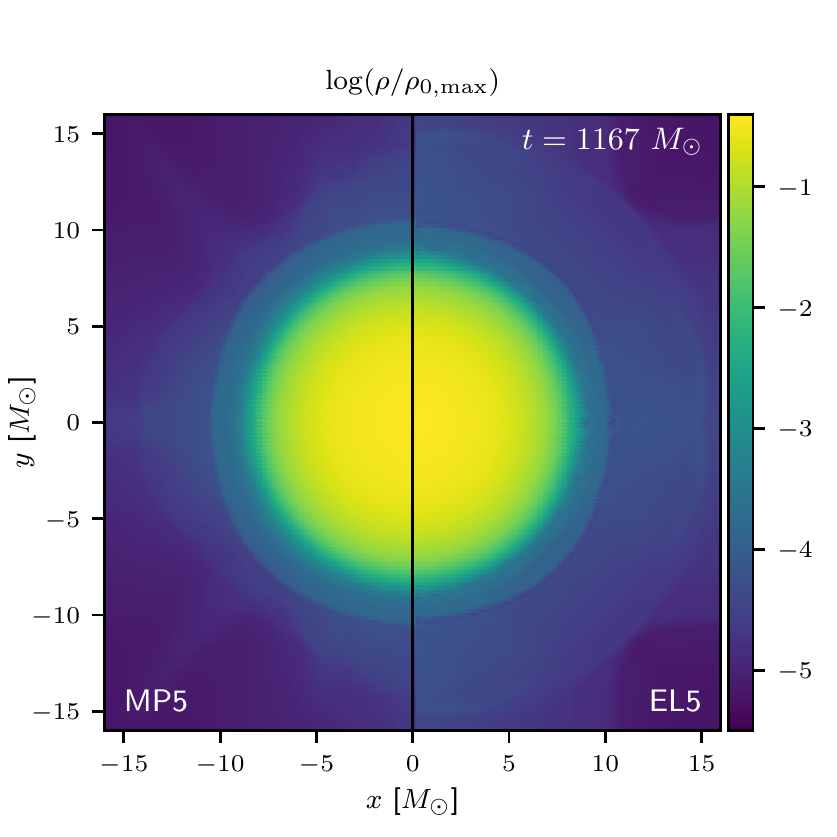}}
    \caption[Migration test: density on $(x,y)$ plane]{Two-dimensional
    rest-mass density distribution relative to the initial data maximum value
    on the equatorial $(x,y)$ plane at time $t=1167\,\msun$ for the migration
    test. Virtually no difference can be detected in the two schemes
    behaviour.}
    \label{fig:mig_2Drho}
\end{figure}

\begin{figure}[!t]
    \centerline{\includegraphics[width=0.9\columnwidth]{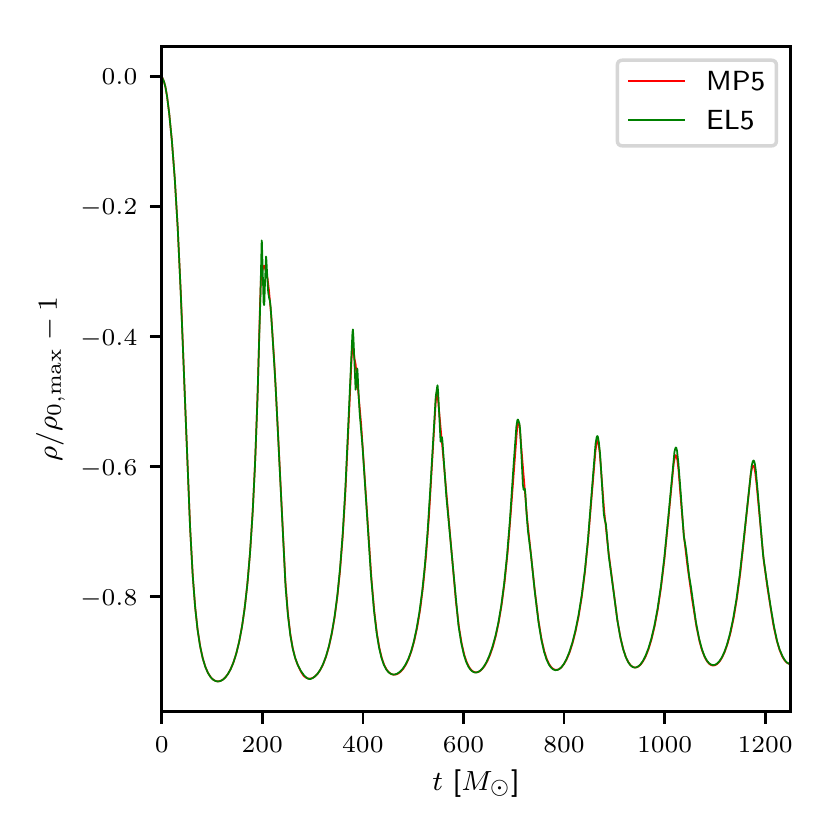}}
    \caption[Migration test: central density evolution]{Central rest-mass
    density in the migration test. The agreement between the two schemes is
    apparent over the whole evolution, apart from high-frequency modes at the
    maxima in EL5 data.}
    \label{fig:mig_r_all}
\end{figure}

\begin{figure}[!t]
    \centerline{\includegraphics[width=0.9\columnwidth]{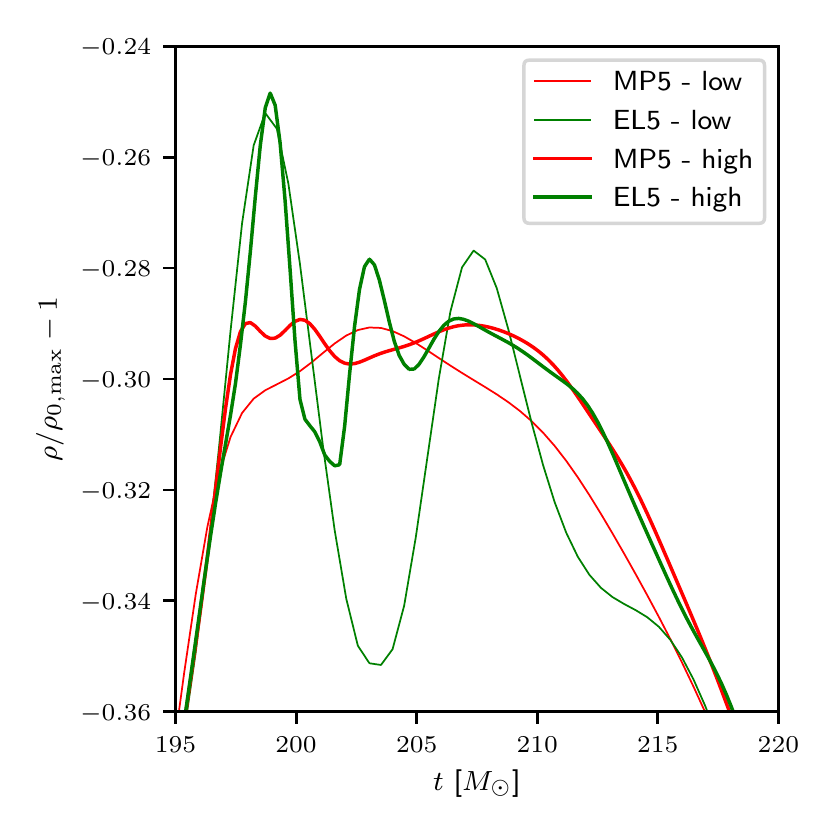}}
    \caption[Migration test: central density evolution at different
    resolutions]{Magnification of the central rest-mass density evolution
    around the first contraction of the migrating star. The low-resolution data
    (thin curves) corresponds to Fig. \ref{fig:mig_r_all} and to a grid spacing
    of $0.2\,\msun$. The high-resolution data (thick curves) corresponds to a
    grid spacing of $0.086\,\msun$. The high-frequency modes present in the EL5
    data at low resolution persist at high resolution in both schemes.}
    \label{fig:mig_r_zoom}
\end{figure}

In Fig. \ref{fig:mig_2Drho} we show the rest-mass density distribution on
the equatorial plane for both schemes and during one of the contractions
of the star, just before the central rest-mass density reaches a maximum
(\cf Fig. \ref{fig:mig_r_all}). The snapshot clearly shows that both the
EL5 and MP5 schemes produce almost identical results for this test. This
is not surprising and mainly due to the matter outflow driven by the
stellar oscillations, which rapidly fills the domain and removes the
sharp feature of the stellar surface, which is the most problematic
structure to resolve and the main difference in the two schemes.

The evolutions of the maximum rest-mass density are shown in
Fig. \ref{fig:mig_r_all}, where they are reported as normalized to the
initial value. The agreement between the two schemes is extremely good
during the entire evolution and the main difference between the two
solutions is the presence of some high-frequency modes near the maxima of
the density in the EL5 data. Such oscillations are the result of
inward-propagating shock waves generated in the outer layers of the star
during the contraction phase. Figure \ref{fig:mig_r_zoom} shows a
magnification of the behaviour of the maximum rest-mass density at the
peak of the first contraction, comparing not only the two schemes but
also the evolutions with two different resolutions. At high resolution,
both the MP5 and the EL5 scheme show small-scale and high-frequency
oscillations that are less pronounced in the low-resolution
data. Interestingly, these oscillations are essentially smoothed out in
the low-resolution run of the MP5 scheme, while they are very visible in
the low-resolution EL5 run. This seems to indicate that the two schemes
tend, with increasing resolution, towards a solution where the
small-scale oscillations are present and therefore physically correct and
not a numerical artefact. Finally, as the evolution progresses, the
contraction/expansion phases become less and less violent as part of the
kinetic energy is converted into internal energy, thereby leading to
milder and milder shocks, and the high-frequency oscillations in the
central rest-mass density all but disappear.

\subsubsection{Isolated rotating neutron star}
\label{sec:RNS}

As the last test case for a stable (or metastable) isolated relativistic
stars we consider the evolution of a rapidly and uniformly rotating
star. More precisely, we set up axisymmetric initial data relative to a
uniformly rotating neutron star governed by a polytropic EOS with
$K=100\,\msun^{-2}$ and $\Gamma=2$, having a central rest-mass density of
$1.28 \times 10^{-3}\,\msun^{-2}$ and a polar to equatorial axis ratio of
$0.8$ using the \verb+RNS+ code \cite{Stergioulas95}. This results in a
star with total rest mass $1.6\,\msun$, radius $10\,\msun$, rotation
frequency $f=673.2~{\rm Hz}$ (about $60\%$ of the mass shedding
frequency) and dimensionless angular momentum $J/M^2=0.46$. Also in this
case the spacetime is evolved in time despite the solution being
stationary.

\begin{figure}[!t]
    \centerline{\includegraphics[width=0.9\columnwidth]{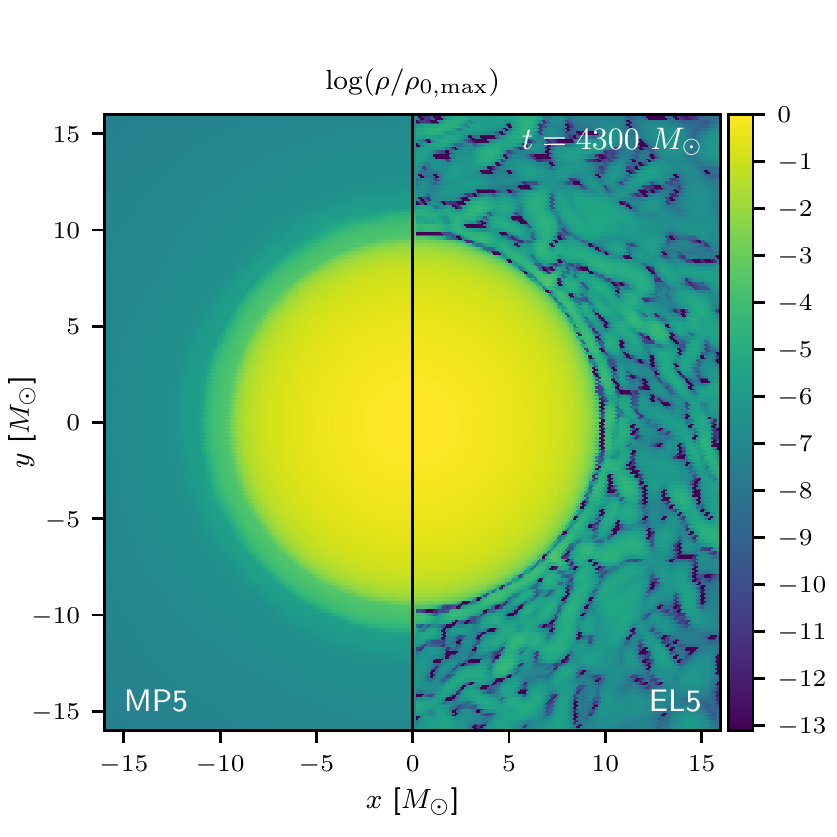}}
    \caption[Rotating star test: density on $(x,y)$ plane]{Two-dimensional
    rest-mass density distribution relative to the initial data maximum value
    on the equatorial $(x,y)$ plane at time $t=4300\,\msun$ for the rotating
    star test. The behaviour of the star exterior is dynamic and chaotic with
    EL5 as compared with MP5.}
    \label{fig:rns_2Drho}
\end{figure}

In Fig. \ref{fig:rns_2Drho} we report again the rest-mass density
distribution on the equatorial plane for both schemes and at time
$t=4300\,M_{\odot}$, that is, after about 14 rotation periods. The figure
clearly illustrates that that both schemes evolve the rotating star with
no noticeable problems and that, as already seen in the case of
nonrotating stars, the part of the domain exterior to the stellar surface
rapidly fills with matter. Also in this case, the behaviour of the two
methods in the low-density regions is rather different, with the MP5
scheme yielding to a volume which is filled by uniform but comparatively
higher-density material, while the EL5 scheme produces a stellar exterior
which has lower-density matter but with small-scale condensations (\cf
Figs. \ref{fig:tovc_2Drho} and \ref{fig:tovd_2Drho} for the equivalent
behaviour in the absence of rotation).

\begin{figure*}[!t]
    \centerline{\includegraphics[width=1.8\columnwidth]{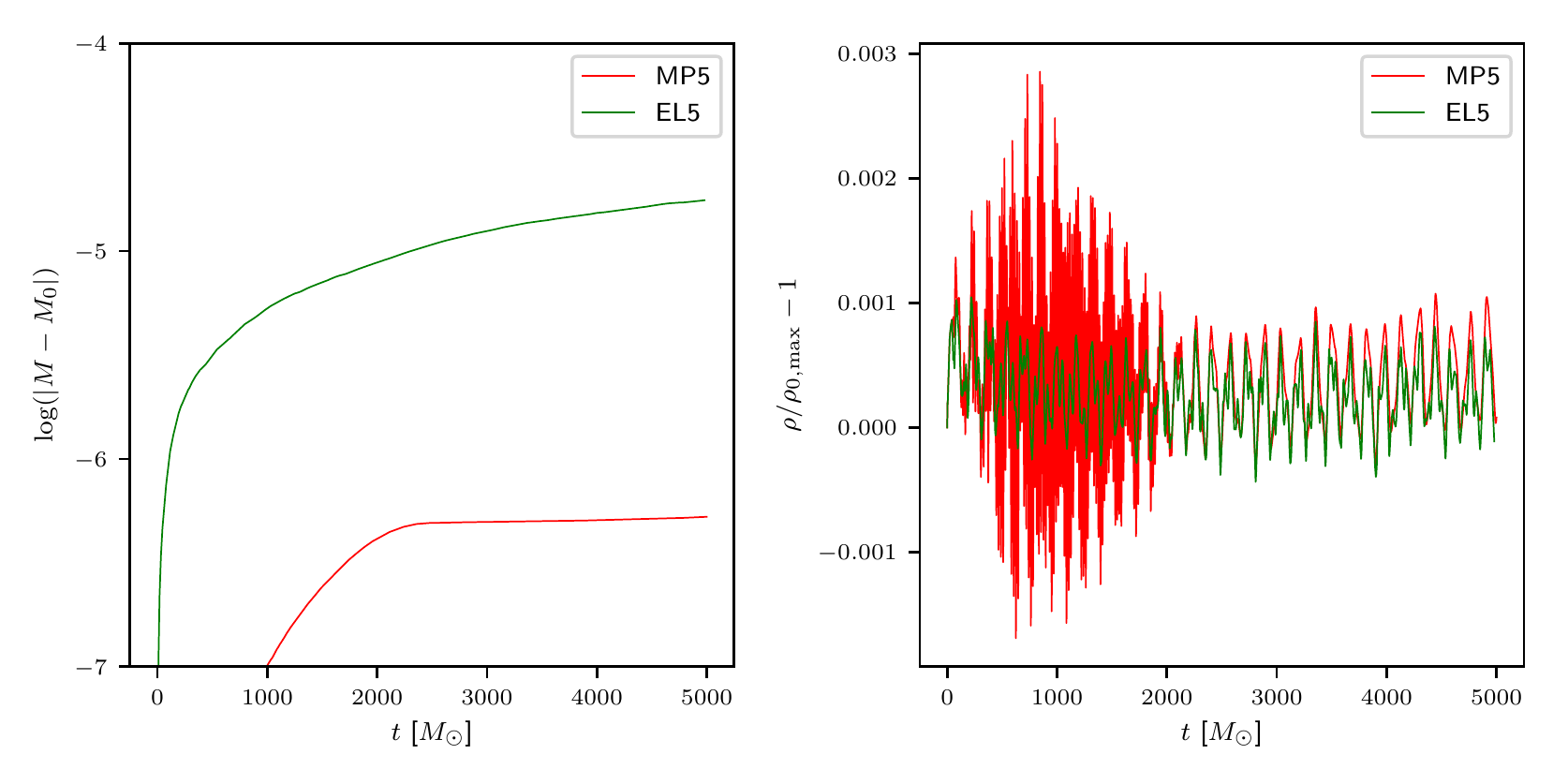}}
    \caption[Rotating star test: total mass and central density
    evolution]{Deviation of the total rest mass (left panel) and central
    rest-mass density (right panel) from the initial values for the
    rotating-star test.}
    \label{fig:rns_mr}
\end{figure*}

However, in contrast with the behaviour seen in the case of nonrotating
stars, the dynamics of the low-density material in the stellar exterior
results in a degradation of the conservation of mass for the EL5 scheme,
as shown in the left panel of Fig. \ref{fig:rns_mr} (\cf the left panels
of Figs. \ref{fig:tovc_mr} and \ref{fig:tovd_mr} for the equivalent
behaviour in the absence of rotation). The deviation of the total
rest-mass density from its initial value is more than one order of
magnitude larger for the EL5 scheme than for MP5 and reaches values of
$\sim\,10^{-5}\,\msun$. This is the result of failures in the conversion
from the conserved variables to the primitive ones, triggered by
oscillations in the solution: the solution can be evolved to an
unphysical state, and in this case the rest-mass density could reach
values below the atmosphere floor value; if so, the conversion routine
resets the affected cells to the atmosphere value, thereby spuriously
creating mass. We speculate that most of these failures result from the
large tangential velocity that is acquired by the shell-like distribution
of matter that builds up in the case of the EL5 scheme and that is
present already in the nonrotating case. While rather innocuous in the
absence of rotation, this shell of matter can fling material to large
distances (but within the computational domain) and lead to a much more
chaotic dynamics of the fluid in the low-density regions (see the
discussion in sec. 3.2.3 of \cite{Radice2013c}).

To assess the impact of the fluid dynamics in the stellar exterior we
report the evolution of the central rest-mass density for the two schemes
in the right panel of Fig. \ref{fig:rns_mr}. We find that the
low-density fluctuations appearing in the stellar exterior with the EL5
scheme do not impact the solution in the stellar interior, with the
low-frequency central density oscillations essentially being in phase for
the two schemes. Also quite apparent is that the EL5 scheme yields
rather constant-amplitude oscillations and this should contrasted with
the MP5 scheme, where the oscillations are comparatively larger in the
first $\sim\,2000\,M_{\odot}$ of the evolution. In both cases, however,
the oscillations are extremely small and below $0.1\%$.

\subsubsection{Grazing collision of neutron stars}
\label{sec:BTOV}

We further test the ELH scheme in another truly dynamical test: the
motion across the numerical grid of two neutron stars in a grazing
collision. This is a setup that is very similar to that of a
binary-neutron star system in quasi-circular orbit, the most obvious
difference being the initial momenta of the two stars do not result in a
quasi-circular orbit and that the initial fluid velocity can be taken to
be arbitrary. In practice, the initial data is set up by generating two
identical TOV models (the same as considered in sections \ref{sec:TOVC}
and \ref{sec:TOVD}), superimposing the two data sets on the computational
grid and imparting suitable initial momenta resulting in a small, but
nonzero, impact parameter. Clearly, such initial data is valid only as a
first approximation since the stars are not in the hydrostatic
equilibrium corresponding to the binary system and the intial metric and
extrinsic curvature do not reflect a solution of the Einstein constraints
equations.

These violations lead to initial oscillations in the evolution (see
\cite{Kastaun2013,Tsatsin2013} for a more detailed discussion of a more
sophisticated setup in which the stars are also subject to a spin up)
which can however be reduced significantly by setting the initial
distance of the two stars to a rather large value. More importantly,
however, these oscillations do not interfere with the main goal of this
test, namely, that of validating the ability of the ELH scheme to
preserve sharply the features of the stellar surface also when the star
moves across the numerical grid.

\begin{figure}[!t]
    \centerline{\includegraphics[width=0.9\columnwidth]{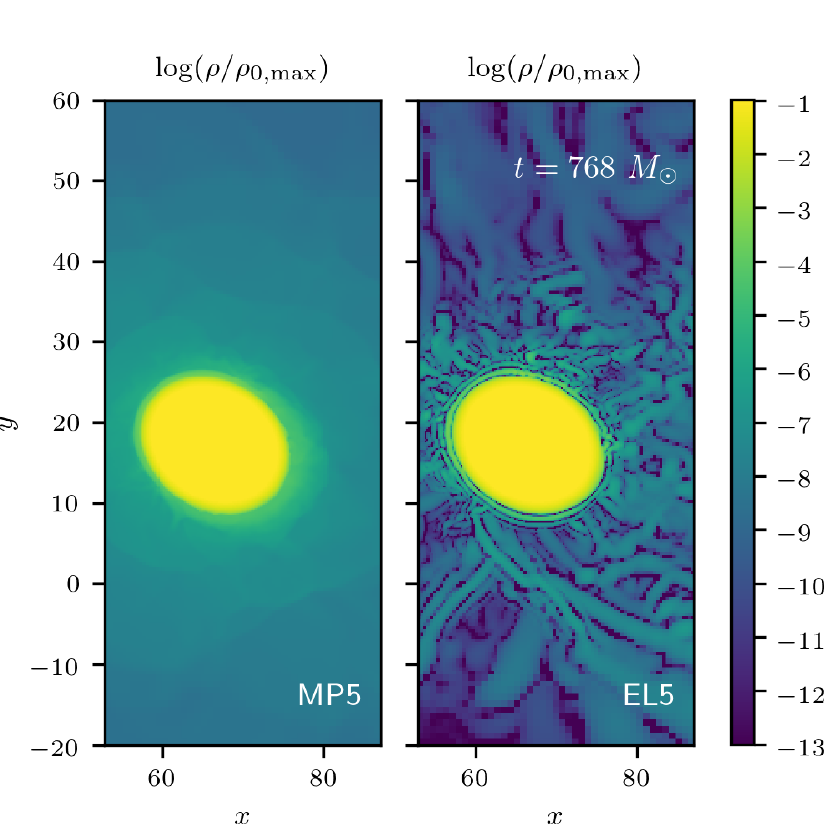}}
    \caption[Grazing collision test: density on $(x,y)$ plane]{Two-dimensional
    rest-mass density distribution relative to the initial data maximum value
    on the equatorial $(x,y)$ plane for the grazing-collision tests at time
    $t=768\,\msun$, \ie after the point of closest approach.}
    \label{fig:btov_2Drho}
\end{figure}

\begin{figure*}[!h]
    \centerline{\includegraphics[width=1.8\columnwidth]{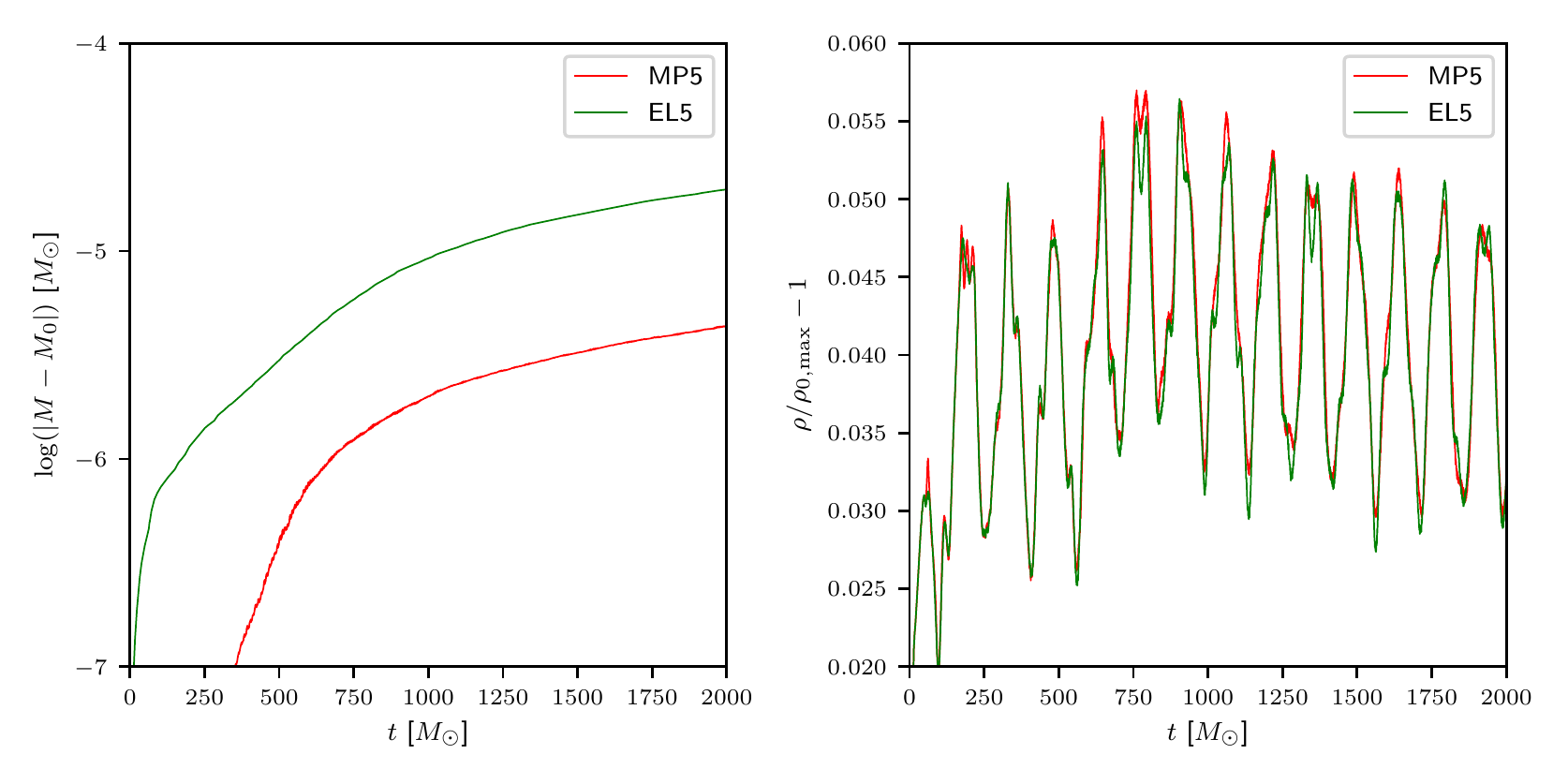}}
    \caption[Grazing collision test: total mass and central density
    evolution]{Deviation of the total rest mass (left panel) and central
    rest-mass density (right panel) from the initial values for the
    grazing-collision test.}
    \label{fig:btov_mr}
\end{figure*}

More in detail, the star centers are set at positions
$(x_1,y_1,z_1)=(50,-50,0)$ and $(x_2,y_2,z_2)=(-50,50,0)$ in units of
$\msun$, \ie symmetric with respect to the grid center on the $(x,y)$
plane and at a distance of $\sim\,141$ $\msun$. The initial 3-velocities
are $(v^x_1,v^y_1,v^z_1)=(0,-0.1,0)$ and $(v^x_2,v^y_2,v^z_2)=(0,0.1,0)$
respectively. We evolve the system on a cubic grid of radius $512$
$\msun$, but employ reflection symmetry boundary conditions across the
$(x,y)$ plane and 180 degrees rotation symmetry boundary conditions
across the $(y,z)$ plane to reduce the computational cost. The grid
structure consists of two identical box-in-box refinement levels
hierarchies with refinement factor 2, each centered on a star and
consisting of 5 cubic levels with radii $12,25,50,100,200$ $\msun$, plus
the coarse base level with radius $512$ $\msun$, so that the grid spacing
in the innermost refinement level is $\Delta^i=0.2\,\msun\simeq0.3$ ${\rm
  km}$.  The refinement levels moved to track the positions of the stars
during the evolution (see also \cite{Radice2016} for further details on
the initial data and grid structure). We set again $\Delta t=0.15 \,
\Delta x$.

The evolution consists in the two stars initially traversing the grid in
the $x$ direction and approaching each other, then bending their
trajectories as in a gravitational scattering process; we do not follow
the dynamics of the process after the first fly-by. Figure
\ref{fig:btov_2Drho} shows the rest-mass density distribution on the
$(x,y)$ plane for this test. The snapshots of one of the two stars are
taken at time $t=768$ $\msun$, when the two stars are past the point of
closest approach and are escaping. One can appreciate the deformation due
to the boost, the acquired spin angular momentum and the tidal
gravitational interaction. From the hydrodynamics point of view the
behaviour of the MP5 and EL5 schemes is consistent with the previous
tests, in particular the TOV in a dynamical spacetime: EL5 shows a
sharper star surface with respect to MP5, as well as an ``emptier''
surrounding region, while the bulk of the star itself is very well
resolved by both schemes.

The conservation of the total rest mass (left panel of
Fig. \ref{fig:btov_mr}) is instead similar to the rotating-star case,
thus showing a better performance by the MP5 scheme. Note however that
the differences between the schemes are far smaller in this case, less
than one order of magnitude. Furthermore, in contrast with the preceding
tests, the grid structure in this case is much more complicated as well
as dynamically updated to track the stars. Interpolation errors at the
refinement level boundaries play therefore a greater role in the
conservation of rest mass.

The evolution of the rest-mass density at the star centers, as shown in
the right panel of Fig. \ref{fig:btov_mr}, is very similar for both
schemes. There is an initial sudden increase in the density of about 4\%
with respect to the initial value, due to the evolution scheme bringing
the star in hydrostatic equilibrium from the initial state. The density
then oscillates around this new value, due to perturbations that in this
case are not only induced by the violation of the constraint equations
but also by the gravitational interaction. Overall, both schemes
reproduce well all of these effects and show a very good agreement.

\subsubsection{Gravitational collapse to a black hole}
\label{sec:COL}

\begin{figure}[!t]
    \centerline{\includegraphics[width=0.9\columnwidth]{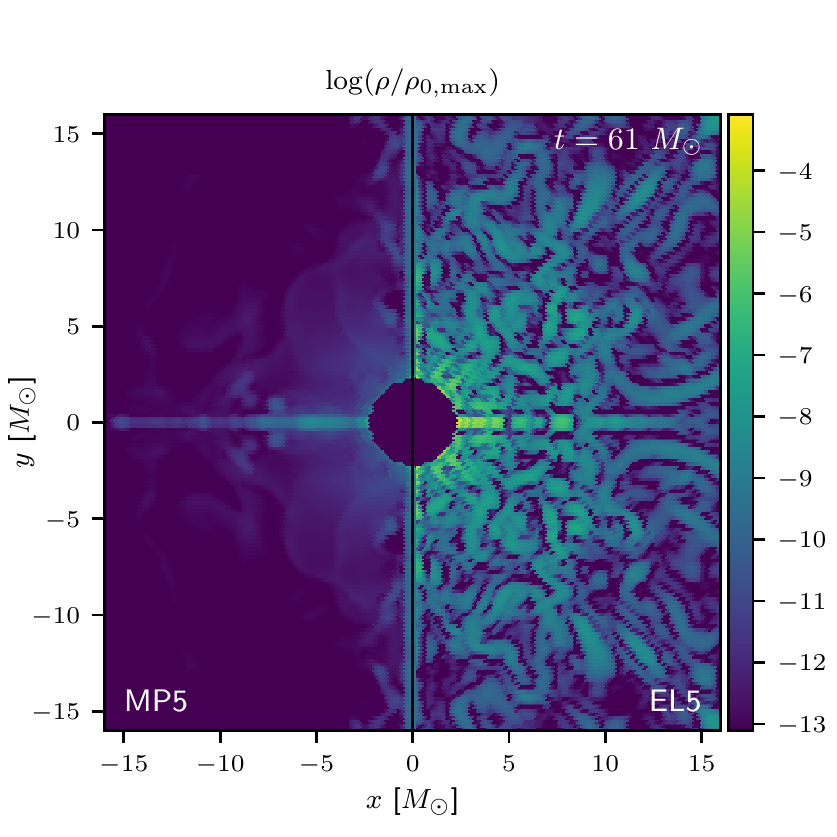}}
    \caption[BH collapse test: density on $(x,y)$ plane]{Two-dimensional
    rest-mass density distribution relative to the initial data maximum value
    on the equatorial $(x,y)$ plane at time $t=61\,\msun$ for the collapse
    test. Streams of matter ejected during collapse and accreting back onto the
    black hole are clearly visible and more prominent in EL5 data.}
    \label{fig:col_2Drho}
\end{figure}

As a final test we evolve the violently dynamical collapse of a star to a
black hole. This is also a common numerical-relativity benchmark (see,
\eg \cite{Font02c, Baiotti04b, Baiotti06, Thierfelder10}), which allows
us to validate ELH in the presence of a physical singularity and of an
apparent horizon. More specifically, we consider a nonrotating star with
central rest-mass density $8 \times 10^{-3}\,\msun^{-2}$, corresponding
to a baryon mass of $1.5\,\msun$ and radius $6\,\msun$, and initiate the
collapse with a velocity perturbation analogous to the one used in the
migration test, but with the opposite sign, \ie radially ingoing.

We define the time of black-hole formation as the instant at which an
apparent horizon is first detected on the numerical domain, which, given
the chosen setup, happens at $t\simeq48\,\msun$. Since we use singularity
avoiding slicing conditions, we do not need to excise the interior
spacetime of the black hole \cite{Baiotti06, Baiotti07,
Thierfelder10}. At the same time, we set the hydrodynamical variables
to their atmosphere values inside a surface with the same shape as the
apparent horizon, but radius $r=0.9\,r{_{\rm AH}}$ in every angular
direction, $r{_{\rm AH}}$ being the radius of the apparent horizon. This
``hydrodynamic excision'' is not strictly necessary as our code can
handle the collapse without it, regardless of the scheme we
employ. However, we have observed that its use improves the accuracy of
the subsequent evolution, most notably, it improves the behaviour of
the rest-mass density and we therefore choose to employ it.

\begin{figure}[!t]
    \centerline{\includegraphics[width=0.9\columnwidth]{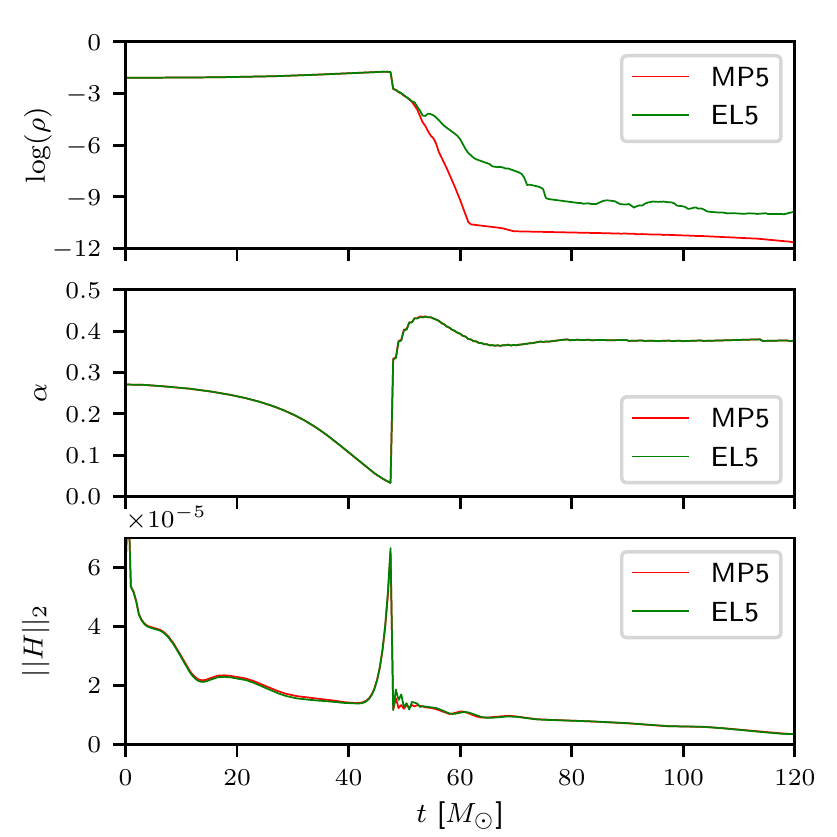}}
    \caption[BH collapse test: density, lapse and Hamiltonian constraint
    evolution]{Maximum rest-mass density (top), minimum lapse (middle) and
    $L_2$-norm of the Hamiltonian constraint (top) for the stellar-collapse
    test. Note that the violation of the Hamiltonian constrain will grow on
    longer timescales as it is typical of BSSNOK evolutions \cite{Alic2013}.}
    \label{fig:col_summary}
\end{figure}

Figure \ref{fig:col_2Drho} shows a snapshot of the rest-mass density on
the equatorial plane just after the collapse. The central area of uniform
low density is where the hydrodynamical variables have been set to
atmosphere inside the horizon, and in this plot appears of identical size
and shape for the two codes. The areas outside the horizon are instead
filled with matter which has been spuriously ejected from the outer layers
of the star star during the collapse. The rest-mass density is evidently
higher in the case of the EL5 scheme, due to a slightly higher proportion
of matter ejected during the collapse, which is in turn triggered by
larger oscillations around the stellar surface for the EL5
scheme. However, in both cases the total rest mass outside the horizon is
tiny, $\sim\,10^{-6}\,\msun$ for the EL5 scheme and $\sim\,10^{-9}\,\msun$
for MP5, and thus dynamically essentially irrelevant on the
properties of the solution. Furthermore, most of this matter is
gravitationally bound and hence accretes back onto the newly formed black
hole, forming streams of infalling matter. This is particularly evident
along the coordinate directions, where the numerical viscosity of the
high-order finite-differences stencil is smaller, independently of the
scheme employed.

The very close agreement between the two solutions is summarised in
Fig. \ref{fig:col_summary}, where the evolution of the central rest-mass
density, minimum lapse and $L_2$-norm of the Hamiltonian constraint is
plotted. The discontinuities in the curves at about $50\,\msun$
correspond to the time of collapse and arise because we exclude points
inside the horizon in the calculation of extrema and norms. In each panel
before the formation of the apparent horizon, the curves corresponding to
the EL5 and MP5 schemes are essentially on top of each other (the largest
differences being of the order of $0.3\%$, $0.6\%$, and $4.7\%$ for each
plot, respectively), showing the very good agreement in the evolution
between the two schemes. Note also that after the apparent horizon
formation and due to our approach of ``hydrodynamic excision'', the upper
panel of Fig. \ref{fig:col_summary} shows the maximum of the density in
the exterior of the horizon rather than the central density. The
disagreement in the EL5 and MP5 curves relates therefore to the tiny
amount of residual matter outside of the black hole, and as such it has
no dynamical impact on the black-hole solution.

\begin{figure}[!t]
    \centerline{\includegraphics[width=0.9\columnwidth]{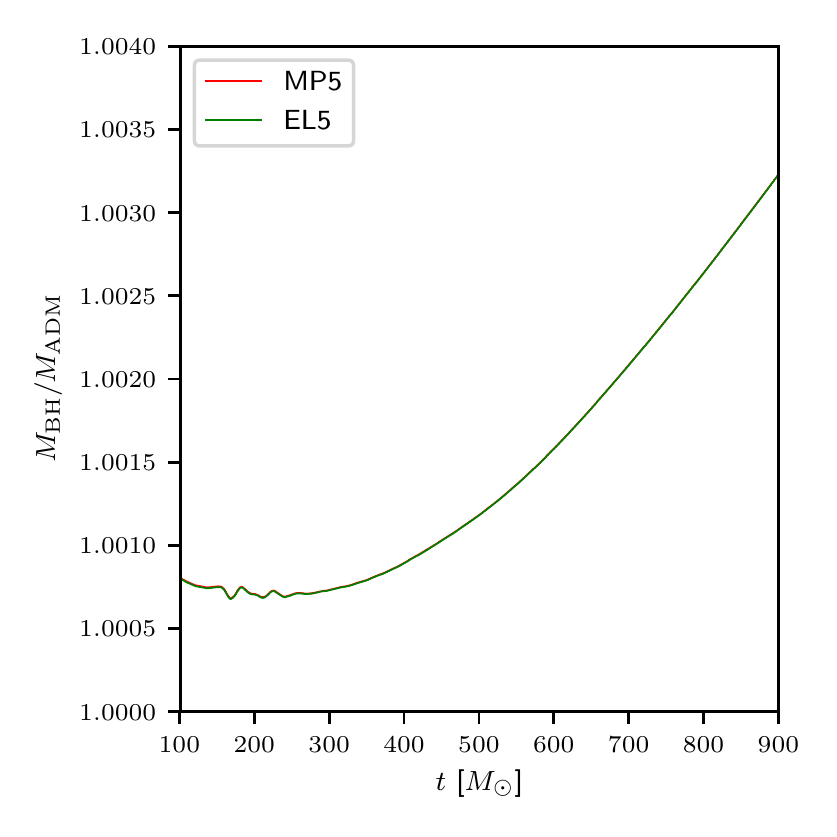}}
    \caption[BH collapse test: BH mass]{Ratio of the apparent-horizon mass to
    the ADM mass in the stellar-collapse test. Note that the growth is actually
    very small and is amplified here to show the difference between the two
    schemes.}
    \label{fig:col_hormass}
\end{figure}

A final confirmation of the equivalence between the two numerical
solutions comes from the comparison of the two black-hole masses computed
using the dynamical horizon formalism \cite{Ashtekar03a} and shown in
Fig. \ref{fig:col_hormass}. As can be seen from the figure, we find a
very close agreement between the two schemes.

\section{Conclusions}
\label{sec:conclusions}

We have presented a new high-order numerical method for the solution of the
Euler equations of general-relativistic hydrodynamics that we name
"entropy-limited hydrodynamics" (ELH). The scheme is of the flux-limiting type,
where a high-order numerical flux is combined with a stable low-order method,
namely the Lax-Friedrichs flux. The flux-limiting is activated and driven by a
shock indicator based on a measure of the entropy generated by the solution.
Such a special and general-relativistic method is inspired by the
entropy-viscosity method proposed recently for the solution of the classical
equations of hydrodynamics \cite{Guermond11}, but it is also importantly
different in that it does not require any change in the equations of
relativistic hydrodynamics.

To assess the robustness and accuracy of our new method, we have
discussed its implementation in the \verb+WhiskyTHCEL+ code, which
exploits the finite-difference capabilities of the \verb+WhiskyTHC+ code,
and tested its validity with an extensive series of tests, comparing the
results of ELH with those obtained with another well-tested and
high-order HRSC scheme: the fifth-order monotonicity-preserving MP5
method \cite{suresh_1997_amp}. Overall, we have found that the scheme is
stable and able to cope with shocks and discontinuities, both in
classical test such as shock-tube tests, as well as in realistic
astrophysical simulations.

Under all of these conditions the scheme has been found to be stable and to
yield accuracy that is comparable, if not better, of that of the MP5 method.
In some tests involving stars that are nonrotating or not moving across the
grid, it also offers definite advantages, such as a sharper resolution of the
surface/vacuum interface. At the same time, however, it also shows a less good
conservation of the rest mass for stars that are rotating or moving across the
computational domain (the opposite is true for stationary nonrotating stars,
where the new method conserves rest mass more accurately). Quite surprisingly,
all of the results presented here were obtained without any fine tuning of the
two arbitrary coefficients that enter the definition of the scheme. Finally,
thanks to its linearity and simplicity, the ELH method can also offer
advantages in terms of performance. In our tests, we have found EL5 to be
$\sim\,50\%$ faster than MP5, even though our implementation was not
particularly optimized. For instance, a definite advantage of ELH, which we did
not exploit, is that it can be easily vectorised. At the same time, we remark
that the exact speed-up that can be achieved with ELH depends also on external
factors, such as the grid setup and number of ghost zones, which may vary for
different applications. An interesting development in this sense would be the
use of this scheme in a discontinuous Galerkin framework, whose superior
scalability properties should decouple the performance of the ELH method from
the grid setup.

The work presented here could be improved in at least two ways. Firstly,
the already good capturing properties of steep gradients as those given
by the stellar surface, could be further enhanced and the full
capabilities of the scheme further exploited by coupling it to truly
multidimensional stencils. Second, the two free coefficients that appear
in the method, and that we have here set to unity for simplicity, could
potentially be tuned to optimise some of the features of the
solution. Both of these aspects will be explored in future work.

In conclusion, we have shown that entropy-limited hydrodynamics is a robust,
stable, and accurate alternative to commonly employed HRSC schemes. Its
performance reaches the level of accuracy and stability necessary to apply it
to realistic astrophysical simulations. Given these encouraging prospects, work
is already in progress to apply this method to realistic simulations of
binaries involving neutron stars and black holes.

\begin{backmatter}

\section*{Competing interests}

The authors declare that they have no competing interests.

\section*{Author's contributions}

The implementation of the EL method in \texttt{WhiskyTHC} and all of the
tests were performed by FG. The project was initiated by DR and LR, and
was closely supervised by both.

\section*{Acknowledgements}
\label{sec:acknowledgements}

We thank Kentaro Takami for providing the stellar oscillation eigenfrequencies,
while Erik Schnetter, Ian Hinder, and Massimiliano Leoni for useful
discussions. The simulations were performed on the SuperMUC cluster at the
LRZ in Garching, on the LOEWE cluster in CSC in Frankfurt, on the HazelHen
cluster at the HLRS in Stuttgart and on the Caltech compute cluster Zwicky.

\section*{Funding}

This research is supported in part by the ERC synergy grant "BlackHoleCam:
Imaging the Event Horizon of Black Holes" (Grant No. 610058), by
``NewCompStar'', COST Action MP1304, by the LOEWE-Program in the Helmholtz
International Center (HIC) for FAIR, and by the European Union's Horizon
2020 Research and Innovation Programme (Grant 671698) (call FETHPC-1-2014,
project ExaHyPE). FG is supported by HIC for FAIR and the graduate school
HGS-HIRe. DR gratefully acknowledges support from the Schmidt Fellowship
and the Max-Planck/Princeton Center (MPPC) for Plasma Physics (NSF
PHY-1523261).

\appendix

\section{General relativistic hydrodynamics formulation}
\label{sec:gr_hydrodynamics}

In this section we summarize the general relativistic formulation of
hydrodynamics, which forms the theoretical background for the neutron star
tests presented above; as well as providing details on the ELH scheme in the
general relativistic case, when they differ from the discussion of section
\ref{sec:ELH}.

\subsection{3+1 Decomposition}
\label{sec:3+1}

We adopt the usual 3+1 decomposition of spacetime, (see
\cite{Alcubierre:2008, Bona2009, Baumgarte2010, Gourgoulhon2012,
Rezzolla_book:2013, Shibata_book:2016} for a detailed discussion), in
which the spacetime is decomposed into spacelike hypersurfaces with normal
$n^\mu = \alpha^{-1}(1, -\beta^i)$, where $\alpha$ is the lapse function
and $\beta^i$ are the components of the shift vector. Within this
formalism, the spacetime metric $?g_\mu\nu?$ is written as
\begin{align}
    ds^2 &= ?g_\mu\nu? dx^\mu dx^\nu =\nonumber\\
         &= -\alpha^2dt^2+?\gamma_ij?(dx^i + \beta^idt)(dx^j + \beta^jdt)\,,
    \label{eq:metric}
\end{align}
where $?\gamma_ij?$ is the spatial three-metric, which together with the
extrinsic curvature $?K_ij?=-\frac{1}{2}\mathcal{L}_n?\gamma_ij?$,
$\mathcal{L}_n$ being the Lie derivative along $n^\mu$, fully determines the
geometry of each leaf of the foliation.

The matter content of the spacetime is described through its
energy-momentum tensor $?T_\mu\nu?$, which within the 3+1 split can be
decomposed in its timelike, spacelike and mixed parts as (see \eg
\cite{Rezzolla_book:2013})
\begin{align}
    E &:= n^\mu n^\nu T_{\mu\nu}\,,\\
    S_i &:= -\gamma^\mu_{\ \,i} n^\nu T_{\mu\nu}\,,\\
    S_{ij} &:= \gamma^\mu_{\ \,i} \gamma^\nu_{\ \,j} T_{\mu\nu}\,.
\end{align}
where $?\gamma^\mu_\nu?:= ?\delta^\mu_\nu? + n_\nu
n^\mu$ is the projection operator orthogonal to $n_{\mu}$, hence yielding
purely spatial tensors.

\subsection{Relativistic hydrodynamics}

The general relativistic generalization of the energy-momentum tensor
\eqref{eq:Tmunu} consists in replacing the Minkowski metric $?\eta_\mu\nu?$
with the generic metric $?g_\mu\nu?$, \plainie
\begin{equation}
    ?T_\mu\nu? = \rho h u_\mu u_\nu + p?g_\mu\nu?\,,
    \label{eq:GRTmunu}
\end{equation}
The corresponding equations of motion for the fluid are the ``conservation''
of the stress-energy tensor
\begin{equation}
    \nabla_\mu ?T^\mu\nu? = 0\,,
    \label{eq:GRconsTmunu}
\end{equation}
and conservation of rest mass
\begin{equation}
    \nabla_\mu (\rho u^\mu) = 0\,,
    \label{eq:GRcontinuity}
\end{equation}
which differ from their special relativistic counterparts \eqref{eq:consTmunu}
and \eqref{eq:continuity} by the use of the $\nabla$ operator, \plainie the
covariant derivative constructed from the metric \eqref{eq:metric}.

Equations \eqref{eq:GRconsTmunu} and \eqref{eq:GRcontinuity} are solved in
conservation form (equation \ref{eq:valencia}),
where the ``conserved'' variables $\boldsymbol{U}$ are defined as
\begin{align}
    \boldsymbol{U} &:= \sqrt{\gamma}
    \left(
    \begin{array}{c}
    D \\ S_j \\ \tau
    \end{array}
    \right) \nonumber\\
    &:=\sqrt{\gamma}\left(
    \begin{array}{c}
    \rho W \\ \rho h W^2 v_j \\ \rho h W^2-p-\rho W
    \end{array}
    \right)\,.
    \label{eq:GRconservatives}
\end{align}
This definitions differ from \eqref{eq:conservatives} by the multiplicative
factor $\gamma$, \plainie the determinant of the 3-metric $?\gamma_ij?$.
The fluxes and sources, containing metric-dependent terms, are given by
\begin{align}
    \boldsymbol{F}^i &= \sqrt{\gamma}
    \left(
    \begin{array}{c}
    (\alpha v^i - \beta^i)D \\ \alpha?S^i_j?-\beta^i?S_j? \\ \alpha(S^i-Dv^i)
        -\beta^i\tau
    \end{array}
    \right)\,,
\end{align}
and
\begin{align}
    \boldsymbol{S} &= \sqrt{\gamma}
    \left(
    \begin{array}{c}
    0 \\ \frac{1}{2}\alpha?S^lm?\partial_j?\gamma_lm?
        + S_k\partial_j\beta^k - E\partial_j\alpha \\
        \alpha?S^ij??K_ij?-S^k\partial_k\alpha
    \end{array}
    \right)\,.
\end{align}
This formulation is known as the ``Valencia formulation'' and was first
proposed by \cite{Banyuls97}. Note that the sources terms for the momentum and
energy equations are non-vanishing, which corresponds to the fact that the
momentum and energy of the fluid are not independently conserved, but the
coupling of the fluid to the spacetime and vice versa has to be taken into
account (see \eg \cite{Rezzolla_book:2013,Shibata_book:2016,Baumgarte2010} for
details). The fluid three-velocity measured by the normal observers is defined
as
\begin{equation}
    v^i := \frac{1}{\alpha}\left(\frac{u^i}{u^t} + \beta^i\right)
    \label{eq:GRv}
\end{equation}
which also contains metric-dependent terms, and the Lorentz factor is $W := (1
- v^i v_i)^{-\frac{1}{2}} = \alpha u^t$. We also have used the fact that
$\sqrt{-g} = \alpha \sqrt{\gamma}$.

\subsection{ELH scheme extension to GR}
\label{sec:GRELH}

The general relativistic formulation of the second law of
thermodynamics, and therefore of the entropy residual $\mathcal{R}$, is
\begin{equation}
    \mathcal{R} := \nabla_\mu(\rho s u^\mu) \geq 0\,,
    \label{eq:GRRinequality}
\end{equation}
once again swapping the partial derivative $\partial$ for the covariant one
$\nabla$. This gets rewritten as
\begin{align}
    \mathcal{R} &= \nabla_\mu(s \rho u^\mu)\nonumber\\
      &= s \nabla_\mu (\rho u^\mu) + \rho u^\mu \nabla_\mu s\nonumber\\
      &= \rho u^\mu \nabla_\mu s
      = \rho u^\mu \partial_\mu s\,,
    \label{eq:GRRs}
\end{align}
which again involves only derivatives of the specific entropy $s$,
and where the continuity equation \eqref{eq:GRcontinuity} was used to remove
the first term on the second line. The 4-velocity $u^\mu$ can again be written
in terms of the fluid three-velocity $v^i$, but it will also contain the lapse
and shift (see Eq. \ref{eq:GRv}), so that the final form of the residual in
the 3+1 split is
\begin{equation}
    \mathcal{R} =
    \frac{\rho W}{\alpha}\left[\partial_t s + (\alpha v^i-\beta^i)
    \partial_i s\right]\,.
    \label{eq:GRR3+1}
\end{equation}

\section{Spacetime evolution}
\label{sec:mclachlan}

The evolution of the spacetime is of course determined by Einstein's equations
\begin{equation}
    ?R_\mu\nu?-\frac{1}{2}R?g_\mu\nu?= 8\pi ?T_\mu\nu?\,,
\end{equation}
where $?R_\mu\nu?$ is the Ricci tensor and $R:=?R^\mu_\mu?$ the Ricci
scalar. The energy-momentum tensor $?T_\mu\nu?$ given by \eqref{eq:GRTmunu}
appears in the right-hand-side of the equations, coupling the spacetime to the
and fluid and vice versa.

We employ the common BSSNOK formulation
\cite{Shibata95,Baumgarte99,Brown09} of the Einstein equations
(extensions to the CCZ4 \cite{Alic:2011a} system are also in progress),
along with the standard 1+log slicing condition and Gamma driver to
specify the evolution of the lapse and shift, respectively (see
\cite{Baiotti2016} for a review of these gauges). Given the following
definitions for the evolved fields
\begin{subequations}
\begin{alignat}{2}
    & \phi                && = \frac{1}{12}\ln(\gamma) \\
    & K                   && = \gamma^{ij}K_{ij} \\
    & \tilde{\gamma}_{ij} && = e^{-4\phi}\gamma_{ij} \\
    & \tilde{A}_{ij}      && = e^{-4\phi}\left(K_{ij}-\frac{1}{3}
    \gamma_{ij}K\right) \\
    & \tilde{\Gamma}^i    && = \tilde{\gamma}^{jk}
    \tilde{\Gamma}^i{}_{jk},
\end{alignat}
\end{subequations}
where $\tilde{\Gamma}^i{}_{jk}$ are the Christoffel symbols computed from
the conformal metric $\tilde{\gamma}_{ij}$, the BSSNOK and gauge equations
take the form:
\begin{subequations}
\begin{alignat}{2}
& \partial_\bot\phi && = \frac{1}{6}\partial_k\beta^k-\frac{1}{6}
    \alpha K \\
& \partial_\bot\tilde{\gamma}_{ij} && = -2\alpha\tilde{A}_{ij}-\frac{2}{3}
    \tilde{\gamma}_{ij}\partial_k\beta^k \\
& \partial_\bot K       && = \alpha\left(\tilde{A}_{ij}\tilde{A}^{ij}
    +\frac{1}{3}K^2\right) - \gamma^{ij}\nabla_i\nabla_j\alpha \\\nonumber
    & && +4\pi(S^k{}_k+E)\\
& \partial_\bot\tilde{A}_{ij} && = e^{-4\phi}\left[\alpha(R_{ij}-8\pi S_{ij}
    )-\nabla_i\nabla_j\alpha\right]^{TF} \\ \nonumber
    &&&-\frac{2}{3}\tilde{A}_{ij}\partial_k
    \beta^k+\alpha\left(K\tilde{A}_{ij}-2\tilde{A}_{ik}
    \tilde{A}^k{}_j\right) \\
& \partial_\bot\tilde{\Gamma}^i && = \tilde{\gamma}^{kl}\partial_k\partial_l
    \beta^i+\frac{2}{3}\tilde{\gamma}^{jk}\tilde{\Gamma}^i{}_{jk}\partial_l
    \beta^l \\\nonumber
&                      && +\frac{1}{3}\tilde{\nabla}^i(\partial_k
    \beta^k)-2\tilde{A}^{ik}\partial_k\alpha+2\alpha\tilde{A}^{kl}
    \tilde{\Gamma}^i{}_{kl} \\\nonumber
&                          && +12\alpha\tilde{A}^{ik}\partial_k\phi
    -\frac{4}{3}\alpha\tilde{\nabla}^iK-16\pi\alpha\tilde{\gamma}^{ij}S_j \\
& \partial_t\alpha             && = -2\alpha K+\beta^k\partial_k\alpha \\
& \partial_t\beta^i                && = \frac{3}{4}B^i+\beta^k\partial_k
    \beta^i \\
& \partial_tB^i                    && = \partial_t\tilde{\Gamma}^i-\eta B^i
    +\beta^k\partial_kB^i,
\end{alignat}
\end{subequations}
where the operator $\partial_\bot$ stands for $\partial_t -
\mathcal{L}_{\boldsymbol{\beta}}$, \ie the derivative with respect to
coordinate time minus the Lie derivative along the shift, and the
notation $[\dots]^{TF}$ indicates terms that are made trace free with
respect to the conformal metric. The covariant derivatives $\nabla$ and
$\tilde{\nabla}$ are constructed from the physical and covariant three
metric respectively, and $\eta\sim\,1/M$ ($M$ being the mass of the
system).

The three dimensional Ricci tensor $?R_ij?$ is split in two parts,
$?R_ij?=\tilde{R}^\phi_{ij}+\tilde{R}_{ij}$, the first
involving the conformal factor $\phi$ and the second the derivatives of the
conformal metric $\tilde{\gamma}_{ij}$:
\begin{subequations}
\begin{alignat}{2}
    & \tilde{R}^\phi_{ij} && = ?\phi^-2?[\phi\left(\tilde{\nabla}_i
    \tilde{\nabla}_j\phi+\tilde{\gamma}_{ij}\tilde{\nabla}^k\tilde{\nabla}_k
    \phi\right) \\\nonumber
    &&&-2\tilde{\gamma}_{ij}\tilde{\nabla}^k\phi\tilde{\nabla}_k\phi] \\
    & \tilde{R}_{ij} && = -\frac{1}{2}\tilde{\gamma}^{lm}\partial_l\partial_m
    \tilde{\gamma}_{ij}+\tilde{\gamma}_{k(i}\partial_{j)}\tilde{\Gamma}^k
     \\\nonumber
    &&&+\tilde{\Gamma}^k\tilde{\Gamma}_{(ij)k}+\tilde{\gamma}^{lm}
    \left[2\tilde{\Gamma}^k{}_{l(i}\tilde{\Gamma}_{j)km}
    +\tilde{\Gamma}^k{}_{im}\tilde{\Gamma}_{kjl}\right].
\end{alignat}
\end{subequations}
In integrating these equations a constrained approach is used, \ie we enforce
the constraints $\text{det}\tilde{\gamma}_{ij}=1$ and
$\text{tr}\tilde{A}_{ij}=0$ at every step.

The spacetime evolution is taken care of by the \verb+McLachlan+ code
\cite{Brown2007b} to evolve the spacetime variables in the BSSNOK
formulation. \verb+McLachlan+ approximates the equations using standard
central finite-difference operators with upwinding of the shift advection
terms and Kreiss-Oliger dissipation \cite{Kreiss73} to ensure
stability. It supports up to eighth-order operators, however since the
major source of errors in our simulations is the hydrodynamical part, we
restrict ourselves here to fourth-order accuracy for the spacetime.


\newcommand{\BMCxmlcomment}[1]{}

\BMCxmlcomment{

<refgrp>

<bibl id="B1">
  <title><p>{Entropy viscosity method for nonlinear conservation
  laws}</p></title>
  <aug>
    <au><snm>Guermond</snm><fnm>J. L.</fnm></au>
    <au><snm>Pasquetti</snm><fnm>R.</fnm></au>
    <au><snm>Popov</snm><fnm>B.</fnm></au>
  </aug>
  <source>J. Comput. Phys.</source>
  <publisher>Elsevier Inc.</publisher>
  <pubdate>2011</pubdate>
  <volume>230</volume>
  <issue>11</issue>
  <fpage>4248</fpage>
  <lpage>-4267</lpage>
  <url>http://linkinghub.elsevier.com/retrieve/pii/S0021999110006583</url>
</bibl>

<bibl id="B2">
  <title><p>Accurate Monotonicity-Preserving Schemes with Runge-Kutta Time
  Stepping</p></title>
  <aug>
    <au><snm>Suresh</snm><fnm>A.</fnm></au>
    <au><snm>Huynh</snm><fnm>H. T.</fnm></au>
  </aug>
  <source>Journal of Computational Physics</source>
  <pubdate>1997</pubdate>
  <volume>136</volume>
  <issue>1</issue>
  <fpage>83</fpage>
  <lpage>99</lpage>
  <url>http://www.sciencedirect.com/science/article/B6WHY-45V7FSX-6/2/d88d3b5c02364ae5aa69b21d0e0787e7</url>
</bibl>

<bibl id="B3">
  <title><p>Numerical Hydrodynamics and Magnetohydrodynamics in General
  Relativity</p></title>
  <aug>
    <au><snm>Font</snm><fnm>J. A.</fnm></au>
  </aug>
  <source>Living Rev. Relativ.</source>
  <pubdate>2008</pubdate>
  <volume>6</volume>
  <fpage>4;http://www.livingreviews.org/lrr</fpage>
  <lpage>2008-7</lpage>
  <url>http://www.livingreviews.org/lrr-2008-7</url>
</bibl>

<bibl id="B4">
  <title><p>Coalescence of Black Hole-Neutron Star Binaries</p></title>
  <aug>
    <au><snm>Shibata</snm><fnm>M</fnm></au>
    <au><snm>Taniguchi</snm><fnm>K</fnm></au>
  </aug>
  <source>Living Rev. Relativity</source>
  <pubdate>2011</pubdate>
  <volume>14</volume>
  <issue>6</issue>
  <url>http://www.livingreviews.org/lrr-2011-6</url>
</bibl>

<bibl id="B5">
  <title><p>Relativistic Hydrodynamics</p></title>
  <aug>
    <au><snm>{Rezzolla}</snm><fnm>L.</fnm></au>
    <au><snm>{Zanotti}</snm><fnm>O.</fnm></au>
  </aug>
  <source>Relativistic Hydrodynamics</source>
  <publisher>Oxford, UK: Oxford University Press</publisher>
  <pubdate>2013</pubdate>
</bibl>

<bibl id="B6">
  <title><p>{Grid-based Methods in Relativistic Hydrodynamics and
  Magnetohydrodynamics}</p></title>
  <aug>
    <au><snm>{Mart{\'{\i}}}</snm><fnm>J. M.</fnm></au>
    <au><snm>{M{\"u}ller}</snm><fnm>E.</fnm></au>
  </aug>
  <source>Living Reviews in Computational Astrophysics</source>
  <pubdate>2015</pubdate>
  <volume>1</volume>
</bibl>

<bibl id="B7">
  <title><p>{Numerical Relativity}</p></title>
  <aug>
    <au><snm>{Shibata}</snm><fnm>M.</fnm></au>
  </aug>
  <source>Numerical Relativity</source>
  <publisher>Singapore: World Scientific</publisher>
  <pubdate>2016</pubdate>
</bibl>

<bibl id="B8">
  <title><p>{Binary neutron-star mergers: a review of Einstein's richest
  laboratory}</p></title>
  <aug>
    <au><snm>{Baiotti}</snm><fnm>L.</fnm></au>
    <au><snm>{Rezzolla}</snm><fnm>L.</fnm></au>
  </aug>
  <source>arxiv:1607.03540</source>
  <pubdate>2016</pubdate>
</bibl>

<bibl id="B9">
  <title><p>{General relativistic simulations of compact binary mergers as
  engines of short gamma-ray bursts}</p></title>
  <aug>
    <au><snm>{Paschalidis}</snm><fnm>V.</fnm></au>
  </aug>
  <source>ArXiv e-prints</source>
  <pubdate>2016</pubdate>
</bibl>

<bibl id="B10">
  <title><p>{Accurate numerical simulations of inspiralling binary neutron
  stars and their comparison with effective-one-body analytical
  models}</p></title>
  <aug>
    <au><snm>{Baiotti}</snm><fnm>L.</fnm></au>
    <au><snm>{Damour}</snm><fnm>T.</fnm></au>
    <au><snm>{Giacomazzo}</snm><fnm>B.</fnm></au>
    <au><snm>{Nagar}</snm><fnm>A.</fnm></au>
    <au><snm>{Rezzolla}</snm><fnm>L.</fnm></au>
  </aug>
  <source>Phys. Rev. D</source>
  <pubdate>2011</pubdate>
  <volume>84</volume>
  <issue>2</issue>
  <fpage>024017</fpage>
</bibl>

<bibl id="B11">
  <title><p>Matter effects on binary neutron star waveforms</p></title>
  <aug>
    <au><snm>{Read}</snm><fnm>J. S.</fnm></au>
    <au><snm>{Baiotti}</snm><fnm>L.</fnm></au>
    <au><snm>{Creighton}</snm><fnm>J. D. E.</fnm></au>
    <au><snm>{Friedman}</snm><fnm>J. L.</fnm></au>
    <au><snm>{Giacomazzo}</snm><fnm>B.</fnm></au>
    <au><snm>{Kyutoku}</snm><fnm>K.</fnm></au>
    <au><snm>{Markakis}</snm><fnm>C.</fnm></au>
    <au><snm>{Rezzolla}</snm><fnm>L.</fnm></au>
    <au><snm>{Shibata}</snm><fnm>M.</fnm></au>
    <au><snm>{Taniguchi}</snm><fnm>K.</fnm></au>
  </aug>
  <source>Phys. Rev. D</source>
  <pubdate>2013</pubdate>
  <volume>88</volume>
  <issue>4</issue>
  <fpage>044042</fpage>
</bibl>

<bibl id="B12">
  <title><p>{High-order fully general-relativistic hydrodynamics: new
  approaches and tests}</p></title>
  <aug>
    <au><snm>{Radice}</snm><fnm>D.</fnm></au>
    <au><snm>{Rezzolla}</snm><fnm>L.</fnm></au>
    <au><snm>{Galeazzi}</snm><fnm>F.</fnm></au>
  </aug>
  <source>Class. Quantum Grav.</source>
  <pubdate>2014</pubdate>
  <volume>31</volume>
  <issue>7</issue>
  <fpage>075012</fpage>
</bibl>

<bibl id="B13">
  <title><p>{Numerical relativity simulations of binary neutron
  stars}</p></title>
  <aug>
    <au><snm>{Thierfelder}</snm><fnm>M.</fnm></au>
    <au><snm>{Bernuzzi}</snm><fnm>S.</fnm></au>
    <au><snm>{Br{\"u}gmann}</snm><fnm>B.</fnm></au>
  </aug>
  <source>Phys. Rev. D</source>
  <pubdate>2011</pubdate>
  <volume>84</volume>
  <issue>4</issue>
  <fpage>044012</fpage>
</bibl>

<bibl id="B14">
  <title><p>{Tidal effects in binary neutron star coalescence}</p></title>
  <aug>
    <au><snm>{Bernuzzi}</snm><fnm>S.</fnm></au>
    <au><snm>{Nagar}</snm><fnm>A.</fnm></au>
    <au><snm>{Thierfelder}</snm><fnm>M.</fnm></au>
    <au><snm>{Br{\"u}gmann}</snm><fnm>B.</fnm></au>
  </aug>
  <source>Phys. Rev. D</source>
  <pubdate>2012</pubdate>
  <volume>86</volume>
  <issue>4</issue>
  <fpage>044030</fpage>
</bibl>

<bibl id="B15">
  <title><p>{Some contributions to the modelling of discontinuous
  flows}</p></title>
  <aug>
    <au><snm>{Roe}</snm><fnm>P. L.</fnm></au>
  </aug>
  <source>Large-Scale Computations in Fluid Mechanics</source>
  <editor>{Lee}, R.~L. and {Sani}, R.~L. and {Shih}, T.~M. and {Gresho},
  P.~M.</editor>
  <pubdate>1985</pubdate>
  <fpage>163</fpage>
  <lpage>193</lpage>
</bibl>

<bibl id="B16">
  <title><p>The Piecewise Parabolic Method (PPM) for gas-dynamical
  simulations</p></title>
  <aug>
    <au><snm>Colella</snm><fnm>P</fnm></au>
    <au><snm>Woodward</snm><fnm>PR</fnm></au>
  </aug>
  <source>Journal of Computational Physics</source>
  <pubdate>1984</pubdate>
  <volume>54</volume>
  <issue>1</issue>
  <fpage>174</fpage>
  <lpage>201</lpage>
  <url>http://www.sciencedirect.com/science/article/B6WHY-4DD1PHM-SJ/2/13d69a59afba3d6a5d6bbf1144d860aa</url>
</bibl>

<bibl id="B17">
  <title><p>{Extension of the Piecewise Parabolic Method to One-Dimensional
  Relativistic Hydrodynamics}</p></title>
  <aug>
    <au><snm>Mart{\'\i}</snm><fnm>J. M.</fnm></au>
    <au><snm>M{\"u}ller</snm><fnm>E.</fnm></au>
  </aug>
  <source>Journal of Computational Physics</source>
  <pubdate>1996</pubdate>
  <volume>123</volume>
  <fpage>1</fpage>
  <lpage>14</lpage>
</bibl>

<bibl id="B18">
  <title><p>{Uniformly High Order Accurate Essentially Non-oscillatory Schemes
  III}</p></title>
  <aug>
    <au><snm>{Harten}</snm><fnm>A.</fnm></au>
    <au><snm>{Engquist}</snm><fnm>B.</fnm></au>
    <au><snm>{Osher}</snm><fnm>S.</fnm></au>
    <au><snm>{Chakravarthy}</snm><fnm>S. R.</fnm></au>
  </aug>
  <source>Journal of Computational Physics</source>
  <pubdate>1987</pubdate>
  <volume>71</volume>
  <fpage>231</fpage>
  <lpage>-303</lpage>
</bibl>

<bibl id="B19">
  <title><p>{Weighted Essentially Non-oscillatory Schemes}</p></title>
  <aug>
    <au><snm>{Liu}</snm><fnm>X. D.</fnm></au>
    <au><snm>{Osher}</snm><fnm>S.</fnm></au>
    <au><snm>{Chan}</snm><fnm>T.</fnm></au>
  </aug>
  <source>Journal of Computational Physics</source>
  <pubdate>1994</pubdate>
  <volume>115</volume>
  <fpage>200</fpage>
  <lpage>212</lpage>
</bibl>

<bibl id="B20">
  <title><p>Efficient Implementation of Weighted ENO Schemes</p></title>
  <aug>
    <au><snm>Jiang</snm><fnm>GS</fnm></au>
    <au><snm>Shu</snm><fnm>CW</fnm></au>
  </aug>
  <source>J. Comput. Phys</source>
  <pubdate>1996</pubdate>
  <volume>126</volume>
  <fpage>202</fpage>
  <lpage>-228</lpage>
</bibl>

<bibl id="B21">
  <title><p>{Modeling equal and unequal mass binary neutron star mergers using
  public codes}</p></title>
  <aug>
    <au><snm>{De Pietri}</snm><fnm>R.</fnm></au>
    <au><snm>{Feo}</snm><fnm>A.</fnm></au>
    <au><snm>{Maione}</snm><fnm>F.</fnm></au>
    <au><snm>{L{\"o}ffler}</snm><fnm>F.</fnm></au>
  </aug>
  <source>Phys. Rev. D</source>
  <pubdate>2016</pubdate>
  <volume>93</volume>
  <issue>6</issue>
  <fpage>064047</fpage>
</bibl>

<bibl id="B22">
  <title><p>{Gravitational waveforms from binary neutron star mergers with
  high-order WENO schemes in numerical relativity}</p></title>
  <aug>
    <au><snm>{Bernuzzi}</snm><fnm>S.</fnm></au>
    <au><snm>{Dietrich}</snm><fnm>T.</fnm></au>
  </aug>
  <source>arXiv:1604.07999</source>
  <pubdate>2016</pubdate>
</bibl>

<bibl id="B23">
  <title><p>{Relativistic Hydrodynamics with Wavelets}</p></title>
  <aug>
    <au><snm>{DeBuhr}</snm><fnm>J.</fnm></au>
    <au><snm>{Zhang}</snm><fnm>B.</fnm></au>
    <au><snm>{Anderson}</snm><fnm>M.</fnm></au>
    <au><snm>{Neilsen}</snm><fnm>D.</fnm></au>
    <au><snm>{Hirschmann}</snm><fnm>E. W.</fnm></au>
  </aug>
  <source>ArXiv e-prints</source>
  <pubdate>2015</pubdate>
</bibl>

<bibl id="B24">
  <title><p>{Discontinuous Galerkin methods for general-relativistic
  hydrodynamics: Formulation and application to spherically symmetric
  spacetimes}</p></title>
  <aug>
    <au><snm>{Radice}</snm><fnm>D.</fnm></au>
    <au><snm>{Rezzolla}</snm><fnm>L.</fnm></au>
  </aug>
  <source>Phys. Rev. D</source>
  <pubdate>2011</pubdate>
  <volume>84</volume>
  <issue>2</issue>
  <fpage>024010</fpage>
</bibl>

<bibl id="B25">
  <title><p>{Solving 3D relativistic hydrodynamical problems with weighted
  essentially nonoscillatory discontinuous Galerkin methods}</p></title>
  <aug>
    <au><snm>{Bugner}</snm><fnm>M.</fnm></au>
    <au><snm>{Dietrich}</snm><fnm>T.</fnm></au>
    <au><snm>{Bernuzzi}</snm><fnm>S.</fnm></au>
    <au><snm>{Weyhausen}</snm><fnm>A.</fnm></au>
    <au><snm>{Br{\"u}gmann}</snm><fnm>B.</fnm></au>
  </aug>
  <source>Phys. Rev. D</source>
  <pubdate>2016</pubdate>
  <volume>94</volume>
  <issue>8</issue>
  <fpage>084004</fpage>
</bibl>

<bibl id="B26">
  <title><p>{Solving the relativistic magnetohydrodynamics equations with ADER
  discontinuous Galerkin methods, a posteriori subcell limiting and adaptive
  mesh refinement}</p></title>
  <aug>
    <au><snm>{Zanotti}</snm><fnm>O.</fnm></au>
    <au><snm>{Fambri}</snm><fnm>F.</fnm></au>
    <au><snm>{Dumbser}</snm><fnm>M.</fnm></au>
  </aug>
  <source>Mon. Not. R. Astron. Soc.</source>
  <pubdate>2015</pubdate>
  <volume>452</volume>
  <fpage>3010</fpage>
  <lpage>3029</lpage>
</bibl>

<bibl id="B27">
  <title><p>{SpECTRE: A Task-based Discontinuous Galerkin Code for Relativistic
  Astrophysics}</p></title>
  <aug>
    <au><snm>{Kidder}</snm><fnm>L. E.</fnm></au>
    <au><snm>{Field}</snm><fnm>S. E.</fnm></au>
    <au><snm>{Foucart}</snm><fnm>F.</fnm></au>
    <au><snm>{Schnetter}</snm><fnm>E.</fnm></au>
    <au><snm>{Teukolsky}</snm><fnm>S. A.</fnm></au>
    <au><snm>{Bohn}</snm><fnm>A.</fnm></au>
    <au><snm>{Deppe}</snm><fnm>N.</fnm></au>
    <au><snm>{Diener}</snm><fnm>P.</fnm></au>
    <au><snm>{H{\'e}bert}</snm><fnm>F.</fnm></au>
    <au><snm>{Lippuner}</snm><fnm>J.</fnm></au>
    <au><snm>{Miller}</snm><fnm>J.</fnm></au>
    <au><snm>{Ott}</snm><fnm>C. D.</fnm></au>
    <au><snm>{Scheel}</snm><fnm>M. A.</fnm></au>
    <au><snm>{Vincent}</snm><fnm>T.</fnm></au>
  </aug>
  <source>ArXiv e-prints</source>
  <pubdate>2016</pubdate>
</bibl>

<bibl id="B28">
  <title><p>{An Operator-Based Local Discontinuous Galerkin Method Compatible
  With the BSSN Formulation of the Einstein Equations}</p></title>
  <aug>
    <au><snm>{Miller}</snm><fnm>J. M.</fnm></au>
    <au><snm>{Schnetter}</snm><fnm>E.</fnm></au>
  </aug>
  <source>ArXiv e-prints</source>
  <pubdate>2016</pubdate>
</bibl>

<bibl id="B29">
  <title><p>Numerical Methods for Conservation Laws</p></title>
  <aug>
    <au><snm>Leveque</snm><fnm>R. J.</fnm></au>
  </aug>
  <publisher>Basel: Birkhauser Verlag</publisher>
  <pubdate>1992</pubdate>
</bibl>

<bibl id="B30">
  <title><p>{wham: a WENO-based general relativistic numerical scheme ? I.
  Hydrodynamics}</p></title>
  <aug>
    <au><snm>Tchekhovskoy</snm><fnm>A</fnm></au>
    <au><snm>McKinney</snm><fnm>JC</fnm></au>
    <au><snm>Narayan</snm><fnm>R</fnm></au>
  </aug>
  <source>Mon. Not. R. Astron. Soc.</source>
  <pubdate>2007</pubdate>
  <volume>379</volume>
  <issue>2</issue>
  <fpage>469</fpage>
  <lpage>-497</lpage>
  <url>http://doi.wiley.com/10.1111/j.1365-2966.2007.11876.x</url>
</bibl>

<bibl id="B31">
  <title><p>{Entropy-based nonlinear viscosity for Fourier approximations of
  conservation laws}</p></title>
  <aug>
    <au><snm>Guermond</snm><fnm>J. L.</fnm></au>
    <au><snm>Pasquetti</snm><fnm>R.</fnm></au>
  </aug>
  <source>C. R. Acad. Sci. Paris</source>
  <pubdate>208</pubdate>
  <volume>346</volume>
  <issue>346</issue>
  <fpage>801</fpage>
  <lpage>-806</lpage>
</bibl>

<bibl id="B32">
  <title><p>{Implementation of the entropy viscosity method with the
  discontinuous Galerkin method}</p></title>
  <aug>
    <au><snm>Zingan</snm><fnm>V</fnm></au>
    <au><snm>Guermond</snm><fnm>JL</fnm></au>
    <au><snm>Morel</snm><fnm>J</fnm></au>
    <au><snm>Popov</snm><fnm>B</fnm></au>
  </aug>
  <source>Computer Methods in Applied Mechanics and Engineering</source>
  <publisher>Elsevier B.V.</publisher>
  <pubdate>2013</pubdate>
  <volume>253</volume>
  <fpage>479</fpage>
  <lpage>-490</lpage>
</bibl>

<bibl id="B33">
  <title><p>{THC: a new high-order finite-difference high-resolution
  shock-capturing code for special-relativistic hydrodynamics}</p></title>
  <aug>
    <au><snm>{Radice}</snm><fnm>D.</fnm></au>
    <au><snm>{Rezzolla}</snm><fnm>L.</fnm></au>
  </aug>
  <source>Astron. Astrophys.</source>
  <pubdate>2012</pubdate>
  <volume>547</volume>
  <fpage>A26</fpage>
</bibl>

<bibl id="B34">
  <title><p>{Beyond second-order convergence in simulations of binary neutron
  stars in full general-relativity}</p></title>
  <aug>
    <au><snm>{Radice}</snm><fnm>D.</fnm></au>
    <au><snm>{Rezzolla}</snm><fnm>L.</fnm></au>
    <au><snm>{Galeazzi}</snm><fnm>F.</fnm></au>
  </aug>
  <source>Mon. Not. R. Astron. Soc. L.</source>
  <pubdate>2014</pubdate>
  <volume>437</volume>
  <fpage>L46</fpage>
  <lpage>L50</lpage>
</bibl>

<bibl id="B35">
  <title><p>{The Einstein Toolkit: a community computational infrastructure for
  relativistic astrophysics}</p></title>
  <aug>
    <au><snm>{L{\"o}ffler}</snm><fnm>F.</fnm></au>
    <au><snm>{Faber}</snm><fnm>J.</fnm></au>
    <au><snm>{Bentivegna}</snm><fnm>E.</fnm></au>
    <au><snm>{Bode}</snm><fnm>T.</fnm></au>
    <au><snm>{Diener}</snm><fnm>P.</fnm></au>
    <au><snm>{Haas}</snm><fnm>R.</fnm></au>
    <au><snm>{Hinder}</snm><fnm>I.</fnm></au>
    <au><snm>{Mundim}</snm><fnm>B. C.</fnm></au>
    <au><snm>{Ott}</snm><fnm>C. D.</fnm></au>
    <au><snm>{Schnetter}</snm><fnm>E.</fnm></au>
    <au><snm>{Allen}</snm><fnm>G.</fnm></au>
    <au><snm>{Campanelli}</snm><fnm>M.</fnm></au>
    <au><snm>{Laguna}</snm><fnm>P.</fnm></au>
  </aug>
  <source>Class. Quantum Grav.</source>
  <pubdate>2012</pubdate>
  <volume>29</volume>
  <issue>11</issue>
  <fpage>115001</fpage>
</bibl>

<bibl id="B36">
  <title><p>{An Introduction to the Einstein Toolkit}</p></title>
  <aug>
    <au><snm>{Zilh{\~a}o}</snm><fnm>M.</fnm></au>
    <au><snm>{L{\"o}ffler}</snm><fnm>F.</fnm></au>
  </aug>
  <source>International Journal of Modern Physics A</source>
  <pubdate>2013</pubdate>
  <volume>28</volume>
  <fpage>40014</fpage>
</bibl>

<bibl id="B37">
  <title><p>{Einstein Toolkit}: {O}pen {S}oftware for {R}elativistic
  {A}strophysics</p></title>
  <note>{\tt http://einsteintoolkit.org}</note>
</bibl>

<bibl id="B38">
  <title><p>High-order conservative finite difference GLM-MHD schemes for
  cell-centered MHD</p></title>
  <aug>
    <au><snm>Mignone</snm><fnm>A.</fnm></au>
    <au><snm>Tzeferacos</snm><fnm>P.</fnm></au>
    <au><snm>Bodo</snm><fnm>G.</fnm></au>
  </aug>
  <source>Journal of Computational Physics</source>
  <pubdate>2010</pubdate>
  <volume>229</volume>
  <fpage>5896</fpage>
  <lpage>5920</lpage>
</bibl>

<bibl id="B39">
  <title><p>{High-Order Numerical-Relativity Simulations of Binary Neutron
  Stars}</p></title>
  <aug>
    <au><snm>{Radice}</snm><fnm>D.</fnm></au>
    <au><snm>{Rezzolla}</snm><fnm>L.</fnm></au>
    <au><snm>{Galeazzi}</snm><fnm>F.</fnm></au>
  </aug>
  <source>Numerical Modeling of Space Plasma Flows ASTRONUM-2014</source>
  <editor>{Pogorelov}, N.~V. and {Audit}, E. and {Zank}, G.~P.</editor>
  <series><title><p>Astronomical Society of the Pacific Conference
  Series</p></title></series>
  <pubdate>2015</pubdate>
  <volume>498</volume>
  <fpage>121</fpage>
</bibl>

<bibl id="B40">
  <title><p>{E}ssentially non-oscillatory and weighted essentially
  non-oscillatory schemes for hyperbolic conservation laws</p></title>
  <aug>
    <au><snm>Shu</snm><fnm>C. W.</fnm></au>
  </aug>
  <source>Lecture notes</source>
  <pubdate>1997</pubdate>
  <issue>ICASE Report 97-65; NASA CR-97-206253</issue>
  <url>http://ntrs.nasa.gov/archive/nasa/casi.ntrs.nasa.gov/19980007543\_1998045663.pdf</url>
</bibl>

<bibl id="B41">
  <title><p>{Positivity-preserving method for high-order conservative schemes
  solving compressible Euler equations}</p></title>
  <aug>
    <au><snm>{Hu}</snm><fnm>X. Y.</fnm></au>
    <au><snm>{Adams}</snm><fnm>N. A.</fnm></au>
    <au><snm>{Shu}</snm><fnm>C. W.</fnm></au>
  </aug>
  <source>Journal of Computational Physics</source>
  <pubdate>2013</pubdate>
  <volume>242</volume>
  <fpage>169</fpage>
  <lpage>180</lpage>
</bibl>

<bibl id="B42">
  <title><p>{Design of provably physical-constraint-preserving methods for
  general relativistic hydrodynamics}</p></title>
  <aug>
    <au><snm>Wu</snm><fnm>K</fnm></au>
  </aug>
  <source>Phys. Rev. D</source>
  <pubdate>2017</pubdate>
  <volume>95</volume>
  <fpage>103001</fpage>
</bibl>

<bibl id="B43">
  <title><p>High Order Strong Stability Preserving Time
  Discretizations</p></title>
  <aug>
    <au><snm>Gottlieb</snm><fnm>S</fnm></au>
    <au><snm>Ketcheson</snm><fnm>D</fnm></au>
    <au><snm>Shu</snm><fnm>CW</fnm></au>
  </aug>
  <source>Journal of Scientific Computing</source>
  <publisher>Springer Netherlands</publisher>
  <pubdate>2009</pubdate>
  <volume>38</volume>
  <fpage>251</fpage>
  <lpage>289</lpage>
  <url>http://dx.doi.org/10.1007/s10915-008-9239-z</url>
  <note>10.1007/s10915-008-9239-z</note>
</bibl>

<bibl id="B44">
  <title><p>{Discrete filters for large eddy simulation}</p></title>
  <aug>
    <au><snm>Sagaut</snm><fnm>P.</fnm></au>
    <au><snm>Grohens</snm><fnm>R.</fnm></au>
  </aug>
  <source>Int. J. Numer. Meth. Fluids</source>
  <pubdate>1999</pubdate>
  <volume>31</volume>
  <fpage>1195</fpage>
  <lpage>-1220</lpage>
</bibl>

<bibl id="B45">
  <title><p>{RAM: A relativistic adaptive mesh refinement hydrodynamics
  code}</p></title>
  <aug>
    <au><snm>Zhang</snm><fnm>W</fnm></au>
    <au><snm>MacFadyen</snm><fnm>A.I.</fnm></au>
  </aug>
  <source>The Astrophysical Journal Supplement Series</source>
  <publisher>IOP Publishing</publisher>
  <pubdate>2006</pubdate>
  <volume>164</volume>
  <fpage>255</fpage>
  <url>http://iopscience.iop.org/0067-0049/164/1/255</url>
</bibl>

<bibl id="B46">
  <title><p>Relativistic Fluids and Magneto-fluids</p></title>
  <aug>
    <au><snm>{Anile}</snm><fnm>A. M.</fnm></au>
  </aug>
  <source>Relativistic Fluids and Magneto-fluids, by A.~M.~Anile, pp.~348.~ISBN
  0521304067.~Cambridge, UK: Cambridge University Press, February
  1990.</source>
  <publisher>Cambridge, UK: Cambridge University Press</publisher>
  <pubdate>1990</pubdate>
</bibl>

<bibl id="B47">
  <title><p>{A survey of several finite difference methods for systems of
  nonlinear hyperbolic conservation laws}</p></title>
  <aug>
    <au><snm>{Sod}</snm><fnm>G. A.</fnm></au>
  </aug>
  <source>Journal of Computational Physics</source>
  <pubdate>1978</pubdate>
  <volume>27</volume>
  <fpage>1</fpage>
  <lpage>31</lpage>
</bibl>

<bibl id="B48">
  <title><p>Numerical Hydrodynamics in Special Relativity</p></title>
  <aug>
    <au><snm>Mart{\'\i}</snm><fnm>J. M.</fnm></au>
    <au><snm>M{\"u}ller</snm><fnm>E.</fnm></au>
  </aug>
  <source>Living Rev. Relativ.</source>
  <pubdate>2003</pubdate>
  <volume>6</volume>
  <fpage>7;http://www.livingreviews.org/lrr</fpage>
  <lpage>2003-7</lpage>
  <url>http://www.livingreviews.org/lrr-2003-7</url>
</bibl>

<bibl id="B49">
  <title><p>Fourier Analysis of Numerical Approximations of Hyperbolic
  Equations</p></title>
  <aug>
    <au><snm>Vichnevetsky</snm><fnm>R.</fnm></au>
    <au><snm>Bowles</snm><fnm>J. B.</fnm></au>
  </aug>
  <publisher>Philadelphia, USA: SIAM</publisher>
  <pubdate>1982</pubdate>
</bibl>

<bibl id="B50">
  <title><p>{Compact Finite Difference Schemes with Spectral-like
  Resolution}</p></title>
  <aug>
    <au><snm>Lele</snm><fnm>SK</fnm></au>
  </aug>
  <source>J. Comput. Phys.</source>
  <pubdate>1992</pubdate>
  <volume>103</volume>
  <fpage>16</fpage>
  <lpage>-42</lpage>
</bibl>

<bibl id="B51">
  <title><p>Quasi-radial modes of rotating stars in general
  relativity</p></title>
  <aug>
    <au><snm>Yoshida</snm><fnm>S.</fnm></au>
    <au><snm>Eriguchi</snm><fnm>Y.</fnm></au>
  </aug>
  <source>Mon. Not. R. Astron. Soc.</source>
  <pubdate>2001</pubdate>
  <volume>322</volume>
  <fpage>389</fpage>
</bibl>

<bibl id="B52">
  <title><p>{A quasi-radial stability criterion for rotating relativistic
  stars}</p></title>
  <aug>
    <au><snm>{Takami}</snm><fnm>K.</fnm></au>
    <au><snm>{Rezzolla}</snm><fnm>L.</fnm></au>
    <au><snm>{Yoshida}</snm><fnm>S.</fnm></au>
  </aug>
  <source>Mon. Not. R. Astron. Soc.</source>
  <pubdate>2011</pubdate>
  <volume>416</volume>
  <fpage>L1</fpage>
  <lpage>L5</lpage>
</bibl>

<bibl id="B53">
  <title><p>{Three-dimensional numerical general relativistic hydrodynamics.
  II. Long-term dynamics of single relativistic stars}</p></title>
  <aug>
    <au><snm>{Font}</snm><fnm>J. A.</fnm></au>
    <au><snm>{Goodale}</snm><fnm>T.</fnm></au>
    <au><snm>{Iyer}</snm><fnm>S.</fnm></au>
    <au><snm>{Miller}</snm><fnm>M.</fnm></au>
    <au><snm>{Rezzolla}</snm><fnm>L.</fnm></au>
    <au><snm>{Seidel}</snm><fnm>E.</fnm></au>
    <au><snm>{Stergioulas}</snm><fnm>N.</fnm></au>
    <au><snm>{Suen}</snm><fnm>W. M.</fnm></au>
    <au><snm>{Tobias}</snm><fnm>M.</fnm></au>
  </aug>
  <source>Phys. Rev. D</source>
  <pubdate>2002</pubdate>
  <volume>65</volume>
  <issue>8</issue>
  <fpage>084024</fpage>
</bibl>

<bibl id="B54">
  <title><p>{Three-dimensional relativistic simulations of rotating
  neutron-star collapse to a Kerr black hole}</p></title>
  <aug>
    <au><snm>{Baiotti}</snm><fnm>L.</fnm></au>
    <au><snm>{Hawke}</snm><fnm>I.</fnm></au>
    <au><snm>{Montero}</snm><fnm>P. J.</fnm></au>
    <au><snm>{L{\"o}ffler}</snm><fnm>F.</fnm></au>
    <au><snm>{Rezzolla}</snm><fnm>L.</fnm></au>
    <au><snm>{Stergioulas}</snm><fnm>N.</fnm></au>
    <au><snm>{Font}</snm><fnm>J. A.</fnm></au>
    <au><snm>{Seidel}</snm><fnm>E.</fnm></au>
  </aug>
  <source>Phys. Rev. D</source>
  <pubdate>2005</pubdate>
  <volume>71</volume>
  <issue>2</issue>
  <fpage>024035</fpage>
</bibl>

<bibl id="B55">
  <title><p>A new three-dimensional general-relativistic hydrodynamics
  code</p></title>
  <aug>
    <au><snm>Baiotti</snm><fnm>L</fnm></au>
    <au><snm>Hawke</snm><fnm>I</fnm></au>
    <au><snm>Montero</snm><fnm>P</fnm></au>
    <au><snm>Rezzolla</snm><fnm>L</fnm></au>
  </aug>
  <source>Computational Astrophysics in Italy: Methods and Tools</source>
  <publisher>Trieste: MSAIt</publisher>
  <editor>R. Capuzzo-Dolcetta</editor>
  <pubdate>2003</pubdate>
  <volume>1</volume>
  <fpage>210</fpage>
</bibl>

<bibl id="B56">
  <title><p>{Improved constrained scheme for the Einstein equations: An
  approach to the uniqueness issue}</p></title>
  <aug>
    <au><snm>{Cordero-Carri{\'o}n}</snm><fnm>I.</fnm></au>
    <au><snm>{Cerd{\'a}-Dur{\'a}n}</snm><fnm>P.</fnm></au>
    <au><snm>{Dimmelmeier}</snm><fnm>H.</fnm></au>
    <au><snm>{Jaramillo}</snm><fnm>J. L.</fnm></au>
    <au><snm>{Novak}</snm><fnm>J.</fnm></au>
    <au><snm>{Gourgoulhon}</snm><fnm>E.</fnm></au>
  </aug>
  <source>Phys. Rev. D</source>
  <pubdate>2009</pubdate>
  <volume>79</volume>
  <issue>2</issue>
  <fpage>024017</fpage>
</bibl>

<bibl id="B57">
  <title><p>{Critical phenomena in neutron stars: I. Linearly unstable
  nonrotating models}</p></title>
  <aug>
    <au><snm>{Radice}</snm><fnm>D.</fnm></au>
    <au><snm>{Rezzolla}</snm><fnm>L.</fnm></au>
    <au><snm>{Kellerman}</snm><fnm>T.</fnm></au>
  </aug>
  <source>Class. Quantum Grav.</source>
  <pubdate>2010</pubdate>
  <volume>27</volume>
  <issue>23</issue>
  <fpage>235015</fpage>
</bibl>

<bibl id="B58">
  <title><p>{Comparing models of rapidly rotating relativistic stars
  constructed by two numerical methods}</p></title>
  <aug>
    <au><snm>{Stergioulas}</snm><fnm>N.</fnm></au>
    <au><snm>{Friedman}</snm><fnm>J. L.</fnm></au>
  </aug>
  <source>Astrophys. J.</source>
  <pubdate>1995</pubdate>
  <volume>444</volume>
  <fpage>306</fpage>
  <lpage>311</lpage>
</bibl>

<bibl id="B59">
  <title><p>{Black hole from merging binary neutron stars: How fast can it
  spin?}</p></title>
  <aug>
    <au><snm>{Kastaun}</snm><fnm>W.</fnm></au>
    <au><snm>{Galeazzi}</snm><fnm>F.</fnm></au>
    <au><snm>{Alic}</snm><fnm>D.</fnm></au>
    <au><snm>{Rezzolla}</snm><fnm>L.</fnm></au>
    <au><snm>{Font}</snm><fnm>J. A.</fnm></au>
  </aug>
  <source>Phys. Rev. D</source>
  <pubdate>2013</pubdate>
  <volume>88</volume>
  <issue>2</issue>
  <fpage>021501</fpage>
</bibl>

<bibl id="B60">
  <title><p>Initial data for neutron star binaries with arbitrary
  spins</p></title>
  <aug>
    <au><snm>Tsatsin</snm><fnm>P</fnm></au>
    <au><snm>Marronetti</snm><fnm>P</fnm></au>
  </aug>
  <source>Phys. Rev. D</source>
  <publisher>American Physical Society</publisher>
  <pubdate>2013</pubdate>
  <volume>88</volume>
  <fpage>064060</fpage>
  <url>http://link.aps.org/doi/10.1103/PhysRevD.88.064060</url>
</bibl>

<bibl id="B61">
  <title><p>{Dynamical Mass Ejection from Binary Neutron Star
  Mergers}</p></title>
  <aug>
    <au><snm>{Radice}</snm><fnm>D.</fnm></au>
    <au><snm>{Galeazzi}</snm><fnm>F.</fnm></au>
    <au><snm>{Lippuner}</snm><fnm>J.</fnm></au>
    <au><snm>{Roberts}</snm><fnm>L. F.</fnm></au>
    <au><snm>{Ott}</snm><fnm>C. D.</fnm></au>
    <au><snm>{Rezzolla}</snm><fnm>L.</fnm></au>
  </aug>
  <source>Mon. Not. R. Astron. Soc.</source>
  <pubdate>2016</pubdate>
  <volume>460</volume>
  <fpage>3255</fpage>
  <lpage>3271</lpage>
</bibl>

<bibl id="B62">
  <title><p>Gravitational-Wave Emission from Rotating Gravitational Collapse in
  three Dimensions</p></title>
  <aug>
    <au><snm>Baiotti</snm><fnm>L</fnm></au>
    <au><snm>Hawke</snm><fnm>I</fnm></au>
    <au><snm>Rezzolla</snm><fnm>L</fnm></au>
    <au><snm>Schnetter</snm><fnm>E</fnm></au>
  </aug>
  <source>Phys. Rev. Lett.</source>
  <pubdate>2005</pubdate>
  <volume>94</volume>
  <fpage>131101</fpage>
</bibl>

<bibl id="B63">
  <title><p>Challenging the paradigm of singularity excision in gravitational
  collapse</p></title>
  <aug>
    <au><snm>Baiotti</snm><fnm>L</fnm></au>
    <au><snm>Rezzolla</snm><fnm>L</fnm></au>
  </aug>
  <source>Phys. Rev. Lett.</source>
  <pubdate>2006</pubdate>
  <volume>97</volume>
  <fpage>141101</fpage>
</bibl>

<bibl id="B64">
  <title><p>{The trumpet solution from spherical gravitational collapse with
  puncture gauges}</p></title>
  <aug>
    <au><snm>{Thierfelder}</snm><fnm>M.</fnm></au>
    <au><snm>{Bernuzzi}</snm><fnm>S.</fnm></au>
    <au><snm>{Hilditch}</snm><fnm>D.</fnm></au>
    <au><snm>{Bruegmann}</snm><fnm>B.</fnm></au>
    <au><snm>{Rezzolla}</snm><fnm>L.</fnm></au>
  </aug>
  <source>Phys. Rev. D</source>
  <pubdate>2010</pubdate>
  <volume>83</volume>
  <fpage>064022</fpage>
</bibl>

<bibl id="B65">
  <title><p>On the gravitational radiation from the collapse of neutron stars
  to rotating black holes</p></title>
  <aug>
    <au><snm>Baiotti</snm><fnm>L</fnm></au>
    <au><snm>Hawke</snm><fnm>I</fnm></au>
    <au><snm>Rezzolla</snm><fnm>L</fnm></au>
  </aug>
  <source>Class. Quantum Grav.</source>
  <pubdate>2007</pubdate>
  <volume>24</volume>
  <fpage>S187</fpage>
  <lpage>S206</lpage>
</bibl>

<bibl id="B66">
  <title><p>{Constraint damping of the conformal and covariant formulation of
  the Z4 system in simulations of binary neutron stars}</p></title>
  <aug>
    <au><snm>{Alic}</snm><fnm>D.</fnm></au>
    <au><snm>{Kastaun}</snm><fnm>W.</fnm></au>
    <au><snm>{Rezzolla}</snm><fnm>L.</fnm></au>
  </aug>
  <source>Phys. Rev. D</source>
  <pubdate>2013</pubdate>
  <volume>88</volume>
  <issue>6</issue>
  <fpage>064049</fpage>
</bibl>

<bibl id="B67">
  <title><p>Dynamical horizons and their properties</p></title>
  <aug>
    <au><snm>Ashtekar</snm><fnm>A</fnm></au>
    <au><snm>Krishnan</snm><fnm>B</fnm></au>
  </aug>
  <source>Phys. Rev. D</source>
  <pubdate>2003</pubdate>
  <volume>68</volume>
  <fpage>104030</fpage>
</bibl>

<bibl id="B68">
  <title><p>Introduction to $3+1$ {N}umerical {R}elativity</p></title>
  <aug>
    <au><snm>Alcubierre</snm><fnm>M</fnm></au>
  </aug>
  <publisher>Oxford, UK: Oxford University Press</publisher>
  <pubdate>2008</pubdate>
</bibl>

<bibl id="B69">
  <title><p>Elements of Numerical Relativity and Relativistic Hydrodynamics:
  From Einstein's Equations to Astrophysical Simulations</p></title>
  <aug>
    <au><snm>Bona</snm><fnm>C.</fnm></au>
    <au><snm>Palenzuela Luque</snm><fnm>C.</fnm></au>
    <au><snm>Bona Casas</snm><fnm>C.</fnm></au>
  </aug>
  <publisher>Berlin Heidelberg: Springer</publisher>
  <series><title><p>Lecture Notes in Physics</p></title></series>
  <pubdate>2009</pubdate>
  <url>http://books.google.co.uk/books?id=KgPGHaCUaAYC</url>
</bibl>

<bibl id="B70">
  <title><p>{Numerical Relativity: Solving Einstein's Equations on the
  Computer}</p></title>
  <aug>
    <au><snm>{Baumgarte}</snm><fnm>T. W.</fnm></au>
    <au><snm>{Shapiro}</snm><fnm>S. L.</fnm></au>
  </aug>
  <source>Numerical Relativity: Solving Einstein's Equations on the Computer by
  Thomas W.~Baumgarte and Stuart L.~Shapiro.~Cambridge University Press,
  2010.~ISBN: 9780521514071</source>
  <publisher>Cambridge, UK: Cambridge University Press</publisher>
  <pubdate>2010</pubdate>
</bibl>

<bibl id="B71">
  <title><p>{3+1 Formalism in General Relativity}</p></title>
  <aug>
    <au><snm>{Gourgoulhon}</snm><fnm>E</fnm></au>
  </aug>
  <source>Lecture Notes in Physics, Berlin Springer Verlag</source>
  <series><title><p>Lecture Notes in Physics, Berlin Springer
  Verlag</p></title></series>
  <pubdate>2012</pubdate>
  <volume>846</volume>
</bibl>

<bibl id="B72">
  <title><p>Numerical {3+1} General-Relativistic Hydrodynamics: A Local
  Characteristic Approach</p></title>
  <aug>
    <au><snm>Banyuls</snm><fnm>F.</fnm></au>
    <au><snm>Font</snm><fnm>J. A.</fnm></au>
    <au><snm>Ib{\'a}{\~n}ez</snm><fnm>J. M.</fnm></au>
    <au><snm>Mart{\'\i}</snm><fnm>J. M.</fnm></au>
    <au><snm>Miralles</snm><fnm>J. A.</fnm></au>
  </aug>
  <source>Astrophys. J.</source>
  <pubdate>1997</pubdate>
  <volume>476</volume>
  <fpage>221</fpage>
</bibl>

<bibl id="B73">
  <title><p>{Evolution of three-dimensional gravitational waves: Harmonic
  slicing case}</p></title>
  <aug>
    <au><snm>{Shibata}</snm><fnm>M.</fnm></au>
    <au><snm>{Nakamura}</snm><fnm>T.</fnm></au>
  </aug>
  <source>Phys. Rev. D</source>
  <pubdate>1995</pubdate>
  <volume>52</volume>
  <fpage>5428</fpage>
  <lpage>5444</lpage>
</bibl>

<bibl id="B74">
  <title><p>{Numerical integration of Einstein's field equations}</p></title>
  <aug>
    <au><snm>{Baumgarte}</snm><fnm>T. W.</fnm></au>
    <au><snm>{Shapiro}</snm><fnm>S. L.</fnm></au>
  </aug>
  <source>Phys. Rev. D</source>
  <pubdate>1999</pubdate>
  <volume>59</volume>
  <issue>2</issue>
  <fpage>024007</fpage>
</bibl>

<bibl id="B75">
  <title><p>{Covariant formulations of Baumgarte, Shapiro, Shibata, and
  Nakamura and the standard gauge}</p></title>
  <aug>
    <au><snm>Brown</snm><fnm>DJ</fnm></au>
  </aug>
  <source>Phys. Rev. D</source>
  <pubdate>2009</pubdate>
  <volume>79</volume>
  <issue>10</issue>
  <fpage>104029</fpage>
</bibl>

<bibl id="B76">
  <title><p>{Conformal and covariant formulation of the Z4 system with
  constraint-violation damping}</p></title>
  <aug>
    <au><snm>{Alic}</snm><fnm>D.</fnm></au>
    <au><snm>{Bona-Casas}</snm><fnm>C.</fnm></au>
    <au><snm>{Bona}</snm><fnm>C.</fnm></au>
    <au><snm>{Rezzolla}</snm><fnm>L.</fnm></au>
    <au><snm>{Palenzuela}</snm><fnm>C.</fnm></au>
  </aug>
  <source>Phys. Rev. D</source>
  <pubdate>2012</pubdate>
  <volume>85</volume>
  <issue>6</issue>
  <fpage>064040</fpage>
</bibl>

<bibl id="B77">
  <title><p>Turduckening black holes: an analytical and computational
  study</p></title>
  <aug>
    <au><snm>Brown</snm><fnm>D</fnm></au>
    <au><snm>Diener</snm><fnm>P</fnm></au>
    <au><snm>Sarbach</snm><fnm>O</fnm></au>
    <au><snm>Schnetter</snm><fnm>E</fnm></au>
    <au><snm>Tiglio</snm><fnm>M</fnm></au>
  </aug>
  <source>Phys. Rev. D</source>
  <pubdate>2009</pubdate>
  <volume>79</volume>
  <fpage>044023</fpage>
  <url>http://arxiv.org/abs/0809.3533</url>
</bibl>

<bibl id="B78">
  <title><p>Methods for the approximate solution of time dependent
  problems</p></title>
  <aug>
    <au><snm>Kreiss</snm><fnm>HO</fnm></au>
    <au><snm>Oliger</snm><fnm>J</fnm></au>
  </aug>
  <publisher>Geneva: GARP publication series No. 10</publisher>
  <pubdate>1973</pubdate>
</bibl>

</refgrp>
} 

\listoffigures

\end{backmatter}
\end{document}